\documentclass[a4paper,11pt]{article}
\pdfoutput=1 

\usepackage[T1,tone,safe]{tipa}
\usepackage{xcolor}
\usepackage{xspace}
\usepackage{lmodern}
\usepackage{latexsym,amsmath,array,amssymb,stmaryrd,theorem,epsfig}
\usepackage{bbm}
\usepackage{pifont}
\usepackage{array}
\usepackage{mathtools}
\usepackage{nicefrac}
\usepackage{slashed}
\usepackage{braket}
\allowdisplaybreaks
\usepackage[utf8]{inputenc}
\usepackage{jheppub}

\newcommand{\tr}{\text{tr}}
\newcommand{\cc}[1]{{#1}^*}

\newcommand{\R}{\mathbb{R}}
\newcommand{\C}{\mathbb{C}}
\newcommand{\diag}{\operatorname{diag}}

\newcommand{\Jfrak}{\mathfrak{J}}
\newcommand{\kfrak}{\mathfrak{K}}

\newcommand{\arccot}{\operatorname{arccot}}
\newcommand{\csch}{\operatorname{csch}}

\def \tr {{\rm tr}}
\def \ha {{\frac{1}{2}}}
\def \N {{\mathcal N}}
\def \diag {\operatorname{diag}}

\def \pp {{\rm p}}

\def \LL {{\mathcal{L}}}
\def \DD {{D}}
\def \DDT {{\tilde{\DD}}}

\def \cX {{\cal X}}

\def \cT {{\cal T}}

\def \cY {{\cal Y}}
\def \cS {{\cal S}}

\def \1   {\mathbf{1}}

\def \YY {{\mathcal{Y}}}
\def \OO {{\mathcal{O}}}
\def \D {{\rm{D}}}
\def \cD {{\mathcal{D}}}
\def \Id {{\mathbbm{1}}}

\def \Z {\Omega}
\def \d {\textrm{d}}
\def \p {\partial}

\def \g {\gamma}
\def \G {\Gamma}
\def \Gh {\hat{\Gamma}}
\def \eps {\epsilon}
\def \hh {{\rm h}}

\def \k {\kappa}
\def \td {\tilde}

\def \qt {\td{q}}
\def \sech{\mathrm{sech}}
\def \no {\nonumber}
\def \vp {\varphi}
\def \vt {\vartheta}
\def \th {\theta}
\def \w{\omega}

\def \c2{\cos^2\!\vp}
\def \s2{\sin^2\!\vp}
\def \fc2{\frac{1}{\c2}}
\def \fs2{\frac{1}{\s2}}
\def \sfc2{\tfrac{1}{\c2}}
\def \sfs2{\tfrac{1}{\s2}}
\newcommand{\AdS}{\textup{AdS}\xspace}
\newcommand{\CFT}{\textup{CFT}}
\newcommand{\Sphere}{\textup{S}}
\newcommand{\Torus}{\textup{T}}

\def \AdsST  {$\AdS_3 \times \Sphere^3 \times \Torus^4$\xspace}
\def \AdsSSS{$\AdS_3 \times \Sphere^3 \times \Sphere^3 \times \Sphere^1$\xspace}
\newcommand{\alg}[1]{\mathfrak{#1}}
\newcommand{\grp}[1]{\mathrm{#1}}

\newcommand{\su}{\alg{su}}
\newcommand{\SU}{\grp{SU}}
\newcommand{\algU}{\alg{u}}

\newcommand{\psu}{\alg{psu}}

\newcommand{\rep}[1]{\mathbf{#1}}
\newcommand{\ce}[1]{#1_{\text{c.e.}}}
\newcommand{\gen}[1]{\mathbf{#1}}
\newcommand{\genQ}{\gen{Q}}
\newcommand{\genS}{\gen{S}}
\newcommand{\genH}{\gen{H}}

\newcommand{\genM}{\gen{M}}
\newcommand{\genC}{\gen{C}}
\newcommand{\genCbar}{\overline{\genC}}
\newcommand{\sL}{\mbox{\tiny L}}
\newcommand{\sR}{\mbox{\tiny R}}

\newcommand{\comm}[2]{[#1,#2]}
\newcommand{\acomm}[2]{\{#1,#2\}}
\newcommand{\fixedspaceL}[2]{\mathrlap{#2}\phantom{#1}}

\title{\boldmath Fermion zero modes for the mixed-flux $\AdS_3$ giant magnon}
\author{Adam Varga}
\affiliation{Department of Mathematics, SMCSE,\\ City, University of London}
\emailAdd{Adam.Varga@city.ac.uk}

\abstract{
We explicitly construct the four and two fermion zero modes for the
mixed-flux generalization of the Hofman-Maldacena giant magnon on
two of the AdS$_3$ backgrounds with maximal amount of supersymmetry,
AdS$_3 \times$S$^3 \times$T$^4$ and 
AdS$_3 \times$S$^3 \times$S$^3 \times$S$^1$.
We also show how to  get the $\mathfrak{psu}(1|1)^4$ and 
$\mathfrak{su}(1|1)^2$ superalgebras from the semiclassically quantized fermion
zero modes.
}

\begin{document} 
\maketitle
\flushbottom


\section{Introduction}

In the study of the holographic correspondence \cite{Maldacena:1997re} 
integrability plays a key role, as it allows us to reduce the dynamics of 
interactions to diffractionless two-body scattering of elementary excitations, 
or magnons. In the context of the best understood example, the $\AdS_5 / \CFT_4$ 
duality, integrable structures were found on both the gauge theory 
($\N=4$ super Yang-Mills) \cite{Minahan:2002ve, Beisert:2003tq, Beisert:2003yb} 
and string theory (type IIB superstring theory on $\AdS_5 \times \Sphere^5$) 
sides \cite{Bena:2003wd, Kazakov:2004qf, Arutyunov:2004vx, Beisert:2005bm, Arutyunov:2004yx}.
This shifted focus to the worldsheet S-matrix \cite{Staudacher:2004tk}, which was determined 
(up to an overall phase) by Beisert, using only the $\SU(2|2) \times  \SU(2|2)$ 
symmetry of the theory \cite{Beisert:2005tm, Beisert:2006qh}. The remaining 
phase factor was then calculated using crossing symmetry 
\cite{Janik:2006dc, Beisert:2006ib, Beisert:2006ez, Dorey:2007xn, Volin:2009uv}. 

An important result of Beisert's analysis was the dispersion relation for the magnon
\begin{equation}\label{Intro:MagnonDisp}
\epsilon = \sqrt{ 1 + 4 \hh^2 \sin^2 \frac{\pp}{2} } \ ,
\end{equation}
where
\begin{equation}
\hh = \frac{\sqrt{\lambda}}{2\pi} \ ,
\end{equation}
and $\pp$ is the momentum of the magnon on the worldsheet. Hofman and Maldacena 
explicitly constructed a classical string configuration on $\R \times \Sphere^2$, 
naming it the giant magnon \cite{Hofman:2006xt}, with dispersion relation 
$\epsilon = 2\hh\sin\frac{\pp}{2}$, in agreement with the large coupling limit 
(this is where the string theory approximation is valid) of \eqref{Intro:MagnonDisp}. 
Subsequently this was generalized to the dyonic giant magnon \cite{Chen:2006gea},
living on $\R \times \Sphere^3$, with the exact dispersion relation already at 
the (semi-)classical level.

The giant magnon is a BPS state, and as such, it should be part of a 
16 dimensional short multiplet of $\SU (2 | 2) \times \SU (2 | 2)$ 
\cite{Beisert:2006qh}. Hofman and Maldacena argued that in order to 
reproduce this representation, the giant magnon should have eight fermion 
zero modes \cite{Hofman:2006xt}. This was later explicitly shown by 
Minahan \cite{Minahan:2007gf}, who constructed these zero modes from the 
quadratic fermion fluctuation piece of the Green-Schwarz action, taking 
the giant magnon as the background. Quantizing these modes, he was also 
able to match them to the fermionic generators of the 
$\SU (2 | 2) \times \SU (2 | 2)$ residual algebra.

Remarkably, integrability persists to other, less symmetric classes of AdS/CFT duals. 
There has been significant progress in $\AdS_4/\CFT_3$, for references see 
\cite{Klose:2010ki}, but this paper is concerned with $\AdS_3/\CFT_2$. 
In particular, we will focus on two\footnote{
	There is a third background with 16 supercharges: 
	$\AdS_3 \times \Sphere^3 \times {\rm K}3$, which should also be
	possible to understand using integrable methods, at least in the
	orbifold limit of K3, and it would then be interesting to see what 
	happens when blow-up modes are turned on.
}
$\AdS_3$ backgrounds with maximal supersymmetry (16 supercharges), 
\AdsST and \AdsSSS. Supergravity equations relate the radii of 
$\AdS_3$ and $\Sphere^3$ components,
for \AdsST they give
\begin{equation}
R_{\AdS_3} = R_{\Sphere^3} \ ,
\end{equation}
while for \AdsSSS the $\AdS$ radius $R$ and the radii of the two 3-spheres 
$R_\pm$ must satisfy \cite{Gauntlett:1998kc}
\begin{equation}
\frac{1}{R_+^2} + \frac{1}{R_-^2} = \frac{1}{R^2} \ .
\end{equation}
Historically, these backgrounds were considered in two different settings:
either supported by Ramond-Ramond (R-R) or Neveu–Schwarz-Neveu–Schwarz 
(NS-NS) fluxes.
They were shown to be classically integrable with pure R-R flux 
\cite{Babichenko:2009dk, OhlssonSax:2011ms, Sundin:2012gc}, and in fact 
remain classically integrable even when supported by mixed R-R and NS-NS 
fluxes \cite{Cagnazzo:2012se}
\begin{align} \label{FZM:mixFluxes}
	\begin{split}
		F &\ =\ 2\td{q} \big( \operatorname{Vol}(\AdS_3) 
									+ \cos\vp \operatorname{Vol}(\Sphere^3_+) 
									+ \sin\vp \operatorname{Vol}(\Sphere^3_-) \big) \ ,
		\\[1em] 
		H &\ =\ 2      q  \big( \operatorname{Vol}(\AdS_3) 
									+ \cos\vp \operatorname{Vol}(\Sphere^3_+) 
									+ \sin\vp \operatorname{Vol}(\Sphere^3_-) \big) \ ,
	\end{split}
\end{align}
where the overall factors satisfy $\td{q}=\sqrt{1-q^2}$, and the range is
given by $q \in [0,1]$. 
Recently, it was found that the mixed-flux action is equivalent to the pure NS-NS 
theory with an R-R modulus turned on \cite{OhlssonSax:2018hgc} upon
identifying $q$ and $\qt$ as 
\begin{equation}
q = k \frac{\alpha'}{R^2}\ , 
\qquad
\qt = - g_s c_0 k \frac{\alpha'}{R^2}\ ,
\end{equation}
where $k$ is integral and $c_0$ is continuous. This is different to the conventional
interpretation of the mixed-flux background as the near-horizon limit of bound states
of D1/D5- and F1/NS5-branes carrying R-R and NS-NS charges, respectively.
Signs of integrability on 
the CFT side have also been identified \cite{Sax:2014mea} in the $\CFT_2$ dual 
to strings on \AdsST, however, finding the $\CFT_2$ dual of \AdsSSS string 
theory proved to be a difficult problem \cite{Boonstra:1998yu, Gukov:2004ym,Tong:2014yna, Eberhardt:2018ouy}.

Assuming that integrability holds for the mixed-flux theory at the quantum level, 
Hoare and Tseytlin first calculated the tree-level S-matrix for the massive spectrum of 
$\AdS_3 \times \Sphere^3 \times \Torus^4$ in uniform light-cone gauge \cite{Hoare:2013pma}, 
then, by analysing the constraints of symmetry, they proposed an exact massive worldsheet 
S-matrix \cite{Hoare:2013ida}, generalizing the result of \cite{Borsato:2013qpa} 
to $q \neq 0$. They also found the magnon dispersion relation to be
\begin{equation} \label{Intro:T4Magnon}
\epsilon_{\pm} = \sqrt{M_{\pm}^2 + 4\, \qt^2\, \hh^2 \sin^2 \frac{\pp}{2}} \ .
\end{equation}
The dyonic giant magnon solutions \cite{Chen:2006gea} on $\R \times \Sphere^3$, 
with two angular momenta $(J_1,J_2)$, were lifted to $q \neq 0$ by Hoare, 
Stepanchuk and Tseytlin \cite{Hoare:2013lja}, fixing\footnote{
	The same mixed-flux classical giant magnon solution was also found as
	a limit of rigidly rotating strings \cite{Banerjee:2014gga}, and 
	using the dressing method \cite{Stepanchuk:2014kza}.
}
\begin{equation}
M_\pm = J_2 \pm q \hh \pp \ .
\end{equation} 
Quantization leads to $M_{\pm} = m \pm  q \hh \pp$ with $m=1$, and in fact, 
using symmetry arguments, the dispersion relation was shown to hold to all loops
for both massive and massless magnons $m = 1, 0$ in \cite{Lloyd:2014bsa},
where the massless and mixed-mass S-matrices were also determined.
In recent developments, integrable methods were used to derive the protected
spectrum of these $\AdS_3$ backgrounds, proving that the dispersion 
relations above receive no corrections to all orders in the sting tension 
\cite{Borsato:2016kbm, Baggio:2017kza}. The same protected spectrum in the 
\AdsSSS background was derived independently using supergravity and WZW 
methods in \cite{Eberhardt:2017fsi}. In the case of the \AdsST background 
the protected spectrum agrees with the older results of \cite{deBoer:1998kjm}.

This paper is concerned with the fermion fluctuations around the $\AdS_3$ 
giant magnon. The residual (off-shell) symmetry algebra of the ground state of  
\AdsST superstring theory is the centrally extended $\psu(1|1)^4$ superalgebra 
\cite{Borsato:2013qpa, Borsato:2014hja, Borsato:2014exa, Lloyd:2014bsa},
while on \AdsSSS the elementary excitations transform under the centrally 
extended $\su(1|1)^2$ algebra \cite{Borsato:2012ud, Borsato:2015mma}. 
Analogously to the case of $\AdS_5 \times \Sphere^5$, 
the giant magnon is a BPS state, and should be part of the 4 and 2 dimensional 
short multiplets of $\psu(1|1)^4$ and $\su(1|1)^2$, respectively. 
To reproduce these representations, the mixed-flux giant magnon on 
\AdsST and  \AdsSSS should have 4 and 2 fermion zero modes, respectively,
and our main objective is to find these zero modes, following the calculation of
Minahan \cite{Minahan:2007gf}. The rest of this paper is structured as follows.

In section \ref{FZM:SecBosonic} we present the two-charge giant magnon
on \AdsST with mixed flux, found by Hoare, Stepanchuk and Tseytlin 
\cite{Hoare:2013lja}, 
and describe a one-parameter family generalizing the Hofman-Maldacena 
magnon of $q=0$, that we call \textit{stationary}.
The reason Minahan \cite{Minahan:2007gf} managed to find zero modes
relatively easily is that he took the HM magnon, rather than the more 
general dyonic magnon, as starting point. Similarly, the stationary magnon
is the bosonic background that will make subsequent calculations most simple.
Finally, we outline how the magnon can be put on \AdsSSS.

In section \ref{FZM:SecFzmEom} we discuss the quadratic fermionic action, 
which is obtained from the GS action by considering perturbations 
around the giant magnon as background. We will look at the zero mode condition 
and kappa-gauge fixing, before arriving at the zero mode equations of motion. 
These equations are then solved in section \ref{FZM:SecMixedFlux}, to 
get the expected number of normalizable zero modes. After semiclassical quantization, we
construct the fermionic generators of the corresponding superalgebras.

In section \ref{FZM:SecQ1} we consider the special case of $q=1$. 
In agreement with the chiral nature of the background,
we find that all of the zero modes are non-normalizable. Since the notion 
of stationary magnon breaks down, we cannot simply take the $q \to 1$ 
limit of the zero modes found for $q<1$, and the issue of semiclassical
quantization also needs further attention. This is a question we hope
to return to in the future. We conclude in section \ref{FZM:SecConclusion}
and present some of the more technical details in appendices.

\section{Bosonic solution}
\label{FZM:SecBosonic}

In the case of the \AdsSSS background, choosing unit radius for $\AdS$, the 
supergravity equations allow a one-parameter family of radii for the the 
two spheres $\Sphere^3_\pm$:
\begin{equation}\label{FZM:radii}
	\frac{1}{R_{+}^2} = \alpha \equiv \c2 \ , 
	\qquad  
	\frac{1}{R_{-}^2} = 1- \alpha \equiv \s2 \ ,
\end{equation}
and in fact this covers \AdsST as well, for $\vp = 0$.
Another parameter of the theory is $q \in [0,1]$, describing the amount 
of NS-NS background flux, or equivalently, the R-R modulus in the pure 
NS-NS theory via $\td{q}=\sqrt{1-q^2}$, as described below \eqref{FZM:mixFluxes}.

Most of this section is a summary of the work done by Hoare, Stepanchuk and 
Tseytlin on the mixed-flux two-charge giant magnon on $\R \times\! \Sphere^3$ 
\cite{Hoare:2013lja}. We make the contribution of pointing out that a certain 
restriction of their solution can be regarded as the mixed-flux generalization of the 
Hofman-Maldacena magnon, which will enable us to identify the fermion 
zero modes on the the $\AdS_3$ backgrounds. Furthermore, we describe the 
corresponding solution on \AdsSSS.

\subsection{Strings on \texorpdfstring{$\R \times\! \Sphere^3$}{R x S3} with mixed flux}

Using Hopf coordinates for the 3-sphere (see appendix \ref{FZM:AppAds3S3})
\begin{equation}\label{FZM:Hopf}
Z_1 = \sin\th \ e^{i\phi_1}, 
\qquad
Z_2 = \cos\th \ e^{i\phi_2},
\end{equation}
the $\R \times\! \Sphere^3$ bosonic string action in static conformal gauge, 
which sets the target-space time (coordinate on $\R$) proportional to the 
worldsheet time
\begin{equation}
t = \k \tau \ ,
\end{equation}
is given by
\begin{align} \label{FZM:HopfActionS3}
\begin{split}
S_1[\Z] = -\frac{\hh}{2} \int_{\mathcal{M}} \d^2 \sigma \Big[ \
		& \p_a\th\p^a\th +  \sin^2\!\th\ \p_a\phi_1 \p^a\phi_1 + \cos^2\!\th\ \p_a\phi_2 \p^a\phi_2
\\
		& \qquad + q(\cos 2\th + c )(\dot{\phi}_1 \phi_2' - \dot{\phi}_2 \phi_1')
	\Big] \ ,
\end{split}
\end{align}
where $a = 0,1$ correspond to $\tau, \sigma$, $\dot{}=\partial_\tau$, 
$'=\partial_\sigma$, the worldsheet metric is $\eta=\diag(-1,1)$,
$\hh$ is the string tension, and we represent maps from the worldsheet 
to  $\Sphere^3$ by $\Z = (\th, \phi_1, \phi_2)$.
The expression proportional to $q$ represents the Wess-Zumino term, and the 
parameter $c$ was introduced by Hoare et al. \cite{Hoare:2013lja}. This $c$-term 
is a total derivative and drops out of the equations of motion, but will affect the 
conserved charges for string solutions with non-periodic boundary conditions,
e.g. the dyonic giant magnon.

From the action it is an easy exercise to derive the equations of motion
\begin{align}\label{FZM:HopfBosEom}
\begin{split}
& \ddot\th  - \th''  - \sin\th\cos\th \left( 
		\dot{\phi}_1^2 - \phi_1'^2 - \dot{\phi}_2^2 + \phi_2'^2 
		+2 q \left( \dot{\phi}_1 \phi_2' - \dot{\phi}_2 \phi_1' \right)
		\right)  \ = \ 0 \ ,
\\[1em]
& \sin^2\!\th \left( \ddot{\phi}_1  - \phi_1'' \right) 
		+2 \sin\th \cos\th \left( \dot\th \dot{\phi}_1 - \th' \phi_1' 
										+q  \left( \th' \dot{\phi}_2 - \dot{\th} \phi_2' \right) 
		\right)  \ = \ 0 \ ,
\\[1em]
& \cos^2\!\th \left( \ddot{\phi}_2  - \phi_2'' \right) 
		- 2 \sin\th \cos\th \left( \dot\th \dot{\phi}_2 - \th' \phi_2' 
										+q  \left( \th' \dot{\phi}_1 - \dot{\th} \phi_1' \right) 
		\right)  \ = \ 0 \ ,
\end{split}
\end{align}
which need to be supplemented with the conformal gauge Virasoro constraints 
\begin{align}\label{FZM:VirasoroS3}
\begin{split}
&V_1[\Z] \equiv \dot{\th}^2 +\th'^2 + \sin^2\!\th(\dot{\phi}_1^2+ \phi_1'^2) 
					+ \cos^2\!\th(\dot{\phi}_2^2+\phi_2'^2)  = \k^2 \ ,
\\[1em]
&V_2[\Z] \equiv \dot{\th} \th'+\sin^2\!\th\ \dot{\phi}_1 \phi_1' 
					+ \cos^2\!\th\ \dot{\phi}_2 \phi_2' = 0 \ .
\end{split}
\end{align}

\paragraph*{Conserved charges.}
Classical string solutions on $\R \times\! \Sphere^3$ will have a number 
of conserved Noether charges, and the ones of particular interest to us are the 
spacetime energy $E$ (due to translational invariance in \AdS time $t$)
and the angular momenta $J_1$ and $J_2$ (due to invariance under shifts
in $\phi_1$ and $\phi_2$)
\begin{align} \label{FZM:S3Charges}
\begin{split}
	E     &= 2\pi\hh\k \ ,
\\[1em]
	J_1[\Z] &= \hh\int_{-\pi}^{\pi} \d\sigma\big[ \sin^2\!\th\ \dot{\phi}_1 
					- \frac{q}{2}(\cos 2\th + c )\phi_2' \big] \ ,
\\[1em]
	J_2[\Z] &= \hh\int_{-\pi}^{\pi} \d\sigma\big[ \cos^2\!\th\ \dot{\phi}_2 
					+ \frac{q}{2}(\cos 2\th + c )\phi_1' \big] \ . 
\\[1em]
\end{split}
\end{align}

\subsection{The \texorpdfstring{$\SU (2)$}{SU(2)} principal chiral model}
\label{FZM:SubSecPCM}

The conformal gauge string action on $\R \!\times\! \Sphere^3$ is equivalent 
to that of the \emph{principal chiral model} (PCM) with a Wess-Zumino term
(proportional to $q \in [0,1]$) and underlying group $\SU(2)$, which has a 
group manifold diffeomorphic to $\Sphere^3$. In terms of the left currents 
$\Jfrak = g^{-1} \d g$, where $g \in \SU (2)$, the action is given by
\begin{equation}
S = -\frac{\hh}{2} \Big[ \int_{\mathcal{M}} \d^2 \sigma\,\tfrac{1}{2}\tr(\Jfrak_+\Jfrak_-) 
					- q\int_{\mathcal{B}}\d^3\sigma\,\tfrac{1}{3}\varepsilon^{abc}\tr(\Jfrak_a\Jfrak_b\Jfrak_c) \Big]\ ,
\qquad \Jfrak_a = g^{-1}\p_a g\ ,
\end{equation}
where $\mathcal{M}$ is the decompactified string worldsheet, $\mathcal{B}$ 
is a 3d manifold with boundary $\mathcal{M}$, and 
$\sigma^\pm = \ha (\tau \pm \sigma)$. Using the parametrization
\begin{equation}\label{FZM:SU2explicit}
g = \left(\begin{array}{cc}
Z_1 & Z_2 \\ -\cc{Z}_2 & \cc{Z}_1
\end{array}\right) \in SU(2)\ ,
\end{equation}
and substituting \eqref{FZM:Hopf} we get the action \eqref{FZM:HopfActionS3}.
Also introducing the right current $\kfrak = \d g g^{-1}$, the PCM equations
of motion can be written in the two equivalent forms
\begin{align}\label{FZM:PCMEom}
\begin{split}
	(1+q)\p_- \Jfrak_+ + (1-q)\p_+ \Jfrak_- &= 0\ ,
\\[1em]
	(1-q)\p_- \kfrak_+ + (1+q)\p_+ \kfrak_- &= 0\ .
\end{split}
\end{align}

From the action we get the left-invariant and right-invariant conserved
$\SU (2)$ currents
\begin{equation}
L_a = \Jfrak_a - q \epsilon_{ab} \Jfrak^b\ ,
\quad
R_a = \kfrak_a + q \epsilon_{ab} \kfrak^b\ ,
\quad
\p_a L^a = \p_a R^a = 0 \ ,
\end{equation}
which give rise to the conserved charges
\begin{equation}
Q_{L} = \hh \int \d\sigma \left( \Jfrak_0 + q \Jfrak_1 \right)\ ,
\quad
Q_{R} = \hh \int \d\sigma \left( \kfrak_0 - q \kfrak_1 \right)\ .
\end{equation}
From these we can define the following pair of scalar charges, 
of particular interest for the case of the giant magnon
\begin{equation}
J = - \frac{i}{4} \left( \tr\left[ Q_L \cdot \sigma_3 \right] 
								 + \tr\left[ Q_R \cdot \sigma_3\right]\right)\ ,
\quad
M = - \frac{i}{4} \left( -\tr\left[ Q_L \cdot \sigma_3 \right] 
								 + \tr\left[ Q_R \cdot \sigma_3\right]\right)\ .
\end{equation}
Substituting in the Hopf parametrization \eqref{FZM:Hopf} we get
\begin{align} \label{FZM:SU2Charges}
\begin{split}
	J &= \hh\int_{-\pi}^{\pi} \d\sigma\big[ \sin^2\!\th\ \dot{\phi}_1 
					- \frac{q}{2}(\cos 2\th + 1 )\phi_2' \big] \ ,
\\[1em]
	M &= \hh\int_{-\pi}^{\pi} \d\sigma\big[ \cos^2\!\th\ \dot{\phi}_2 
					+ \frac{q}{2}(\cos 2\th  - 1 )\phi_1' \big] \ . 
\\[1em]
\end{split}
\end{align}
Comparing these $\SU (2)$ charges to those in \eqref{FZM:S3Charges} we have
\begin{equation}
J_1 = J - \frac{c-1}{2}\hh q \Delta\phi_2\ ,
\quad
J_2 = M + \frac{c+1}{2}\hh q \Delta\phi_1\ ,
\end{equation}
where $\Delta\phi_i = \phi_i (\pi)- \phi_i (-\pi)$. We see that non-zero
boundary twists $\Delta\phi_i$ break the $\SU (2)$ symmetry, there is no 
choice of $c$ for which $J, M$ are obtained as Noether charges of the 
local action \eqref{FZM:HopfActionS3}.

\subsection{Two-charge giant magnon on \texorpdfstring{$\R \times\! \Sphere^3$}{R x S3}}

The giant magnon is a solution in the Hofman-Maldacena limit \cite{Hofman:2006xt},
where $E$ and $J_1$ are taken to infinity---i.e. $\k \to \infty$---with their difference 
held fixed (thus finite)
\begin{equation}\label{HMlimit}
E,\ J_1 \to \infty \ , \qquad \qquad
 E- J_1, \ J_2 \ = \ \textrm{fixed}\ .
\end{equation}
We can decompactify the worldsheet by rescaling\footnote{
	Slightly abusing notation, in static gauge the target-space 
	time is functionally the same as our rescaled $\tau$.}
\begin{equation}\label{decompactify}
x = \k \sigma\ , \ t = \k \tau \ , \qquad \qquad \k \to \infty \ , \ x \in ( -\infty, +\infty)\ ,
\end{equation}
which essentially describes an open string with non-trivial boundary conditions.
With this, the giant magnon on \AdsST with mixed flux \cite{Hoare:2013lja} is given by
\begin{align}\label{FZM:HoareMagnon}
	\begin{split}
		Z_1 &= \frac{e^{i t} \left[ b + i \, \tanh U \right]}{\sqrt{1+b^2}}\ ,
		\qquad
		Z_2 = \frac{e^{i V} \, \sech{U}}{\sqrt{1+b^2}}\ ,
	\end{split}
\end{align}
where, with $\g = (1-u^2)^{-1/2}$,
\begin{align}
	\begin{split}
		U 		&= \cos\rho\ \qt\, \g ( x - u t) ,
		\\[0.5em]
		V 		&=  \sin\rho\ \qt\, \g ( t - u x) - q x ,
	\end{split}
\end{align}
and the parameters\footnote{
	Throughout \cite{Hoare:2013lja} another parameter $v$ is used,
	which is the magnon speed only in the $q=0$ limit, and related to 
	our parameter $u$ by the relativistic boost $u = \frac{v-q}{1-vq}$.
} are related via
\begin{equation}\label{magnon:pWithRhoU}
b  =  \qt\, \g u\, \sec\rho + q \tan\rho .
\end{equation}

Hoare et al. also found that the worldsheet momentum of the magnon
is related to the opening angle between the two endpoints on the equator 
$\pp = \Delta \phi_1$. Looking at \eqref{FZM:HoareMagnon} this is
\begin{equation}
\pp = 2 \arccot b\ .
\end{equation}
Furthermore, the boundary term in the local action \eqref{FZM:HopfActionS3}
was fixed to be $c=1$ to keep the difference $E-J_1$ finite, in particular
\begin{align}
	\begin{split}
		E - J_1 &=2 \hh \qt \g \sec\rho \, \sin^2\tfrac{\pp}{2} ,
		\\[1em]
		     J_2 &= M + \hh q \pp , \qquad M = 2 \hh \sin^2\tfrac{\pp}{2} \left( \tan\rho - q \cot\tfrac{\pp}{2} \right) .
	\end{split}
\end{align}
Using \eqref{magnon:pWithRhoU}, it is easy to see that the magnon 
satisfies the dispersion relation
\begin{equation} \label{magnon:dispersion}
	E - J_1 = \sqrt{\left(J_2 - \hh q \pp \right)^2 + 4 \hh^2 \qt^2\, \sin^2\tfrac{\pp}{2}} .
\end{equation}

\subsection{Stationary magnon on \texorpdfstring{$\R \times\! \Sphere^3$}{R x S3}}

The parameter $u$ can be regarded as the velocity of the magnon, and 
the boosted worldsheet coordinates
\begin{equation}\label{FZM:qCoords}
 \cX = \g (x - u t) , \qquad \cT = \g (t - u x)
\end{equation}
naturally appear in the solution \eqref{FZM:HoareMagnon}. In the next section 
we want to obtain the fermion zero modes of the giant magnon. These are,
in some loose sense, independent of time, but also require the bosonic solution
to be stationary, i.e. have a time-independent shape. In other words, apart 
form the $e^{i t}$ term in $Z_1$, we want the solution to only depend on 
$\cX$. This fixes the value of $\rho$
\begin{equation}\label{magnon:statCond}
\sin\rho = \frac{q u}{\sqrt{1-q^2}\sqrt{1-u^2}} = \frac{\g u q}{\qt} ,
\end{equation}
and the solution becomes
\begin{align}\label{FZM:StationaryMagnon}
	\begin{split}
		Z_1 &= \frac{ e^{i t}
					\left[ b + i \, \tanh\!\left(\g \sqrt{\qt^2-u^2} \cX \right) \right]}
					{\sqrt{1+b^2}},
		\\[1em]
		Z_2 &= \frac{e^{-i q \g \cX}\, \sech\! \left(\g \sqrt{\qt^2-u^2} \cX \right)}
					{\sqrt{1+b^2}},
		\qquad \ \
		b		= \frac{u}{\sqrt{\qt^2-u^2}}.
	\end{split}
\end{align}

It is worth noting that $M=0$ for this choice of the parameter $\rho$,
so the stationary magnon is a natural restriction from the perspective
of the PCM, and the dispersion relation \eqref{magnon:dispersion} takes 
the simpler form
\begin{equation}
	E - J_1 = 2 \hh \qt\, \sin\tfrac{\pp}{2}.
\end{equation}
This is much like the dispersion relation of the single-charge magnon of
Hofman and Maldacena, for which Minahan found the fermion zero modes in 
the $\AdS^5$ geometry \cite{Minahan:2007gf}. This further justifies
taking the stationary magnon as the starting point of the zero-mode analysis.

\subsubsection{Parameter ranges}

Since $u$ is a worldsheet speed, one might expect it to take values in the range 
$(-1,1)$. This is certainly true for the general solution \eqref{FZM:HoareMagnon}, 
but the stationary condition \eqref{magnon:statCond} further restricts
\begin{equation}
\sin^2\rho \leq 1
\quad \Rightarrow \quad
|u| \leq \qt .
\end{equation}
It might look like we are missing some solutions, but in fact there will be a 
stationary magnon for each value of the worldsheet momentum $p$. This 
becomes obvious once we rewrite the stationary condition \eqref{magnon:statCond} 
using \eqref{magnon:pWithRhoU} as
\begin{equation} \label{FZM:upRel}
u = \qt \cos\frac{\pp}{2},
\qquad
\tan\rho = q \cot\frac{\pp}{2}.
\end{equation}

\subsection{Mixed-flux giant magnon on \texorpdfstring{\AdsSSS}{AdS3 x S3 x S3 x S1}}

We can construct the \AdsSSS dyonic magnon using the prescription of appendix 
\ref{FZM:AppRxS3xS3solutions}, by putting the magnon \eqref{FZM:HoareMagnon}
on $\Sphere^3_+$ and the BMN string on $\Sphere^3_-$. In Hopf coordinates for
the two spheres $\Sphere^3_\pm$
\begin{equation}\label{FZM:HopfS3S3}
Z_1^\pm  = \sin\th^\pm \ e^{i\phi_1^\pm} \ , 
\quad 
Z_2^\pm= \cos\th^\pm \ e^{i\phi_2^\pm} \ ,
\end{equation}
the solution is given by
\begin{align}\label{FZM:S3S3DyonicMagnon}
	\begin{split}
		&\th^+ = \arccos \left( \frac{\sech\left[A \cos\rho\ \qt \cX \right]}
											{\sqrt{1+b^2}}  \right) \ ,
		\\[1em]
		&\phi_1^+ = A t  
		+   \arctan\big(b^{-1}\tanh\left[A \cos\rho\ \qt \cX \right]\big)\ ,  
		\qquad
		\phi_2^+ = A \sin\rho\ \qt\, \cT - A q x \ ,
		\\[1em]
		&\th ^- = \frac{\pi}{2} \ , 
			\qquad 	\phi_1^- = B t \ ,
			\qquad 	\phi_2^- =0 \ .
		\\[1em]
		&\g^2= \frac{1}{1-u^2} \ , 
			\qquad 	b=  \qt\, \g u\, \sec\rho + q \tan\rho \ , 
			\qquad 	u \in (0,1) \ ,  
	\end{split}
\end{align}

The parameters $A$ and $B$ determine the angle at which the ends
of the string move in the $(\phi_1^+, \phi_1^-)$ plane, and they 
satisfy the Virasoro constraint
\begin{equation}\label{ABVirasoro}
\frac{A^2}{\c2}  + \frac{B^2}{\s2} = 1 \ .
\end{equation}

\paragraph*{Noether charges.}
The \AdsSSS action is invariant under four different angular shifts
leading to four conserved angular momenta on top of the conserved 
energy, in terms of the $\R\!\times\! \Sphere^3$ charges \eqref{FZM:S3Charges}
\begin{align} \label{FZM:S3S3Charges}
	E  = 2\pi\hh\k \ , 
\quad 
	J^+_{1,2} = \fc2 J_{1,2} [\Z_+] \ , 
\textrm{ and } \quad 
	J^-_{1,2} = \fs2 J_{1,2} [\Z_-] \ .
\end{align}
The physical angular momenta relevant to the giant magnon are
\begin{equation}
	J_1 \equiv A J_1^+ + B J_1^-\ ,
\qquad
	J_2 = J_2^+\ ,
\end{equation}
and with these the giant magnon has
\begin{align}
	\begin{split}
			E - J_1 &= \frac{A}{\c2} 2 \hh \qt \g \sec\rho \, \sin^2\tfrac{\pp}{2} ,
		\\[1em]
		    J_2 &= \fc2\left( M + \hh q \pp \right) , 
		\qquad 
			M = 2 \hh \sin^2\tfrac{\pp}{2} \left( \tan\rho - q \cot\tfrac{\pp}{2} \right) .
	\end{split}
\end{align}
The dispersion relation is therefore
\begin{equation}
	E - J_1 = \frac{A}{\c2} \sqrt{\left(\c2 J_2 - \hh q \pp \right)^2 
												+ 4 \hh^2 \qt^2\, \sin^2\tfrac{\pp}{2}} .
\end{equation}

There are two conclusions to be made. Firstly, to match the correct 
dispersion relation derived from symmetry \cite{Borsato:2015mma}, we 
need to take $A=\c2$, a choice that is also physically motivated
if we recall that the giant magnon is an excitation above the BMN vacuum.
The true vacuum of the theory should preserve maximal amount of 
supersymmetry, and this condition leaves the (up to signs) unique 
choice \cite{Babichenko:2009dk} $A=\c2$, $B=\s2$, which we will 
refer to as \textit{maximally SUSY} solution. Secondly, we see that we have 
found one of the light magnons with mass $m = \c2$. We can get the other 
light magnon of mass $\s2$ by switching the two spheres, but we have 
not been able to find the massless ($m=0$) or heavy ($m=1$) magnons 
with this construction.

\paragraph*{Stationary magnon.}

As the starting point of our fermion zero mode analysis we will take the
maximally SUSY \AdsSSS generalization of the stationary magnon 
\eqref{FZM:StationaryMagnon} given by
\begin{align}\label{FZM:s3s3StationaryMagnon}
	\begin{split}
		&\th^+ = \arccos \left( \frac{\sech\cY}{\sqrt{1+b^2}}  \right) \ ,
		\\[1em]
		&\phi_1^+ = \c2\ t  +   \arctan\big(b^{-1}\tanh\cY\big)\ ,  
		\qquad
		\phi_2^+ = - \frac{ q\, \cY}{\sqrt{\qt^2-u^2}} \ ,
		\\[1em]
		&\th ^- = \frac{\pi}{2} \ , 
			\qquad 	\phi_1^- = \s2\ t \ ,
			\qquad 	\phi_2^- =0 \ .
		\\[1em]
		&\g^2= \frac{1}{1-u^2} \ , 
			\qquad 	b=  \frac{u}{\sqrt{\qt^2-u^2}} \ , 
			\qquad 	u \in (-\qt,\qt ) \ ,  
	\end{split}
\end{align}
where we further defined the scaled and boosted worldsheet coordinate
\begin{equation}\label{FZM:magnonYcoord}
	\cY = \c2\ \g \sqrt{\qt^2-u^2} \cX .
\end{equation}

\section{Fermion zero mode equations}
\label{FZM:SecFzmEom}

In this section we look at the equations of motion describing fermion 
perturbations around the stationary giant magnon 
\eqref{FZM:s3s3StationaryMagnon}. Note that this treatment includes both the
\AdsSSS and \AdsST (for $\vp = 0$) cases. We explain what is meant by zero 
modes, and describe in some detail the fixing of fermionic kappa-gauge.
Finally, we write down the zero mode equations for kappa-fixed spinors, 
that will be solved in the next section.

\subsection{Fermionic equations of motion}

The quadratic fermionic action in conformal gauge is given by 
\cite{Cvetic:1999zs}
\begin{equation}\label{FZM:FermLagr}
	S_{\text{F}} = \hh \int \d^2\sigma\ \LL_{\text{F}}\ ,
\qquad
	\LL_{\text{F}}= -i\left(\eta^{ab}\delta^{IJ} 
											+  \eps^{ab}\sigma_3^{IJ}\right)\;
 								\bar{\vt}^I\rho_a\cD_b\,\vt^J \ , 
\end{equation}
where $I,J=1,2$, the $\vt^I$ are ten-dimensional Majorana-Weyl spinors, 
and $ \rho_a$ are projections of the ten-dimensional Dirac matrices\footnote{
	The matrices $\rho_a$ are not to be confused with the parameter $\rho$
	of the dyonic giant magnon \eqref{FZM:HoareMagnon}, which has been fixed
	for the stationary magnon, and will not appear in this section.
}
\begin{equation}
	\rho_a \equiv e_a^A\, \G_A \ , 
	\qquad 
	e_a^A\equiv \p_a X^\mu E_\mu ^A (X) \ .
\end{equation}
$X^\mu$ are the coordinates of the target spacetime  \AdsSSS and will be 
evaluated on the classical solution \eqref{FZM:s3s3StationaryMagnon}.
The giant magnon solution has non-constant components for 
$\mu=t,\th^+,\phi_1^+,\phi_2^+,\phi_1^-$ corresponding to the tangent 
space components $A=0,3,4,5,7$ respectively. The covariant derivative 
is given by
\begin{equation}
	\cD_a\vt^I =  \big( \D_a \delta^{IJ} 
									+ \frac{1}{48}\slashed{F}\rho_a \sigma_1^{IJ} 
									+ \frac{1}{8}\slashed{H}_a \sigma_3^{IJ} 
							\big) \ \vt^J \ ,
\end{equation}
where $\D_a =\p_a+\frac{1}{4}\omega_a^{AB}\G_{AB}$ with the 
pullback of the spin connection 
$\omega_a^{AB}\equiv \p_a  X^\mu  \omega_\mu^{AB}$. 
For a detailed review of the vielbein and spin connection the reader is 
referred to appendix \ref{FZM:AppAds3S3}, while explicit expressions for 
the pullbacks $e_a^A$, $\omega_a^{AB}$ can be found in appendix \ref{FZM:AppPullBacks}.

The tangent space components of the fluxes (\ref{FZM:mixFluxes}) are 
given by
\begin{align}
	F_{012}   & = 2 \td{q} \ , \qquad 
	F_{345}= 2 \td{q} \cos\vp \ , \qquad 
	F_{678}= 2 \td{q} \sin\vp \ ,
	\\[0.5em]
	H_{012}   & = 2 q \ , \qquad 
	H_{345}= 2 q \cos\vp \ , \qquad 
	H_{678}= 2 q \sin\vp \ ,
\end{align}
and they appear in the action as $\slashed{F} \equiv F_{ABC} \G^{ABC}$, 
$\slashed{H}_a \equiv e_a^A H_{ABC} \G^{BC}$. Introducing
\begin{align}
	\G_* & \equiv \quad \G^{012} \ , \qquad 
	(\G_*)^2 = \quad \Id \ ,
	\\[0.5em]
	\G_+ & \equiv \quad \G^{345} \ , \qquad 
	(\G_+)^2 = -\Id \ ,
	\\[0.5em]
	\G_- & \equiv \quad \G^{678} \ , \qquad 
	(\G_-)^2 = -\Id \ ,
\end{align}
the contractions of the fluxes with the Dirac matrices are
\begin{align}\label{FZM:slashes}
	\begin{split}
	\slashed{F} &=  12 \td{q} \ \big(\G_* + \cos\vp\ \G_+ 
															  	+ \sin\vp\  \G_-   \big) \ ,
	\\[1em]
	\slashed{H} &=  12       q  \ \big(\G_* + \cos\vp\ \G_+  
															  	+ \sin\vp\ \G_-    \big)  \ .
	\end{split}
\end{align}
 
The equations of motion derived from \eqref{FZM:FermLagr} are
\begin{align}
\begin{split}
(\rho_0 + \rho_1)(\cD_0 - \cD_1)\ \vt^1    &=   0 \ ,
\\[1em]
(\rho_0 - \rho_1)(\cD_0 + \cD_1)\ \vt^2    &=   0 \ .
\end{split}
\end{align} 
After expanding the covariant derivatives $\cD_a$ we get
\begin{align}\label{FZM:EoMexpanded}
	\begin{split}
		(\rho_0 + \rho_1) \left[  (\D_1 - \D_0) \ \vt^1
					- \frac{1}{48}\slashed{F}(\rho_0 - \rho_1)\ \vt^2 
					- \frac{1}{8}(\slashed{H}_0 - \slashed{H}_1) \ \vt^1 
					\right] &= 0 \ ,
		\\[1em]
		(\rho_0 - \rho_1) \left[  (\D_1 + \D_0) \ \vt^2  
					+ \frac{1}{48}\slashed{F}(\rho_0 + \rho_1)\ \vt^1
					- \frac{1}{8}(\slashed{H}_0 + \slashed{H}_1) \ \vt^2 
					\right] &= 0 \ .
	\end{split}
\end{align}

At this point it is natural to change variables to the scaled and boosted 
worldsheet coordinates \eqref{FZM:magnonYcoord}
\begin{equation}\label{FZM:boostedCoords}
	\cY = \c2\, \zeta \cX, \quad \cS = \c2\, \zeta \cT, 
	\qquad \zeta =  \g \sqrt{\qt^2-u^2} ,
\end{equation} 
satisfying
\begin{equation}
	\p_1 \mp \p_0 = \c2\, \zeta  (1 \pm u) \g (\p_\cY \mp \p_\cS).
\end{equation}
With this, the equations become
\begin{align}\label{FZM:EomNoZeroMode}
	\begin{split}
		(\rho_0 + \rho_1) \left[ \zeta (1+ u) \g  
												\big( \DD  - \p_\cS \big) \ \vt^1 
												+ \OO \vt^2  \right] &= 0 \ ,
		\\[1em]
		(\rho_0 - \rho_1) \left[ \zeta (1- u) \g  
												\big( \DDT + \p_\cS \big) \ \vt^2 
												+ \td{\OO} \vt^1 \right] &= 0 \ ,
	\end{split}
\end{align} 
where \vspace{-0.5em}
\begin{align} \label{FZM:OOdef}
\OO          = -  \frac{1}{48 \c2} \slashed{F} (\rho_0 - \rho_1) \ ,
\qquad
\td{\OO}  =     \frac{1}{48 \c2} \slashed{F} (\rho_0 + \rho_1) \ ,
\end{align}
and the fermion derivatives are
\begin{align} \label{FZM:DDdef}
	\begin{split}
		\DD   &=  \p_\cY + \ha G\ \G_{34} + \ha Q\ \G_{35}  
						-  \frac{(1-u) \g}{48 \c2\, \zeta} 
							\left( \slashed{H} (\rho_0 - \rho_1) 
									+ (\rho_0 - \rho_1) \slashed{H} \right)\ ,
	\\[1em] 
		\DDT &=  \p_\cY + \ha \tilde{G}\ \G_{34}  + \ha Q\ \G_{35} 
						-  \frac{(1+u) \g}{48 \c2\, \zeta} 
							\left( \slashed{H} (\rho_0 + \rho_1) 
									+ (\rho_0 + \rho_1) \slashed{H} \right)\ .
	\end{split}					
\end{align} 
A detailed derivation can be found in appendix 
\ref{FZM:AppFermionDerivatives}, together with explicit expressions for the 
scalar functions $G, \tilde{G}, Q$ in \eqref{FZM:ExplGQ}.

The operators in front of the equations \eqref{FZM:EomNoZeroMode} are nilpotent
\begin{equation}\label{FZM:nilpotent}
	(\rho_0 + \rho_1)^2  = (\rho_0  - \rho_1)^2 = 0 \ ,
\end{equation}
since they are evaluated on the classical solution, which satisfies the Virasoro constraints. If we further define 
\begin{equation}\label{FZM:rho0bar}
	\bar{\rho}_0 \equiv e_0^0\, \G_0 - e_0^3\, \G_3 - e_0^4\, \G_4 - e_0^5\, \G_5 + e_0^7\, \G_7 \ ,
\end{equation}
which turns out to be $\bar{\rho}_0 = - \rho_0^{\dagger}$ for 
the gamma matrices described in appendix \ref{FZM:AppDirac},
we get another set of nilpotent operators $(\bar{\rho}_0 + \rho_1)^2 = (\bar{\rho}_0 - \rho_1)^2 = 0$.
However, the two sets differ by the nonsingular operator $\bar{\rho}_0 - \rho_0$, which squares to
\begin{align} \label{barMinusSq}
(\bar{\rho}_0 - \rho_0)^2  &=  4 \c2\, \qt^{-2}\left( \zeta^2 \tanh^2\!\cY + q^2 u^2 \g^2 \right) \ \Id \ .
\end{align}
The kernel of a $2m$-dimensional nilpotent operator is of at least $m$ dimensions since all its eigenvalues are zero. If the sum of two nilpotent operators is full-rank, as above, the kernels must be disjoint, therefore the sum of their nullities is at most the full $2m$. From this we see that the $(\rho_0 \pm \rho_1)$ are half-rank, an important observation for subsection \ref{FZM:subsecKappa}.

\subsection{Zero mode condition}

Note that the fermion Lagrangian \eqref{FZM:FermLagr} has a 
dependence on the worldsheet coordinates only through the vielbein and 
spin connection. These quantities, on the other hand, depend only on $\cY$, 
i.e. the Lagrangian is independent of the temporal coordinate $\cS$
\begin{equation}
	\LL_{\text{F}} 
		= \LL_{\text{F}} \left( \cY , \vt^J, \p_{\td{a}} \,\vt^J \right) \ ,
\end{equation}
where $\td{a}=\td 0, \td 1$ correspond to the variables $\cS$ and $\cY$, 
respectively.

Translations in $\cS$ can be equivalently described as
a transformations of the fields
\begin{equation}
	\delta \vt^J = 	\varepsilon\ \p_{\cS} \vt^J  \ , 
	\qquad  
	\delta (\p_{\td{a}}\,\vt^J )	= 
							\varepsilon\ \p_{\cS} (\p_{\td{a}}\,\vt^J ) \ ,
\end{equation}
and accordingly
\begin{align}
	\delta \LL_{\text{F}}	& = \varepsilon\ 
			\left(  \frac{\p \LL_{\text{F}}}{\p \vt^J} \p_{\cS} \vt^J 
					+ \frac{\p \LL_{\text{F}}}{\p \left( \p_{\td{a}} \,\vt^J \right)} 
						\p_{\cS} (\p_{\td{a}}\,\vt^J ) \right) 
	\\[1em]
					&	= \varepsilon\ \p_{\cS}\LL_{\text{F}} 
						= \varepsilon\ \p_{\td{a}} 
							\left( \delta^{\td{a}}_{\td{0}}\LL_{\text{F}}\right) \ .
\end{align}
The change in the Lagrangian is a total derivative, and applying 
Noether's theorem we get a conserved current
\begin{equation}
	j^{\td{a}} = \frac{\p \LL_{\text{F}}}
								{\p \left( \p_{\td{a}} \,\vt^J \right)}\ \p_{\cS} \, \vt^J 
						-  \delta^{\td{a}}_{\td{0}}\LL_{\text{F}} \ ,
\end{equation}
where summation over $J=1,2$ is understood. However, for the 
fermionic action we have $\LL_{\text{F}}=0$ on-shell, 
and the current simply reduces to
\begin{equation}
	j^{\td{a}} = \frac{\p \LL_{\text{F}}}
								{\p \left( \p_{\td{a}} \,\vt^J \right)} \  
								\p_{\cS} \, \vt^J \ ,
\end{equation}
The explicit form of this current is unimportant for the present argument. 

Since $\cS$ is a time-like worldsheet coordinate, we might interpret the
corresponding conserved quantity as the energy of the fermionic 
perturbation above the giant magnon background
\begin{equation}
	E_{\text{F}}	= \int \d \cX \ j^{\td{0}} 
					 	= \int \d \cX \ \frac{\p \LL_{\text{F}}}
					 									{\p \left( \p_{\cS} \,\vt^J \right)} \  
					 									\p_{\cS} \, \vt^J \ .
\end{equation}
Zero modes, by definition, are zero energy fluctuations above the giant
magnon, i.e. $E_{\text{F}} = 0$. Henceforth, we will take the zero mode 
condition to be
\begin{equation}\label{FZM:ZeroModeCond}
	\p_{\cS} \, \vt^J = 0 \ ,
\end{equation}
and with this, the equations for the fermion zero modes are
\begin{align}\label{FZM:ZeroModeEqn}
	\begin{split}
		(\rho_0 + \rho_1) \left[ \zeta (1+ u) \g   \DD \ \vt^1 
												+ \OO \vt^2  \right] &= 0 \ ,
	\\[1em] 
		(\rho_0 - \rho_1) \left[ \zeta (1- u) \g   \DDT \ \vt^2 
												+ \td{\OO} \vt^1 \right] &= 0 \ .
	\end{split}
\end{align} 

\subsection{Fixing kappa symmetry}
\label{FZM:subsecKappa}

The Green-Schwarz superstring has a local fermionic symmetry, 
the so-called kappa-symmetry, that ensures spacetime supersymmetry of 
the physical spectrum. Let us take another look at the quadratic fermionic Lagrangian 
(\ref{FZM:FermLagr})
\begin{align}
	\LL_{\text{F}} &= \ -i\, \left( \eta^{ab}\delta^{IJ}
													+\eps^{ab}\sigma_3^{IJ}\right)\;
 								   			\bar{\vt}^I  \rho_a  \cD_b\,  \vt^J \ , 
	\\[1.5em]
							&= \ i\, \bar{\vt}^1(\rho_0 + \rho_1)(\cD_0 - \cD_1) \vt^1
						    	+  i\, \bar{\vt}^2(\rho_0 - \rho_1)(\cD_0 + \cD_1) \vt^2 \ ,
	\\[1.5em]
	\begin{split}
							&= \ - i \c2\, \bar{\vt}^1(\rho_0 + \rho_1)
												\Big( \zeta (1+ u) \g \big( \DD  - \p_\cS \big) 
															\vt^1  +  \OO \vt^2   \Big) 
	\\[1em]
							& \qquad  +  i \c2\, \bar{\vt}^2(\rho_0 - \rho_1)
						    					\Big( \zeta (1- u) \g \big( \DDT + \p_\cS \big)  
						    								\vt^2 + \td{\OO} \vt^1  \Big) \ ,					
	\end{split}  
\end{align}
where $\DD, \DDT, \OO$ and $\td{\OO}$ are defined in \eqref{FZM:OOdef}
--\eqref{FZM:DDdef}. We see the nilpotent operators $(\rho_0 \pm \rho_1)$ 
acting on the conjugate spinors: components of $\vt^1$ and $\vt^2$ that 
are projected out by $(\rho_0 + \rho_1)$ and $(\rho_0 - \rho_1)$, respectively, 
do not contribute to the action, we can consider them non-dynamical. 

To fully fix kappa-gauge, however, not only do we need to project out 
non-dynamical degrees of freedom, but also specify what happens to the rest, 
i.e. we need actual projectors:
\begin{align} 
	K_1 =  \frac{1}{2} \Pi ( \rho_0 + \rho_1) \ ,
 	\qquad  
 	K_2 =  \frac{1}{2} \Pi ( \rho_0 -  \rho_1) \ ,
\end{align}
for some invertible $\Pi$, that has to satisfy a number of conditions. 
A straightforward, albeit somewhat cumbersome,\footnote{
	One can easily convince themselves that it is sufficient to check the 
	$\G_{3}$, $\G_{4}$ and $\G_{5}$ components of the operator equations, 
	simplifying matters a great deal.
} calculation gives 
$[\rho_0 + \rho_1 , \DD] = [\rho_0 - \rho_1 , \DDT] = 0$ , so we have
\begin{align} \label{FZM:KappaD}
[   K_1  ,  \DD  ] = 0 \ ,
 \qquad  
[   K_2  ,  \DDT  ] = 0 \ ,
\end{align}
provided $[ \Pi , \DD ] = [ \Pi , \DDT ] = 0 $. Another condition of course, 
is that the $K_J$ have to be genuine projectors --- i.e. $K_J^2 = K_J$ --- , 
which, with \eqref{FZM:OOdef}, would imply that
\begin{align} \label{FZM:KappaO}
\OO          =        \OO  K_2 \ ,
\qquad
\td{\OO}  =  \td{\OO} K_1 \ .
\end{align}

The most obvious choice would be $\Pi = \G^0$, but taking this route 
one encounters technical difficulties when considering the 
\AdsSSS geometry, arising from the appearance of $\G_7$ in 
$(\rho_0 \pm \rho_1)$. Noting that in both of these operators $\G_7$ 
only appears in the combination $\G_0 + \sin\vp\, \G_7$, it is tempting to 
``boost'' our gamma matrices in the 0-7 directions
\begin{equation}\label{FZM:newGamma}
	\Gh^0 = \sec\vp \left(  \G^0 - \sin\vp\ \G^7   \right) \ ,
	\qquad
	\Gh^7 = \sec\vp \left(  \G^7 - \sin\vp\ \G^0   \right) \ ,
\end{equation}
leaving unchanged all the others $\Gh^A = \G^A$, $A\neq 0, 7$. 
One can easily check that these satisfy the Clifford algebra. We lower
the index on $\Gh^A$ with the Minkowski metric, in particular 
$\Gh_0 =  -\Gh^0  = \sec\vp (  \G_0 + \sin\vp\ \G_7 )$ 
soaks up all the $\G_7$ dependence in $(\rho_0 \pm \rho_1)$
\begin{equation}
\rho_0  \pm  \rho_1 	= 	\cos\vp \left(\Gh_0 + \hat{e}_\pm^3\ \Gh_3
															+ \hat{e}_\pm^4\ \Gh_4
															+ \hat{e}_\pm^5\ \Gh_5 \right)
\end{equation}
where $\hat{e}_\pm^A = \sec\vp\, (e_0^A \pm e_1^A  )$. All of this is 
good motivation for the choice of $\Pi =  \sec\vp\, \Gh^0$, which can be 
easily shown to satisfy our conditions. Henceforth, we will take
\begin{align} \label{FZM:KappaProj}
	 K_1 =  \ha \sec\vp\ \Gh^0( \rho_0 + \rho_1) \ ,
 	\qquad  
 	K_2 =  \ha \sec\vp\ \Gh^0( \rho_0 - \rho_1) \ .
\end{align}
The advantages of this choice will become obvious in the next subsection.

If we take a basis of gamma matrices such that $\Gh^A$ have definite
hermiticity, e.g. the one described in appendix \ref{FZM:AppDirac}, 
the projectors are Hermitian $ K_J^\dag = K_J$. Furthermore, in such a 
basis the Hermitian conjugate intertwiner (see app. \ref{FZM:AppDirac}) 
is given by $\Gh^0$, hence the Dirac conjugate is $\bar{\vt} = \vt^\dag\ \Gh^0$. 
With this, and the properties listed above, we can write the Lagrangian as
\begin{align}\label{FZM:KappaFixedLagr}
	\begin{split}
		\LL_{\text{F}}  &= \ - 2 i \cos^3\!\vp\, (\Psi^1)^\dag  \Big( \zeta (1+ u) \g  
													\big(  \DD  - \p_{\cS} \big)\ \Psi^1 
												+ \OO \Psi^2  \Big)
	\\[1em] 
					   	& \qquad  + 2 i \cos^3\!\vp\,  (\Psi^2)^\dag \Big( \zeta (1- u) \g  
					   								\big(  \DDT  +  \p_{\cS} \big)\ \Psi^2 
					   							+ \td{\OO} \Psi^1  \Big) \ ,			
	\end{split}						    					    
\end{align}
where we introduced the notation $\Psi^J = K_J \vt^J$ for the projected
spinors, and we indeed see that only these components are dynamical.

Using the kappa-projectors, the zero mode equations 
\eqref{FZM:ZeroModeEqn} can be written as
\begin{align}
	\begin{split}
		K_1 \left[ \zeta (1+ u)\g \DD\   \vt^1 +       \OO \vt^2  \right] &= 0 \ ,
	\\[1em] 
		K_2 \left[ \zeta (1- u) \g \DDT\ \vt^2 + \td{\OO} \vt^1 \right] &= 0 \ .
	\end{split}
\end{align} 
For the  kappa-fixed spinors $\Psi^J = K_J \vt^J$, using \eqref{FZM:KappaD}
--\eqref{FZM:KappaO}, these equations become
\begin{align}\label{FZM:EomEquivFixed}
	\begin{split}
		\zeta (1+ u)\g \DD\  \Psi^1 + K_1 		\OO \Psi^2 	&= 0 \ ,
	\\[1em] 
		\zeta (1- u)\g \DDT\ \Psi^2 + K_2 \td{\OO} \Psi^1	&= 0 \ .
	\end{split}
\end{align}

\subsection{Zero mode equations}
\label{FZM:subsecDelta0} 

With the choice of kappa projectors \eqref{FZM:KappaProj} we get a 
commuting 6d chirality projector for free\footnote{
	In any spinor operator $M$, replace $\G^A$
	by $\Gh^A$ to get $\hat{M}$.}
\begin{equation}
	P_\pm = \ha \left( \Id   \pm  \Gh_*\Gh_+ \right) \ ,
	\qquad
	[ P_\pm , K_J ] = 0 \ .
\end{equation}
Using this, the contraction of the background fluxes $\slashed{F}$, 
$\slashed{H}$ \eqref{FZM:slashes} can be written as
\begin{align}\no
	\G_* +   \cos\vp\ \G_+  + \sin\vp\ \G_-  
     &=   \cos\vp   \Big( \left( \sec\vp\ \G^0 + \tan\vp\ \G^{1268} \ \G^7 \right)
 										 \G^{12} + \G_+  \Big)
	\\[0.5em] \no
	&=   \cos\vp   \left( \Gh^{0} \G^{12} + \G_+  \right) - 2 \cos\vp\ \Delta\ \G^{12} 
 	\\[0.5em] \label{FZM:slashRewrite}
	&=   2 \cos\vp \left( \Gh_{*}\  P_{+}  -  \Delta\ \Gh^{12} \right) ,													 
\end{align}
where
\begin{equation}\label{FZM:DeltaDef}
\Delta = - \ha \tan\vp \left( \Gh^{1268} + \Id \right) \G^7 
		  \equiv\ \Delta_0\, \Gh^{0} + \Delta_7\, \Gh^{7} \ ,
\end{equation}
with \vspace{-0.5em}
\begin{align}
	\begin{split}
		\Delta_0 =  - \ha \tan^2\!\vp  \left( \Gh^{1268} + \Id \right)\ ,
		\qquad
		\Delta_7 =  \csc\vp\, \Delta_0\ .
	\end{split}
\end{align}
Even though $\Delta_0$ and $\Delta_7$ are matrices, we can essentially 
treat them as scalars, since they commute with the equations of motion.

Recalling $\bar{\rho}_0$ from \eqref{FZM:rho0bar}, which also satisfies 
$\rho_0\, \Gh^0 = \Gh^0\, \bar{\rho}_0 $, we can define an  
invertible operator from \eqref{barMinusSq}
\begin{equation}\label{FZM:R}
	R = \ha \sec\vp\ \Gh_* (\bar{\rho}_0 - \rho_0)      
	\quad  :  \quad      
	R^2 = -  \qt^{-2}\left( \zeta^2 \tanh^2\!\cY + q^2 u^2 \g^2 \right) \ \Id \ .
\end{equation}
With all of this, the fermion derivatives \eqref{FZM:DDdef} can be 
rewritten as (see appendix \ref{FZM:AppFermionDerivatives})
\begin{align} \label{FZM:DDfinal}
\begin{split}
\DD   &=  \p_{\cY} + \ha G\ \Gh_{34} + \ha Q\ \Gh_{35}  
					+ \frac{q (1-u) \g}{\zeta}
						\left(  R P_{-} - (R + \Gh_{12})\,  P_{+}  + \Delta_0\, \Gh_{12} \right)\ ,
\\[1em] 
\DDT &=  \p_{\cY} + \ha \tilde{G}\ \Gh_{34}  + \ha Q\ \Gh_{35} 
					+ \frac{q (1+u) \g}{\zeta} 
						\left(  R P_{-} - (R + \Gh_{12})\,  P_{+}  + \Delta_0\, \Gh_{12}\right)\ ,
\end{split}					
\end{align} 
however, these expressions are only valid when acting on 
kappa-fixed spinors, i.e. in the form $\DD K_1$ and $\DDT K_2$.
As for the terms \eqref{FZM:OOdef} mixing the two spinors 
in the equations of motion, we have
\begin{align}
	\begin{split}
		\OO       \ & =      -     \qt \left(  \Gh^{12} P_{-}   +  \Delta\ \Gh_{*} \right) K_2 \ ,
	\\[1em] 
		\td{\OO}\ & =  \ \ \,  \qt \left(  \Gh^{12} P_{-}   +  \Delta\ \Gh_{*} \right) K_1 \ .
	\end{split}
\end{align}
Using the nilpotency relations $(\rho_0 \pm \rho_1)^2 = 0$, 
it is easy to see that
\begin{align} \label{FZM:KKR}
	\G^{12} K_1 K_2 = - R K_2 \ , \qquad  
	\G^{12} K_2 K_1 = - R K_1 \ ,
\end{align}
and the equations of motion \eqref{FZM:EomEquivFixed} become
\begin{align} \label{FZM:EomRDelta}
	\begin{split}
		\zeta (1+ u)\g \DD\ \Psi^1  
			+ \qt \left ( R\ P_{-}  -  K_1 \Delta\ \Gh_{*} \right) \Psi^2  &= 0 \ ,
	\\[1em] 
		\zeta (1- u)\g  \DDT\ \Psi^2
		 	-  \qt \left( R\ P_{-} -  K_2 \Delta\ \Gh_{*} \right) \Psi^1   &= 0 \ .
\end{split}
\end{align}

\subsubsection{Equations for \texorpdfstring{$\Delta = 0$}{Delta = 0}}
\label{FZM:secDelta0}

Equation \eqref{FZM:slashRewrite} might seem arbitrary at first,
so let us elaborate on the advantages of this rearrangement. Our goal was to have 
$( \Gh_*   +  \Gh_+ )$ --- instead of $\slashed{F}$ --- in the equations, 
since $P_{\pm}$ commutes with $K_J$. After this rewriting we are 
left with an extra term $K \Delta K$, which does not in general commute 
with $P_{\pm}$. However, in the following two cases we have $\Delta = 0$
\begin{itemize}
	\item  $\vp = 0$ : corresponding to the \textbf{\AdsST geometry}.
	\item  ``$\Gh^{1268} =-1$'' : i.e. the \textbf{\AdsSSS geometry, 
		with the fermions restricted to the $-1$ eigenspace of $\Gh^{1268}$}.
		Note that this is compatible with the equations, since $\Gh^{1268}$ 
		commutes with all the terms.
\end{itemize}

Assuming $\Delta = 0$, the fermion derivatives take the simpler form
\begin{align} \label{FZM:DDonDelta0}
\begin{split}
\DD   &=  \p_{\cY} + \ha G\ \Gh_{34} + \ha Q\ \Gh_{35}  
					+ \frac{q (1-u) \g}{\zeta}
						\left(  R P_{-} - (R + \Gh_{12})\,  P_{+} \right)\ ,
\\[1em] 
\DDT &=  \p_{\cY} + \ha \tilde{G}\ \Gh_{34}  + \ha Q\ \Gh_{35} 
					+ \frac{q (1+u) \g}{\zeta} 
						\left(  R P_{-} - (R + \Gh_{12})\,  P_{+} \right)\ .
\end{split}					
\end{align} 
Also note that the equations of motion have no explicit 
dependence on $\vp$, only an implicit one via the rescaled variable 
$\cY$ \eqref{FZM:boostedCoords}. In other words, the following equations apply in both
geometries
\begin{align} \label{FZM:EomUniversalPsi}
	\begin{split}
		\zeta (1+ u)\g  \DD   \ \Psi^1  + \qt\ R\ P_{-} \Psi^2  &= 0 \ ,
	\\[1em] 
		\zeta (1-  u)\g  \DDT \ \Psi^2  -  \qt\ R\ P_{-} \Psi^1  &= 0 \ .
	\end{split}
\end{align}
as long as we impose the extra condition $\Gh^{1268} \vt^J = - \vt^J$ 
in the $\Sphere^1$ geometry.

\subsubsection{The case of \texorpdfstring{$\Delta \neq 0$}{nonzero Delta}}

As we have seen above, we can treat the \AdsSSS fermion zero modes 
in much the same way as those of the \AdsST giant magnon, provided 
$\Delta = 0$. In section \ref{FZM:SecMixedFlux} this will allow us to find 
solutions for both geometries and general values of $q$ in a single 
calculation. However, we need to make sure there are no zero modes
that we are missing by restricting to $\Delta = 0$.

We can get an intuition for why this must be the case by looking at the 
near BMN spectrum of the \AdsSSS superstring. Since in the BMN limit the zero modes become 
the fermion superpartners of the magnon background, they must all 
have the same mass, hence definite chirality under $\Gh_{1268}$, according 
to equation (2.33) in \cite{Borsato:2015mma}\footnote{
	We are grateful to Bogdan Stefanski for pointing out this relation.
}. Given that in the next section we find normalizable solutions for $\Delta = 0$ 
($\Gh_{1268} = -1$), we expect no zero modes for $\Delta \neq  0$ ($\Gh_{1268} = +1$).
In appendix \ref{FZM:AppDeltaNonZero} we show that there are in fact no
normalizable solutions to \eqref{FZM:EomRDelta} for $\Delta \neq 0$.

\section{Mixed-flux fermion zero modes}
\label{FZM:SecMixedFlux}

In this section we find exact solutions for the ($\Delta = 0$) zero mode 
equations \eqref{FZM:EomUniversalPsi}. Our main aim is to write down the 
normalizable solutions, representing the perturbative zero modes over the 
giant magnon background. Using these normalizable zero modes, we then perform semiclassical 
quantization, and reproduce the the algebra that the fermion excitations 
must satisfy. 

\subsection{Fixing kappa-gauge}

We start by noting that the kappa-projectors \eqref{FZM:KappaProj} 
can be written as
\begin{align} \label{FZM:KappaRewrite}
\begin{split}
	K_1			& = \ha \Big(	\Id 
							-  \sin (2 \chi )	\cos\upsilon_+\, \Gh_{03} \, 
							-  \cos (2 \chi )	\cos\upsilon_+\, \Gh_{04} 
							+  					\sin\upsilon_+\, \Gh_{05}
							\Big) \ ,
	\\[1em] 
	K_2			& = \ha \Big( 	\Id 
							+  \sin(2 \td{\chi} )	\cos\upsilon_-\,	\Gh_{03} \, 
							+  \cos(2\td{\chi} )	\cos\upsilon_-\, \Gh_{04} 
							-  						\sin\upsilon_-\, \Gh_{05}
							\Big) \ ,
\end{split}
\end{align}
where
\begin{align} \label{FZM:logPhasesq}
	\begin{split}
		\chi (\cY)	    & = \ha \left( 
								\arccot \left( \frac{u \csch \cY}{\qt} \right) 
							- 	\arcsin \left( \frac{\tanh\cY}
													{\sqrt{1-Q_+^2\, \sech^2\!\cY}}
							\right) \right) \ ,
	\\[1em] 
		\td{\chi} (\cY)	& = \ha \left( 
								\arccot \left( \frac{u \csch \cY}{\qt} \right) 
							+ 	\arcsin \left( \frac{\tanh\cY}
													{\sqrt{1-Q_-^2\, \sech^2\!\cY}}
							\right) \right) \ ,
	\end{split}
\end{align}
and we also introduced
\begin{equation}
	Q_\pm = \frac{q\sqrt{\qt^2-u^2}}{\qt (1\pm u)} , 
	\qquad
	\upsilon_\pm = \arcsin\left( Q_\pm\ \sech\cY \right) \ .
\end{equation}

\paragraph*{Ansatz.}

Since $K_J$, $\Gh_{12}$ and $\Gh_* \Gh_+$ all mutually commute, as our 
starting point we can take shared eigenvectors $U$
\begin{equation}
\Gh_{12} U^{J} = \lambda_{12} U^{J} \ , 
\qquad
\Gh_* \Gh_+ U^{J} = \lambda_P U^{J} \ ,
\end{equation}
where $\lambda_{12} = \pm i$, and $\lambda_P = \pm 1$ correspond to the
$P_{\pm}$ projections. Accordingly, there are no restrictions on these eigenvalues 
for the kappa-fixed spinor. The operator $\Gh_{34}$ does not commute with $K_J$, 
hence a suitable combination of its opposite eigenvectors makes a good candidate 
for the general gauge-fixed spinor. This motivates the further restriction of 
$\Gh_{34} U^{J} = i U^{J}$ and the ansatz
\begin{equation}\label{FZM:KappaFixedAnsatz}
\Psi^{J} = \left( \alpha^{J}_{+}(\cY) + \alpha^{J}_{-}(\cY)\, \Gh_{45} \right) U^{J}
\end{equation}

\paragraph*{Solution.}

Substituting this into the equations $K_1 \Psi^1 = \Psi^1$, and using the various 
eigenvector relations of $U^1$, we get one equation for each eigenspace of $\Gh_{34}$
\begin{align}
	\begin{split}
		\lambda e^{2 i \chi}\cos\upsilon_{+}\, \alpha^{1}_{-} - 
			\lambda \sin\upsilon_{+}\, \alpha^{1}_{+} 
			 & = \alpha^{1}_{+} \ ,
		\\[0.5em]
		 \lambda e^{-2 i \chi}\cos\upsilon_{+}\, \alpha^{1}_{+} +
		 	\lambda \sin\upsilon_{+}\, \alpha^{1}_{-} 
		 	 & = \alpha^{1}_{-} \ ,
	\end{split}
\end{align}
where $\lambda = i \lambda_{12} \lambda_{P} = \pm 1$. What we have here
are two equations for the single variable $\alpha_{-}/\alpha_{+}$, corresponding to the 
fact that the norm of the eigenvector is not fixed. The equations are consistent, and a symmetric 
solution is given by
\begin{equation}\label{FZM:Alphas1}
	\alpha^{1}_{+} = e^{i \chi}          \sqrt{1 - \lambda Q_{+}\, \sech\cY} \ , 
	\qquad
	\alpha^{1}_{-} = e^{-i \chi} \lambda \sqrt{1 + \lambda Q_{+}\, \sech\cY} \ .
\end{equation}
A similar calculation gives
\begin{equation}\label{FZM:Alphas2}
	\alpha^{2}_{+} =    e^{i \td{\chi}}         \sqrt{1 + \lambda Q_{-}\, \sech\cY} \ , 
	\qquad
	\alpha^{2}_{-} = - e^{-i \td{\chi}} \lambda \sqrt{1 - \lambda Q_{-}\, \sech\cY} \ .
\end{equation}
Written in a single expression, the most general gauge-fixed spinors are
\begin{align}\label{FZM:KappaFixed}
	\begin{split}
		\Psi^1 = \sum_{\lambda = \pm} 
				\left( e^{i \chi}          \sqrt{1 - \lambda Q_{+}\, \sech\cY} + 
					   e^{-i \chi} \lambda \sqrt{1 + \lambda Q_{+}\, \sech\cY}\, \Gh_{45}\right)  
				U^{1}_{\lambda} \ ,
		\\[0.5em]
		\Psi^2 = \sum_{\lambda = \pm} 
				\left( e^{i \td{\chi}}         \sqrt{1 + \lambda Q_{-}\, \sech\cY} - 
					   e^{-i \td{\chi}} \lambda \sqrt{1 - \lambda Q_{-}\, \sech\cY}\, \Gh_{45}\right)  
				U^{2}_{\lambda} \ ,
	\end{split}
\end{align}
where $\Gh_{34} U^{J}_\pm = +i U^{J}_\pm$ and 
$i \Gh_{12} \Gh_{*} \Gh_{+} U^{J}_\pm = i \Gh_{0345} U^{J}_\pm = \pm U^{J}_\pm$. 
The above analysis shows that these are kappa-fixed eigenvectors, and by counting 
the degrees of freedom (components of $U^{J}$) we see that there are no others.

\subsection{Zero mode solutions}

The projectors $P_\pm$ commute with the equations of motion
\eqref{FZM:EomUniversalPsi}, therefore we can consider solutions 
of definite $P_{\pm}$ ``chirality''. In the following we obtain solutions 
on the two subspaces in turn,  by letting $U^{J}_\pm$ depend on 
$\cY$, and substituting \eqref{FZM:KappaFixed} into the equations. 
The identities listed in appendix \ref{FZM:AppPhase} were useful in 
simplifying some of the more complicated expressions.

\subsubsection{Solutions on the \texorpdfstring{$P_{+}$}{P+} subspace}

For this projection the spinors decouple
\begin{align}
	\DD   \ \Psi^1   = 0 \ ,
	\qquad
	\DDT  \ \Psi^2   = 0 \ .
\end{align}
Substitution gives
\begin{align}\label{FZM:EomPplusU}
\begin{split}
	\sum_{\lambda = \pm} 
			\left( \alpha^{1}_{+} + \alpha^{1}_{-}\, \Gh_{45} \right)  
			\left( \p_\cY + C_{+} \right)	U^{1}_{\lambda}  &= 0 \ ,
	\\
	\sum_{\lambda = \pm} 
			\left( \alpha^{2}_{+} + \alpha^{2}_{-}\, \Gh_{45} \right)  
			\left( \p_\cY + C_{-} \right)	U^{2}_{\lambda}  &= 0 \ ,
\end{split}
\end{align}
with c-numbers \vspace{-1em}
\begin{align}\label{FZM:Cpm}
	\begin{split}
		C_\pm \ = \ 
				\frac{ i\lambda q}{2 \sqrt{\qt^2-u^2}}
				+ \frac{ i\lambda Q_\pm \sqrt{1-Q_\pm^2}}
						  {2 \left(\cosh^2\!\cY - Q_\pm^2\right)} \ ,
	\end{split}
\end{align}
and this simple form of the equations is a consequence (or proof in itself) 
of the fact that kappa-fixing commutes with the fermion derivative operators.
The solution is
\begin{align}
	\begin{split}
		U^{1}_{\lambda}  &= e^{  - \frac{i \lambda q}{2 \sqrt{\qt^2 - u^2}} \cY  
				- \frac{i}{2} \lambda \arctan \left( \frac{Q_+\tanh\cY}
																{\sqrt{1- Q_+^2}}\right) 
													} \ V^{1}_{\lambda} \ ,
		\\[0.5em]
		U^{2}_{\lambda}   &= e^{  - \frac{i \lambda q}{2 \sqrt{\qt^2 - u^2}} \cY
				- \frac{i}{2} \lambda \arctan \left( \frac{Q_-\tanh\cY}
																{\sqrt{1- Q_-^2}}\right) 
													} \ V^{2}_{\lambda} \ ,
	\end{split}
\end{align}
where $V^{J}$ are (independent) constant MW spinors with 
$\Gh_{34} V^J_\pm = +i V^J_\pm$, $\Gh_{12} V^J_\pm = \mp i V^J_\pm$ and 
$P_+ V^J_\pm = V^J_\pm$. However, with these, the spinors \eqref{FZM:KappaFixed} 
are not normalizable and we discard them as perturbative zero modes.

\subsubsection{Solutions on the \texorpdfstring{$P_{-}$}{P-} subspace} 

The equations on this subspace become
\begin{align}
\begin{split}
\zeta (1+ u)\g \DD   \ \Psi^1  + \qt R\  \Psi^2 &= 0 \ ,
\\[1em] 
\zeta (1- u)\g \DDT   \ \Psi^2 -  \qt R\ \Psi^1  &= 0 \ ,
\end{split}
\end{align}
with fermion derivatives
\begin{align}\label{FZM:qD}
\begin{split}
\DD   &=  \p_{\cY} + \ha G\ \Gh_{34} + \ha Q\ \Gh_{35} 
				+ \frac{q (1-u) \g}{\zeta}\ R\ ,
\\[1em]
\DDT &=  \p_{\cY} + \ha \tilde{G}\ \Gh_{34}  + \ha Q\ \Gh_{35} 
				+  \frac{q (1+u) \g}{\zeta}\ R\ .
\end{split}
\end{align}
After substitution, and a considerable amount of simplification, we get
\begin{align}\label{FZM:EomPminU}
\begin{split}
	\sum_{\lambda = \pm} 
			\left( \alpha^{1}_{+} + \alpha^{1}_{-}\, \Gh_{45} \right)  
			\left[ \left(\p_\cY + C_{11}\right) U^{1}_{\lambda} + C_{12} U^{2}_{\lambda} \right] &= 0 \ ,
	\\
	\sum_{\lambda = \pm} 
			\left( \alpha^{2}_{+} + \alpha^{2}_{-}\, \Gh_{45} \right)  
			\left[ \left(\p_\cY + C_{21}\right) U^{2}_{\lambda} + C_{22} U^{1}_{\lambda} \right] &= 0 \ ,
\end{split}
\end{align}
with \vspace{-0.5em}
\begin{align}\label{FZM:Cij}
	\begin{split}
		C_{11} \ &= \ 
				- \frac{ i\lambda q (1-2u)}{2 \sqrt{\qt^2-u^2}}
				+ \frac{ i\lambda Q_{+}\sqrt{1-Q_{+}^2}}
						  {2 \left(\cosh^2\!\cY - Q_{+}^2\right)} \ ,
	\\[0.5em]
		C_{21} \ &= \ 
				- \frac{ i\lambda q (1+2u)}{2 \sqrt{\qt^2-u^2}}
				+ \frac{ i\lambda Q_{-}\sqrt{1-Q_{-}^2}}
						  {2 \left(\cosh^2\!\cY - Q_{-}^2\right)} \ ,
	\\[0.5em]
		C_{12} \ &= \  (1-u) \g\ e^{\int (C_{21} - C_{11}) \d \cY}\ 
								e^{+i 2 \lambda \xi \cY}\ \left( \lambda \tanh\cY - i \xi \right) \ ,
	\\[0.5em]
		C_{22} \ &= \  (1+u) \g\ e^{\int (C_{11} - C_{12}) \d \cY}\ 
								e^{-i 2 \lambda \xi \cY}\ \left( \lambda \tanh\cY + i \xi \right) \ ,
	\end{split}
\end{align}
where we also defined
\begin{equation}
\xi = \frac{q u}{\sqrt{\qt^2-u^2}} \ .
\end{equation}

The motivation for writing $C_{12}$ and $C_{22}$ in the above form becomes clear once we make the ansatz
\begin{align}
	\begin{split}
		U^{1}_{\lambda}  \ &=\ \frac{e^{- \int C_{11}\d\cY}}{\sqrt{1+u}}  \ \td{U}^{1}_{\lambda}  \ = \
										       \frac{1}{\sqrt{1+u}} e^{   \frac{i \lambda q (1-2u)}{2 \sqrt{\qt^2 - u^2}} \cY  
													- \frac{i}{2} \lambda \arctan \left( \frac{Q_+\tanh\cY}
																{\sqrt{1- Q_+^2}}\right) 
													} \ \td{U}^{1}_{\lambda} \ ,
		\\[0.5em]
		U^{2}_{\lambda}   \ &=\ \frac{e^{- \int C_{12}\d\cY} }{\sqrt{1-u}} \ \td{U}^{2}_{\lambda}  \ = \
										         \frac{1}{\sqrt{1-u}} e^{   \frac{i \lambda q (1+2u) }{2 \sqrt{\qt^2 - u^2} } \cY
													- \frac{i}{2} \lambda \arctan \left( \frac{Q_-\tanh\cY}
																{\sqrt{1- Q_-^2}}\right) 
													} \ \td{U}^{2}_{\lambda} \ ,
	\end{split}
\end{align}
and the equations in brackets \eqref{FZM:EomPminU} reduce to
\begin{align}
	\begin{split}
		&\p_{\cY} \td{U}^{1}_{\lambda} 	 \ +\  
				e^{+i 2 \lambda \xi \cY} \left( \lambda \tanh\cY - i \xi \right)  \td{U}^{2}_{\lambda} \ =\ 0 \ ,
	\\[1em]
		&\p_{\cY}  \td{U}^{2}_{\lambda}	 \ +\    
				e^{-i 2 \lambda \xi \cY} \left( \lambda \tanh\cY + i \xi \right)  \td{U}^{1}_{\lambda}  \ =\ 0 \ .
	\end{split}
\end{align}

Inverting the first equation and substituting into the second we get a 
second-order ODE for $\td{U}^{1}_{\lambda}$
\begin{equation}\label{FZM:MixedPminusSndOrdEom}
\p^2_{\cY}  \td{U}^{1}_{\lambda} - 
\left( 2 i \lambda \xi + \frac{\sech^2\!\cY}{\tanh\cY - i \lambda \xi} \right) \p_{\cY}  \td{U}^{1}_{\lambda}
 - \left( \tanh^2\!\cY + \xi^2 \right) \td{U}^{1}_{\lambda} = 0
\end{equation}
with solutions
\begin{equation}
	\td{U}^{1}_{\lambda} = \left( \sech\cY\ V_{\lambda}+ 
			(\cosh\cY - i \lambda \xi \sinh\cY - i \lambda \xi\, \cY\, \sech\cY )\ \td{V}_{\lambda} \right)\, 
			e^{i \lambda \xi \cY}.
\end{equation}
Taking $\td{V}_{\lambda} = 0$, we obtain the normalizable solutions
\begin{equation}
	\td{U}^{1}_{\lambda} = \sech\cY\ e^{i \lambda \xi \cY}\ V_{\lambda}\ ,
	\qquad
	\td{U}^{2}_{\lambda} = \lambda\ \sech\cY\  e^{- i \lambda \xi \cY}\ V_{\lambda}\ ,
\end{equation}
and the (kappa-fixed) fermion zero modes are given by
\begin{align}\label{FZM:NormZeroModes}
	\begin{split}
		\Psi^1 &= \sum_{\lambda = \pm} 
						\frac{\sech\cY}{4 \sqrt{1+u}}\ e^{ i \lambda \w_{+} }
				\left( e^{i \chi}          \sqrt{1 - \lambda Q_{+}\, \sech\cY} + 
					   e^{-i \chi} \lambda \sqrt{1 + \lambda Q_{+}\, \sech\cY}\, \Gh_{45}\right)  
				V_{\lambda} \ ,
		\\[0.5em]
		\Psi^2 &= \sum_{\lambda = \pm} 
						\frac{\lambda\ \sech\cY}{4 \sqrt{1-u}}\ e^{i \lambda \w_{-} }
				\left( e^{i \td{\chi}}         \sqrt{1 + \lambda Q_{-}\, \sech\cY} - 
					   e^{-i \td{\chi}} \lambda \sqrt{1 - \lambda Q_{-}\, \sech\cY}\, \Gh_{45}\right)  
				V_{\lambda} \ ,
	\end{split}
\end{align}
where
\begin{equation}
\w_\pm(\cY) = \frac{q\, \cY}{2 \sqrt{\qt^2 - u^2}}
						- \frac{1}{2}\arctan \left( \frac{Q_\pm \tanh\cY}
												{\sqrt{1- Q_\pm^2}}\right)  \ ,
\end{equation}
and the constant MW spinors $V_\pm$ satisfy $P_{-} V_\pm = V_\pm$,
$\Gh_{34} V_\pm = +i V_\pm$, and  $\Gh_{12} V_\pm = \pm i V_\pm$.

\paragraph*{Counting the zero modes.}
The normalizable zero modes above are parametrized by the constant 
spinor $V = V_{+} + V_{-}$. An unconstrained 
10-d MW spinor has 16 real degrees of freedom, but kappa-fixing 
(which in our parametrisation translates to $\Gh_{34} V = +i V$) and 
6d-chirality ($P_{-} V = V$) both reduce the number of components 
by half. Recalling the further restriction $\Gh^{1268} V = -V$ for 
the $\Sphere^1$ case, we conclude that there are 4 and 2 normalizable 
solutions for the \AdsST and \AdsSSS backgrounds, respectively, 
i.e. we get the expected number of fermion zero modes.

\subsection{Zero mode action}

Now letting $V = V_{+} + V_{-}$ depend on $\cT$ and substituting these 
zero modes into \eqref{FZM:KappaFixedLagr} we get \vspace{-1em}
\begin{align}
\LL_{\text{F,0}} 	 &= \   2 i \cos\vp\, (1+ u) \g \, {\Psi^1}^\dag \p_{\cT} \Psi^1	
						       			 + 2 i \td{A} (1- u) \g \, {\Psi^2}^\dag \p_{\cT}  \Psi^2	 \ ,
\\[1em]
							 &= \  \frac{i \cos\vp\, \g}{2} \, \sech^2\!\YY \ V^\dag \p_{\cT} V	\ ,
\end{align}
where, going to the second line, we implicitly used the fact that 
$V = \frac{1}{2}(\Id - i \G_{34})\, V $, and 
$(\Id - i \G_{34}) \G_{45} (\Id - i \G_{34}) = 0$. 
Integrating over $\cX$ we get the zero mode action
\begin{align}
	S_{\text{F,0}}  = \  
			\hh \td{\g} \sec\vp \int \d \cT \, \Big( \, i \, V^\dag \p_{\cT} V \,  \Big) \ ,
\end{align}
with \vspace{-0.5em}
\begin{equation}
	\td{\g} = \frac{\g}{\zeta} = \frac{1}{\sqrt{\qt^2 - u^2}} \ .
\end{equation}

We can further simplify this by considering a Majorana basis, where all 
(boosted) gamma-matrices are purely imaginary $\Gh_A^* = - \Gh_A^*$, 
and the Majorana condition reduces to reality of the spinors ${\Psi^I}^* = \Psi^I$. 
Applying this to the solutions \eqref{FZM:NormZeroModes}, we get
\begin{equation}
	V_{-} = {V_{+}}^* \quad \Rightarrow \quad V^* = V \ ,
\end{equation}
and the zero mode action becomes
\begin{align}
	S_{\text{F,0}}  = \   
			\hh \td{\g} \sec\vp \int \d \cT \, \Big( \, i \, V^T \p_{\cT} V \,  \Big)  \ .
\end{align}

As we have noted above, there are 2~and~4 real fermion zero modes 
for the giant magnons on \AdsSSS and \AdsST respectively. Quantization 
of these real fermions leads to the anticommutators
\begin{equation}
	\acomm{V_{\alpha a}}{V_{\beta b}} = 
		\delta_{\alpha\beta}\, \delta_{ab}\ \frac{\cos\vp}{\hh \td{\g}} \ ,
\end{equation}
where $a, \alpha = 1, 2$, and for $\vp \neq 0$ only the $a=1$ 
modes are present. After complexifying 
\begin{equation}
	V_{\sL a} = \frac{1}{\sqrt{2}} \left( V_{1 a} + i\, V_{2 a} \right) \ ,
\quad
	V_{\sR a} = \frac{1}{\sqrt{2}} \left( V_{1 a} - i\, V_{2 a} \right) \ ,
\end{equation}	
the only non-trivial zero-mode anticommutator is
\begin{equation}\label{FZM:zeroModeAntiComm}
\acomm{V_{\sL a}}{V_{\sR b}} = \delta_{ab}\ \frac{\cos\vp}{\hh \td{\g}} \ .
\end{equation}
In the remaining part of this section we will see, for both geometries, 
how the symmetry superalgebra of the ground state (BMN vacuum) 
arises from these zero modes.

\subsection{Zero-mode algebra for \texorpdfstring{\AdsSSS}{AdS3 x S3 x S3 x S1}}

By considering the corresponding spin-chain, it was argued that the 
fundamental excitations transform in the 2 dimensional short representations 
of the centrally extended $\su(1|1)^2$ algebra \cite{Borsato:2012ud}. 
This superalgebra has 4 fermionic generators and 4 central charges 
satisfying\footnote{
	For a detailed description of $\ce{\su(1|1)^2}$ and its 
	representations see appendix \ref{App:S1Alg}.}
\begin{equation}\label{FZM:su(1|1)^2ce}
  \begin{aligned}
    \acomm{\genQ_{\sL}}{\genS_{\sL}} &= \genH_{\sL} \ , \qquad &
    \acomm{\genQ_{\sL}}{\genQ_{\sR}} &= \genC  \ , \\
    \acomm{\genQ_{\sR}}{\genS_{\sR}} &= \genH_{\sR} \ , &
    \acomm{\fixedspaceL{\genQ_{\sL}}{\genS_{\sL}}}{\fixedspaceL{\genQ_{\sR}}{\genS_{\sR}}} &= \genCbar \ .
  \end{aligned}
\end{equation}
Consequently, the symmetry algebra of light-cone gauge superstring theory 
on \AdsSSS was shown to take the same form, after lifting the level-matching 
condition \cite{Borsato:2015mma}. Thus, it is important to see how the 
supercharges of the algebra can be constructed from the 
zero modes~\eqref{FZM:zeroModeAntiComm}.

For an off-shell one-particle representation the values of the central charges are given by
\begin{equation}\label{FZM:CCvalues}
\begin{aligned}
	&\genH_{\sL} =  \ha ( \epsilon + M ) \ , \qquad
	\genC		=  \frac{\hh  \varsigma}{\td{\g}}  \ ,
	\\[0.5em]	
	&\genH_{\sR} = \ha ( \epsilon - M ) \ ,	 \qquad
	\genCbar	= \frac{\hh }{\td{\g} \varsigma} \ ,
\end{aligned}
\end{equation}
where $M = m \pm q h p$, with mass $m$,  $\epsilon$ is the energy of the magnon 
\begin{equation}
\epsilon = \sqrt{M^2 + \frac{4 \hh^2}{\td{\g}^2} } \ ,
\end{equation}
and $\varsigma$ can be removed by rescaling for a one-particle state, 
but plays an important role in constructing multi-particle representations 
\cite{Arutyunov:2006yd}. Note that the momentum of the excitation enters into 
these expressions through \eqref{FZM:upRel}
\begin{equation}
(\qt \td{\g})^{-1} = \sin\frac{p}{2} \ .
\end{equation}

These values satisfy the shortening condition 
	$\genH_{\sL} \genH_{\sR} - \genC\genCbar = 0 $, 
therefore on this representation the supercharges must be related to each other. 
Assuming only $\acomm{\genQ_{\sL}}{\genQ_{\sR}} = \frac{\hh \varsigma}{\g} $, 
it is not too hard to justify\footnote{
	In doing so, one might find useful the fact that acting on the short representation 
	\eqref{alg:chiral-rep}, the supercharges satisfy: 
	$\comm{\genQ_{\sL}}{\genQ_{\sR}} = - (-1)^F \genC$, 
	$\comm{\genS_{\sL}}{\genS_{\sR}} = (-1)^F \genCbar$.
} that the rest of \eqref{FZM:su(1|1)^2ce} will follow from
\begin{equation}\label{FZM:SfromQ}
	\genS_{\sL, \sR} \ =\ 
		\varsigma^{-1} \left( \sqrt{\frac{\td{\g}^2 M^2}{4 \hh^2} + 1} 
											+ \frac{\td{\g} M}{2\hh} (-1)^F  \right)\, \genQ_{\sR, \sL} \ ,
\end{equation}
where $F$ is the fermion number operator, i.e. $(-1)^F$ anticommutes with the 
supercharges. This leaves us with the task of expressing $\genQ_{\sL, \sR}$ 
in terms of the zero modes. We can make the general ansatz
\begin{equation}
\genQ_{\sL,\sR} \ = \ 
				\varsigma^{1/2} \left( \mathcal{A} - \mathcal{B} (-1)^F \right) \, V_{\sL,\sR} \ ,
\end{equation}
where $\mathcal{A}$ and $\mathcal{B}$ are some c-numbers, and 
\eqref{FZM:zeroModeAntiComm} guarantees that the condition 
$\acomm{\genQ_{\sL}}{\genQ_{\sR}} = \frac{h \varsigma}{\td{\g}} $ 
will be satisfied as long as \mbox{$\mathcal{A}^2 - \mathcal{B}^2 = \sec\vp\, \hh^2$}. 
Our freedom in choosing $\mathcal{A}$ is just basis dependence, 
and a symmetric identification is given by
\begin{equation}\label{FZM:QUsymmBasis}
	\mathcal{A} = \sqrt{\frac{\sec\vp\, \hh^2}{2}} 
	\left( \sqrt{\frac{\td{\g}^2 M^2}{4 \hh^2} + 1} + 1  \right)^{1/2} \, ,
\quad
	\mathcal{B} = \sqrt{\frac{\sec\vp\, \hh^2}{2}} 
	\left( \sqrt{\frac{\td{\g}^2 M^2}{4 \hh^2} + 1} - 1  \right)^{1/2} \, ,
\end{equation}
with the supercharges taking the form
\begin{equation}
\begin{aligned}
	\genQ_{\sL,\sR} \ &=  &\varsigma^{1/2}\ &\left( \mathcal{A} - \mathcal{B} (-1)^F \right) \, V_{\sL,\sR} \ ,
	\\
	\genS_{\sL,\sR} \ &=  &\varsigma^{-1/2}\ &\left( \mathcal{A} + \mathcal{B} (-1)^F \right) \, V_{\sR,\sL} \ .
\end{aligned}
\end{equation}
	
\subsection{Zero-mode algebra for \texorpdfstring{\AdsST}{AdS3 x S3 x T4}}

The off-shell symmetry algebra of superstring theory on this background 
is the centrally extended $\psu(1|1)^4$ \cite{Lloyd:2014bsa}, which is 
essentially a tensor product of two $\ce{\su(1|1)^2}$ algebras with matching 
central charges.\footnote{
	See appendix \ref{App:T4Alg} for the construction and short representations 
	of $\ce{\psu(1|1)^4}$.
} The giant magnon is part of a 4 dimensional short representation, and we should 
be able to match the supercharges to the zero modes.

Having noted the tensor product structure of the algebra, the construction is trivial, 
since \eqref{FZM:zeroModeAntiComm} gives us two non-interacting copies of 
$U_{\sL,\sR}$. The central charges take the same values as in \eqref{FZM:CCvalues}, 
hence everything from the previous subsection holds for each copy of $\ce{\su(1|1)^2}$, 
and the supercharges of $\ce{\psu(1|1)^4}$ are simply 
\begin{equation}
\begin{aligned}
	\genQ_{\sL,\sR\ a} \ &=  &\varsigma^{1/2}\ &\left( \mathcal{A} - \mathcal{B} (-1)^F \right) \, V_{\sL,\sR\ a} \ ,
	\\
	\genS_{\sL,\sR\ a} \ &=  &\varsigma^{-1/2}\ &\left( \mathcal{A} + \mathcal{B} (-1)^F \right) \, V_{\sR,\sL\ a} \ .
\end{aligned}
\end{equation}
where $\mathcal{A}$ and $\mathcal{B}$ are still given by \eqref{FZM:QUsymmBasis}.

\subsection{Zero modes in the \texorpdfstring{$\alpha \to 0, 1$}{alpha -> 0, 1} limits}

The parameter $\alpha \in [0,1]$ determines the radii of the 3-spheres in the
\AdsSSS geometry \eqref{FZM:radii}, and in the limits $\alpha \to 0,1$, 
blowing up either of the spheres, we are left with---up to compactification of 
the flat directions---\AdsST. It is interesting to see what happens to the 
fermion zero modes \eqref{FZM:NormZeroModes} in the process.

Taking $\alpha \to 1$ (or $\vp \to 0$) blows up $\Sphere^3_-$, the sphere on 
which we have the BMN-like leg of the magnon \eqref{FZM:s3s3StationaryMagnon}. 
In this limit $\cY \to \g \sqrt{\qt^2-u^2} \cX$, i.e. the magnon becomes the 
$\Torus^4$ magnon, and the zero modes reduce to two of the 
four real $\Torus^4$ zero modes, the ones on the $\G_{1268} = -1$ 
subspace. The remaining two we will find on the $\G_{1268} = +1$ 
eigenspace, where $\Delta$ also becomes zero \eqref{FZM:DeltaDef}.

On the other hand, $\alpha \to 0$ (or $\vp \to \frac{\pi}{2}$) blows up $\Sphere^3_+$ with the stationary
magnon on it, and the bosonic solution becomes a BMN string on $\Sphere^3_-$. 
Since the rescaled coordinate
\begin{equation}
 \cY = \cos^2\!\vp\, \g \sqrt{\qt^2-u^2} \cX \to 0
\end{equation}
for all points on the string, the zero mode solution \eqref{FZM:NormZeroModes} reduces to
constant spinors. The highest weight state of the massless magnon is fermionic
\cite{Lloyd:2014bsa} and should correspond to the limit of our fermion fluctuations,
but it appears we are unable to learn more about these modes from a semiclassical
analysis. This shows that some aspects of the massless modes can only be captured 
by exact in $\alpha'$ results, in agreement with similar findings in the spin chain
limit \cite{Sax:2012jv}.

\section{Fermion zero modes for \texorpdfstring{$q=1$}{q=1}}
\label{FZM:SecQ1}

In this section we take a look at the special case of $q=1$, as
there are some subtleties not captured by our general discussion.
The $q=1$ fermion zero modes on the two $\AdS_3$ backgrounds 
are more closely related than for $q<1$, hence we will first focus on 
the \AdsST case, then briefly describe the differences for \AdsSSS.

\subsection{Bosonic solution}

For $q = 1$, the giant magnon \vspace{-0.5em}
\begin{align}\label{FZM:Q1magnon}
	\begin{split}
		Z_1 &= e^{i t} \left[ \cos \tfrac{\pp}{2} + i \sin \tfrac{\pp}{2} \, \tanh U \right],
		\\[1em]
		Z_2 &= e^{i V} \sin \tfrac{\pp}{2} \ \sech{U},
	\end{split}
\end{align}
found in \cite{Hoare:2013lja} is still a valid solution,
but the magnon speed on the worldsheet is actually 
fixed to be the speed of light, and we will use a different parametrization\footnote{
	In the $q \to 1$ limit the parameter $v$ of \cite{Hoare:2013lja} 
	is meaningless, instead we will use $\beta =\sqrt{\frac{1-v}{1+v}}$. }
\begin{align}\label{FZM:Q1UV}
	\begin{split}
		U 		&= \cos\rho\ \beta \, ( x + t) ,
		\\[0.5em]
		V 		&=  \sin\rho\ \beta \, ( x + t) - x ,
	\end{split}
\end{align}
where  $\beta > 0$, $\rho \in [0, 2\pi)$, and the parameters are related via
\begin{equation}
b \equiv \cot\tfrac{\pp}{2}  =  \frac{\sin\rho - \beta}{\cos\rho} .
\end{equation}

Already from this representation of the solution it seems like the main 
dependence is on the light-cone coordinate $x^+ = \ha (t + x)$. This is hinting at 
the magnon having a definite chirality, not completely unexpectedly considering 
that bosonic theory reduces to the conformal WZW model 
at $q=1$. This statement will be made more precise shortly.

\paragraph*{Conserved charges.}
For the above solution the conserved charges are
\begin{align}
	\begin{split}
		E - J_1 &= M =2 \hh \sin^2\tfrac{\pp}{2} \left(\tan \rho  - \cot\tfrac{\pp}{2} \right) ,
		\\[1em]
		     J_2 &= M + \hh \pp , 
	\end{split}
\end{align}
with dispersion relation
\begin{equation} 
	E - J_1= J_2  - \hh\pp .
\end{equation}

\paragraph*{The WZW model.}

The $\SU (2)$ PCM with WZ term (see section \ref{FZM:SubSecPCM}) simplifies significantly
for the case of $q=1$, with the equations of motion \eqref{FZM:PCMEom} now reading 
\begin{equation}
\p_-\Jfrak_+ = 0 \ , \quad \p_+\kfrak_- = 0 \ .
\end{equation}
The degrees of freedom separate based on chirality: the left-movers are 
described by $\Jfrak_+ (x^+)$, while $\kfrak_- (x^-)$ describes right-movers. 
Looking at the magnon's $\SU (2)$ currents, listed in appendix \ref{FZM:AppQ1Currents},
we note that $\kfrak_-$ is in fact constant with no dynamical information 
(i.e. it can be gauged away). It is in this sense that the classical bosonic solution 
has a definite chirality. 

\subsection{Zero mode equations for \texorpdfstring{\AdsST}{AdS3 x S3 x T4}}

The derivation of the fermion equations of motion is analogous to 
the $q \in [0,1)$ case presented in section \ref{FZM:SecFzmEom}, and we
omit the details here. In terms of the light-cone coordinates 
$x^\pm = \ha (t \pm x)$, we have 
\begin{align}
	\begin{split}
		\left(  \p_-  - 2 \beta \cos\rho \,       M (x^+) \right)  \Psi^1   &= 0 \ ,
	\\[1em]
		\left(  \p_+ + 2 \beta \cos\rho\, \td{M}(x^+) \right)  \Psi^2   &= 0 \ ,
	\end{split}
\end{align}
with \vspace{-1em}
\begin{align}
	\begin{split}
		M         &=  \frac{1}{2 \beta \cos\rho }\left(  \frac{1}{2} G\ \G_{34} + \frac{1}{2} Q\ \G_{35} 
																				+   R\, P_{-} - (R + \G_{12})\, P_{+} \right)
	\\[1em]
		\td{M} &=  \frac{1}{2 \beta \cos\rho} \left( \frac{1}{2} \td{G}\ \G_{34}  + \frac{1}{2} \tilde{Q}\ \G_{35} 
															                  +   R\, P_{-} - (R + \G_{12})\, P_{+} \right)
	\end{split}
\end{align}
where all the dependence is on $x^{+}$ via \vspace{-0.5em}
\begin{equation}
	\cY = 2 \beta \cos\rho \, x^{+} \ .
\end{equation}
The expressions for $G, \td{G}, Q$ and $\td{Q}$, along with the pullbacks of the 
vielbein and spin connection can be found in Appendix \ref{FZM:AppQ1}.
These equations are the $q=1$ versions of \eqref{FZM:EomNoZeroMode},
but also after commuting the kappa projectors through. 
Note however, that they cannot be obtained as limits 
of the $q<1$ analogues. In this general setting for $q=1$ surely not 
(there are two parameters $\beta, \rho$ here versus the one parameter 
$u$ in section \ref{FZM:SecFzmEom}), but not even for any special case,
since there is no $q=1$ stationary magnon (see towards the end of this section).

\paragraph*{Zero mode condition.}
As we have seen above, the bosonic background is itself chiral ($\p_-\Jfrak_+ = 0$), 
and it is reasonable to expect this to carry through to the fermionic zero modes, 
i.e. $\p_-\vt^J = 0$. This can be viewed as the extension of the zero mode condition 
for $q \in [0,1)$, and forces the first spinor to be trivial
\begin{equation}
\Psi^1 = 0 \ .
\end{equation}
Changing to the variable $\cY$, the remaining equation for $\Psi^2$ reads
\begin{equation}\label{FZM:Q1eqnY}
	\left(  \p_\cY + \td{M} \right) \Psi^2  = 0 \ .
\end{equation}

\subsection{Zero mode solutions for \texorpdfstring{\AdsST}{AdS3 x S3 x T4}}

We can find the solutions for $\Psi^2$ in much the same way we did
in section \ref{FZM:SecMixedFlux}. First we solve for the general kappa-fixed
spinor, then substituting it into \eqref{FZM:Q1eqnY} we get a set of simpler
equations on the $P_\pm$ subspaces, that we can easily solve.

\paragraph*{Fixing kappa-gauge.}
The projector can be written as
\begin{equation}
K_2	 = \frac{1}{2} \left( \Id  \,  -  \cos\upsilon\, \G_{04} -  \sin\upsilon\, \G_{05}\right) \ ,
\end{equation}
with 
\begin{equation}
\upsilon = \arcsin\left( \frac{\sech\cY}{\sqrt{1+b^2}}\right) \ .
\end{equation}
Making the ansatz
\begin{equation}
\Psi^2 = \left( \alpha_{+}(\cY) + \alpha_{-}(\cY)\, \G_{45} \right) U\ ,
\end{equation}
with $\G_{34} U = i U$ and $i \G_{12} \G_{*} \G_{+} U = \lambda U$, 
the equation $K_2 \Psi^2 = \Psi^2$ reduces to
\begin{align}
	\begin{split}
		\lambda \sin\upsilon\, \alpha_{+} + \lambda \cos\upsilon\, \alpha_{-} & = \alpha_{+} \ ,
		\\[0.5em]
		\lambda \cos\upsilon\, \alpha_{+} - \lambda \sin\upsilon\,  \alpha_{-}  & = \alpha_{-} \ .
	\end{split}
\end{align}
A symmetric solution is given by
\begin{equation}
	\alpha_{+} = \sqrt{1 + \lambda \sin\upsilon} \ , 
	\qquad
	\alpha_{-} = \lambda \sqrt{1 - \lambda \sin\upsilon} \ ,
\end{equation}
and the most general gauge-fixed spinor is 
\begin{equation}\label{FZM:Q1KappaFixed}
\Psi^2 = \sum_{\lambda = \pm} \left( \sqrt{1 + \lambda \sin\upsilon} + 
												 \lambda \sqrt{1 - \lambda \sin\upsilon}\, \G_{45}\right)  U_{\lambda} \ ,
\end{equation}
where still $\G_{34} U_\pm = +i U_\pm$ and 
$i \G_{12} \G_{*} \G_{+} U_\pm = i \G_{0345} U_\pm = \pm U_\pm$.

\paragraph*{Solutions on the $P_{\pm}$ subspaces.}
Now letting $U_{\lambda}$ depend on $\cY$ and substituting 
\eqref{FZM:Q1KappaFixed} into the $P_{\pm}$ projections of 
\eqref{FZM:Q1eqnY}, after a considerable amount of simplification, we get
\begin{align}
\begin{split}
& \sum_{\lambda = \pm} \left( \alpha_{+} + \alpha_{-}\, \G_{45}\right)  
									   \left( \p_\cY +  C_{+} \right) U_{\lambda} = 0
\qquad \text{on}\ \ P_+	\ ,							   
\\[0.5em]
&\sum_{\lambda = \pm} \left( \alpha_{+} + \alpha_{-}\, \G_{45}\right)  
									   \left( \p_\cY +  C_{-} \right) U_{\lambda} = 0
\qquad \text{on}\ \ P_-	\ ,
\end{split}
\end{align}
with the scalars\footnote{
	Note that these are different from \eqref{FZM:Cpm}.
} \vspace{-0.5em}
\begin{equation}
C_{\pm} = - \frac{i \lambda}{4} 
							\left( \frac{2 b\, \sech^2\!\cY}{b^2 + \tanh^2\!\cY} 
										\pm \frac{\sec^2\!\rho}{b - \tan\rho}  +  2 \tan\rho   \right) \ .
\end{equation}
It is now a simple exercise to arrive at the solutions $\Psi^{2+}, \Psi^{2-}$ 
on the $P_+$ and $P_-$ subspaces, respectively,
\begin{equation}\label{FZM:T4Q1Sln}
\Psi^{2\pm} = \sum_{\lambda = \pm} e^{i \lambda \omega_{\pm}(\cY)}
				\left( \sqrt{1 + \lambda \sin\upsilon} + 
								\lambda \sqrt{1 - \lambda \sin\upsilon}\, \G_{45}\right)  V^{\pm}_{\lambda} \ ,
\end{equation}
where
\begin{equation}
\omega_{\pm}(\cY) = \frac{1}{2}  \arctan\left(\frac{\tanh\cY}{b}\right)  + 
				\frac{1}{4}\left( 2 \tan\rho \pm \frac{\sec^2\!\rho}{b - \tan\rho} \right) \cY \ ,
\end{equation}
and the constant spinors $V^a_\lambda$ satisfy $\G_{34} V^a_\lambda = +i V^a_\lambda$,
$P_{\pm} V^{\pm}_\lambda=  V^{\pm}_\lambda$  and $i \G_{0345} V^{a}_\pm = \pm V^{a}_\pm$.
Starting with 16 (unconstrained) real MW spinors, these conditions leave us with 
4+4 real zero modes on the $P_{+}$ and $P_{-}$ subspaces. We see that none 
of the solutions are normalizable, which is to be expected given the chiral nature 
of the background. However, only looking at the solutions, and not extrapolating from
the $q<1$ case, it is unclear which 4 of these should be included in 
semiclassical quantization and the construction of the algebra.

\subsection{Zero modes for \texorpdfstring{\AdsSSS}{AdS3 x S3 x S3 x S1}}

We can put the magnon \eqref{FZM:Q1magnon} on \AdsSSS the same way 
we did in \eqref{FZM:S3S3DyonicMagnon}. Just like above, the zero modes satisfy
$\Psi^1 = 0$ and 
\begin{equation}
	\left(  \p_\cY + \td{M} \right) \Psi^2  = 0 \ ,
\end{equation}
where, similarly to \eqref{FZM:DDfinal}
\begin{equation}
	\td{M} = \frac{1}{2 \beta \cos\rho} 
		\left( \frac{1}{2} \td{G}\ \Gh_{34}  + \frac{1}{2} \tilde{Q}\ \Gh_{35} 
													+   R\, P_{-} - (R + \Gh_{12})\, P_{+}  + \Delta_0\, \Gh_{12}\right)\ .
\end{equation}
The boosted worldsheet coordinate is $\cY = 2 \c2\, \beta \cos\rho \, x^{+}$,
the scalar functions $G, \td{G}, Q, \td{Q}$ are still as given in Appendix 
\ref{FZM:AppQ1}, and from \eqref{FZM:DeltaDef}
\begin{equation}
\Delta_0 =  - \frac{\k^2}{2} \left(\Gh^{1268} + \Id \right)\ , 
\qquad
\k \equiv \tan\vp\ .
\end{equation}

\paragraph*{$\mathbf{\Gh^{1268} = - 1}$.}
On the $-1$ eigenspace of $\Gh^{1268}$ the solutions are the same 
as for \AdsST  \eqref{FZM:T4Q1Sln}, with all $\G_A$ replaced by 
$\Gh_A$  (including $P_\pm = \ha (\Id \pm \Gh_* \Gh_+)$) and imposing 
the extra condition $\Gh^{1268} V^a_\lambda = -  V^a_\lambda $.
 
\paragraph*{$\mathbf{\Gh^{1268} = + 1}$.}
On this subspace $\Delta_0 =  - \k^2$, and after making
the ansatz \eqref{FZM:Q1KappaFixed} we get 
\begin{align}
\begin{split}
& \left( \p_\cY +  C_{+} 
+ \frac{i \lambda \k^2}{2 \beta\cos\rho}\right) U_{\lambda} = 0
\qquad \text{on}\ \ P_+	\ ,							   
\\[1em]
&\left( \p_\cY +  C_{-} 
- \frac{i \lambda \k^2}{2 \beta\cos\rho}\right) U_{\lambda} = 0
\qquad \text{on}\ \ P_-	\ .	
\end{split}
\end{align}
The zero mode solutions are
\begin{equation}
\Psi^{2\pm} = \sum_{\lambda = \pm} e^{i \lambda \td\omega_{\pm}(\cY)}
				\left( \sqrt{1 + \lambda \sin\upsilon} + 
								\lambda \sqrt{1 - \lambda \sin\upsilon}\, \G_{45}\right)  V^{\pm}_{\lambda} \ ,
\end{equation}
where
\begin{equation}
\td\omega_{\pm}(\cY) = \frac{1}{2}  \arctan\left(\frac{\tanh\cY}{b}\right)  + 
				\frac{1}{4}\left( 2 \tan\rho \pm \frac{(1+\k^2)\sec^2\!\rho}{b - \tan\rho} \right) \cY \ ,
\end{equation}
and $\G_{34} V^a_\lambda = +i V^a_\lambda$, $P_{\pm} V^{\pm}_\lambda=  V^{\pm}_\lambda$,
$i \G_{0345} V^{a}_\pm = \pm V^{a}_\pm$, $\Gh^{1268} V^a_\lambda = + V^a_\lambda $.
\vspace{1em}

We have 8 real solutions in total, 2+2 for $P_{\pm}$ on each eigenspace of $\Gh^{1268}$.
Once again, all of these zero modes are non-normalizable, and without extrapolating from
the $q<1$ analysis, we have not been able to find any distinguishing features of the 2
that would enter into canonical quantization.

\subsection{The \texorpdfstring{$q \to 1$}{q->1} limit}

We can go from the $q<1$ dyonic magnon \eqref{FZM:HoareMagnon}
to the $q=1$ solution \eqref{FZM:Q1magnon} by taking
\begin{equation}\label{FZM:toQ1}
\qt \to 0,    \quad   u \to -1, \qquad \text{with} \quad \qt\g = \beta \ \text{fixed.}
\end{equation}
However, to compare the zero modes above to those found in section 
\ref{FZM:SecMixedFlux}, we need the $q=1$ version of the stationary magnon
\eqref{FZM:StationaryMagnon} we used as a background for the $q<1$ fermions.
There are two natural ways of taking the $q \to 1$ limit, let us look at them in turn.

Our first instinct would be to take the same limit \eqref{FZM:toQ1} for the
stationary magnon \eqref{FZM:StationaryMagnon}, but this is not compatible 
with the condition \eqref{FZM:upRel}, restricting $|u| \leq \qt$.
Equivalently, we cannot make $V$ in \eqref{FZM:Q1UV}
only depend on $x^+$ (technically one could take $\beta \to \infty$,
but this results in a discontinuous bosonic solution).

Alternatively, we can impose the second form of the stationary condition,
and require the $\SU(2)$ charge $M$ to be zero. This would mean $\beta =0$, 
and then $U \equiv 0$, with the endpoints not on the equator any more.
Furthermore, the parameter $\pp$ in \eqref{FZM:Q1magnon} would not be 
the worldsheet momentum, as $\Delta\phi_1 = 0$. 

Lacking a suitable generalization of the Hofman-Maldacena magnon for $q=1$,
it is not immediately clear how we can apply the analysis of previous sections.
It would be interesting to further investigate the relation between the $q \to 1$
limit of zero modes found in section \ref{FZM:SecMixedFlux}, to the $q=1$
fermion fluctuations found here.

\section{Conclusions}
\label{FZM:SecConclusion}

In this paper we wrote down the stationary giant magnon solution on
\AdsSSS, which is the $q \in [0,1)$ mixed-flux generalization of the 
Hofman-Maldacena magnon. We then explicitly constructed the fermion 
zero modes of this bosonic string solution from the quadratic action, 
and found that there are 4 and 2 zero modes for the \AdsST and 
\AdsSSS magnons, respectively, in agreement with the algebraic
structure. We also showed how to get the generators of the centrally
extended $\psu(1|1)^4$ and $\su(1|1)^2$ algebras from the 
semiclassically quantized fermion zero modes.

We treated the $q=1$ limit separately, and found that there is no 
stationary magnon in this case. As expected from the chiral nature 
of the magnons at the $q=1$ point, all of the zero modes we found
are non-normalizable. We have the same number of fermionic generators
in the off-shell algebra as for $q<1$, and with the excess number of solutions, 
the issue of canonical quantization needs to be further addressed.

Our understanding of the $\AdS_3 / \CFT_2$ duality is far from complete, and it is an 
area of active research. The $\CFT$ dual of the WZW model at $k=1$, i.e. \AdsST string 
theory with the smallest amount of quantized NS-NS flux, has been recently argued to be 
the limit of a symmetric product orbifold \cite{Giribet:2018ada, Gaberdiel:2018rqv,Eberhardt:2018ouy}. 
The complete Yangian algebra of the mixed-flux \AdsSSS superstring has also 
been found \cite{Pittelli:2017spf}, and it would be interesting to see how it 
relates to our semi-classical quantization.
Semiclassical methods continue to be useful in probing the string theory side, present 
paper being one example, or the one-loop corrections to rigid spinning string dispersion 
relations \cite{Nieto:2018jzi}. It seems, however, that massless modes cannot be 
captured in the semiclassical limit. Considering that in the $\Torus^4$ theory 
the massless modes' highest weight state is a fermion \cite{Borsato:2014hja}, 
taking the $\alpha \to 0$ limit of the fermion zero modes might have been a 
good way to arrive at the solutions. The fact that this did not work indicates 
that the fermionic massless mode is inherently non-perturbative in nature. 
This is also in agreement with \cite{Sax:2012jv}, where it was found that the 
$\alpha \to 0$ limit fails to capture the non-perturbative nature of the massless 
mode at the spin chain point (i.e. at the opposite limit of the duality).
Furthermore, when the worldsheet is compactified, the presence of massless
particles does not allow for perturbative computations of wrapping corrections
\cite{Abbott:2015pps}. Instead, such wrapping corrections can be computed 
using a non-perturbative TBA which allows for an alternative low-momentum
expansion \cite{Bombardelli:2018jkj}, based on the earlier observation of 
non-trivial massless scattering in the BMN limit \cite{Borsato:2016xns}.

We propose two directions for future research. Firstly, we want to follow up
on the $q=1$ limit, to get a deeper understanding of the magnon and its
fermion zero modes. Secondly, we would like to generalize the dressing method
to the case of $\R \times\! \Sphere^3 \times\! \Sphere^3$ and carry out 
an analysis of dressing phases for the mixed-flux \AdsSSS background,
similar to what was done for \AdsST in \cite{Stepanchuk:2014kza}.

\acknowledgments

AV would like to thank B. Stefanski and O. Ohlsson-Sax for useful discussions. 
AV acknowledges the support of the George Daniels postgraduate scholarship.

\appendix

\section{Geometry of \texorpdfstring{$\AdS_3$}{AdS3} and \texorpdfstring{$\Sphere^3$}{S3}}\label{FZM:AppAds3S3}

To understand string motion on $\AdS_3 \!\times\! \Sphere^3 \!\times\! \Sphere^3$ 
one must first study its geometry. Let us take the time here to describe the 
components of this space in so-called Hopf coordinates, which are well-suited 
for describing the main object of our discussion, the giant magnon. 

We can describe both $\AdS_3$ and $\Sphere^3$ as hypersurfaces in 
four flat dimensions (with different signatures) and by parametrizing these 
embeddings we derive the metrics $g_{\mu\nu}$ in Hopf coordinates. 
From the metric we can easily read off (the natural choice for) the 
\textit{vielbein} $E_\mu ^A$ satisfying
\begin{equation}
E_\mu ^A  E_\nu ^B \eta_{AB} = g_{\mu\nu} \ ,
\end{equation}
where $A$, $B$ are tangent-space indices and $\eta_{AB}$ is the flat 
(Minkowski or Euclidean) metric. The vielbein provides the most tractable 
construction of curved-space Dirac matrices ($\G_\mu$) from those of 
flat space ($\G_A$):
\begin{equation}
	\G_\mu \equiv E_\mu^A \G_A 
\quad \Rightarrow \quad
	\{ \G_\mu , \G_\nu \} = E_\mu ^A  E_\nu ^B\  
			\underbrace{\{ \G_A , \G_B \}}_{2 \eta_{AB}} = 2 g_{\mu\nu} \ ,
\end{equation}
hence its appearance in the fermionic Lagrangian. Another object of similar 
importance is the \textit{spin connection} $\omega_\mu^{AB}$, as it 
appears in the construction of the covariant derivative for spinors.
 It is given by the formula
\begin{equation}
	\omega_\mu^{AB} = E_\nu^A \p_\mu E^{\nu B} 
					+ E_\nu^A \G^\nu_{\sigma\mu} E^{\sigma B} \ ,
\end{equation}
where the Greek indices are raised by the inverse metric $g^{\mu\nu}$, 
and the Christoffel symbols are given by the usual 
$ \G^\nu_{\sigma\mu} = \frac{1}{2} g^{\nu\rho} 
\left( \p_\sigma g_{\mu\rho} +\p_\mu g_{\sigma\rho} - \p_\rho g_{\sigma\mu}  \right)$.

Embedding into four flat dimensions $(X_1, X_2, X_3, X_4)$ is 
equivalent to embedding into $\C^2$ via
\begin{equation}
Z_1 = X_1 + i X_2\ , \quad  Z_2 = X_3 + i X_4 \ ,
\end{equation}
and we will use this correspondence below.
\begin{figure}[ht]
\centerline{\includegraphics[scale=0.5]{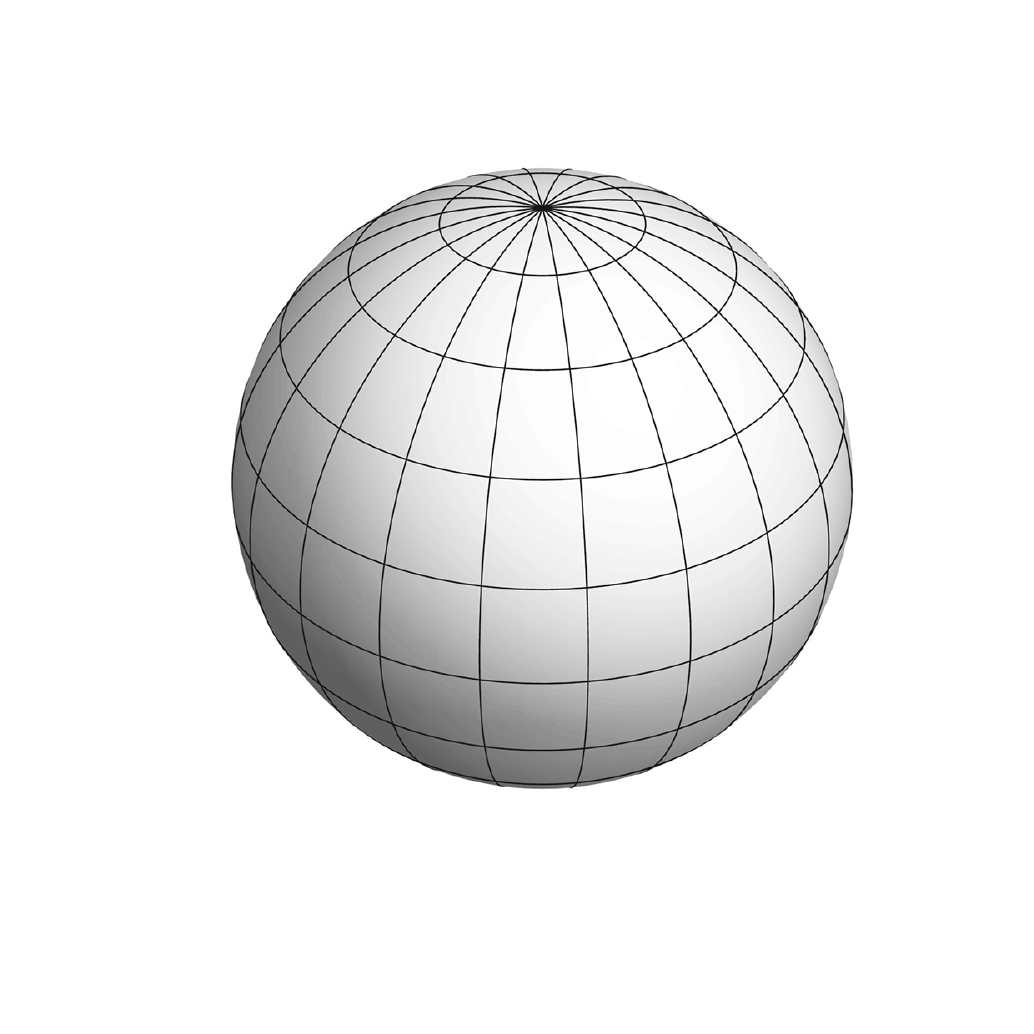}
\includegraphics[scale=0.5]{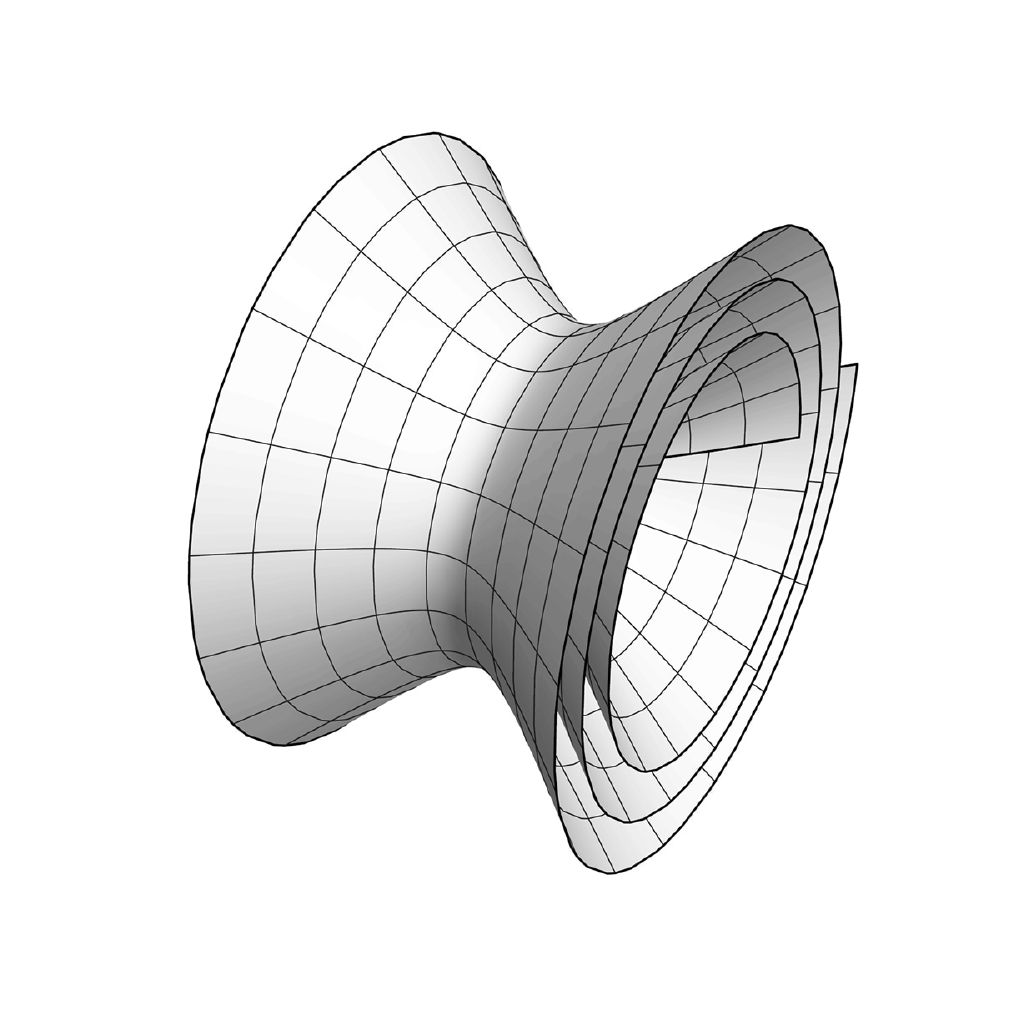} }
\caption{Images of a sphere and of a universal cover of $\AdS$ space
\label{sfel}}\nonumber
\end{figure}

\subsection{\texorpdfstring{$\AdS_3$}{AdS3}}

The three dimensional anti-de Sitter space $\AdS_3$ can be represented 
as a hyperboloid (a constant negative curvature quadric)
\begin{equation}
\eta_{PQ} X^P X^Q = - X_1^2 - X_2^2 + X_3^2 + X_4^2 = -1
\end{equation}
in $\R^{2,2}$ with the metric
\begin{equation}
\d s^2 = \eta_{PQ}\ \d X^P \d X^Q \ , \qquad \eta_{PQ} = \diag (-1,-1,+1,+1)  \ .
\end{equation}
Equivalently in $\C^{1,1}$ 
\begin{equation}
| Z_1 | ^2 - | Z_2 |^2 = 1 \ ,
\end{equation}
which has the  global solution (the analogue of the Hopf coordinates for the sphere)
\begin{equation}
Z_1 = \cosh \rho\ e^{i t} \ , \quad  Z_2 = \sinh \rho\ e^{i \psi} \ ,
\end{equation}
where the radius takes values $\rho \in [0,\infty)$ and $\psi \in [0,2\pi)$. 
Note that $t \in [0,2\pi)$ already covers the hyperboloid once. In the 
context of $\AdS / \CFT$ however, it is standard to decompactify the $t$ 
direction (to avoid closed time-like curves), i.e. to assume 
$t \in (-\infty , \infty)$. In the case of $\AdS_2$ this ``cutting open'' 
along the circular time direction is depicted in Figure \ref{sfel}. 

Substituting into the $\R^{2,2}$ metric we get
\begin{equation}
\d s^2 = - \cosh^2\!\rho\ \d t^2 + \d \rho^2 + \sinh^2\!\rho\ \d \psi^2 \ ,
\end{equation}
from which we can immediately read off the $\AdS_3$ metric and the 
vielbein in coordinates $(t,\rho,\psi)$
\begin{align}
g_{\mu\nu} &= \diag(- \cosh^2\!\rho, \ 1, \ \sinh^2\!\rho) \ ,
\\[1em]
E_\mu^A     &=  \diag( \quad \cosh\rho\ , \ 1, \ \sinh\rho  ) \ .
\end{align}
Straightforward calculation yields the only non-zero Christoffel symbols:
\begin{equation}
\G^\tau_{\tau\rho} = \G^\tau_{\rho \tau} = 
		\tanh\rho \ , \qquad \G^\psi_{\psi\rho} = 
		\G^\psi_{\rho \psi} = \coth\rho \ ,
\qquad 
\G^\rho_{\tau\tau} = - \G^\rho_{\psi\psi} = \cosh\rho\ \sinh\rho \ ,
\end{equation}
while some further crunching at the formulae gives the only non-zero 
components of the spin connection:
\begin{align}
\omega_\tau^{0 1} &= - \omega_\tau^{1 0} = \sinh\rho \ ,
\\[1em]
\omega_\psi^{2 1} &= - \omega_\psi^{1 2} =  \cosh\rho \ .
\end{align}
Note that the tangent space indices corresponding to 
$(t,\rho,\psi)$ run from $0$ to $2$.

\subsection{\texorpdfstring{$\Sphere^3$}{S3}}

The 3-sphere is the hypersurface
\begin{equation}
\eta_{PQ} X^P X^Q =  X_1^2 + X_2^2 + X_3^2 + X_4^2 =  1
\end{equation}
in $\R^4$ with the metric
\begin{equation}
\d s^2 = \eta_{PQ}\ \d X^P \d X^Q \ , 
\qquad 
\eta_{PQ} = \diag (+1,+1,+1,+1)  \ .
\end{equation}
Equivalently in $\C^2$ 
\begin{equation}
| Z_1 | ^2 + | Z_2 |^2 = 1 \ ,
\end{equation}
which motivates the choice of the so-called \textit{Hopf coordinates}
\begin{equation}
Z_1 = \sin\th \ e^{i\phi_1} \ , \quad  Z_2 = \cos\th \ e^{i\phi_2} \ ,
\end{equation}
where the range of $\th$ is $[0,\nicefrac{\pi}{2}]$, while  $\phi_1$ 
and $\phi_2$ take values in $[0,2\pi)$ with periodic identification at 
the endpoints. Substituting into the $\R^4$ metric we get
\begin{equation}
\d s^2 =  \d \th^2 + \sin^2\!\th\ \d \phi_1^2 + \cos^2\!\th\ \d \phi_2^2 \ ,
\end{equation}
hence the  $\Sphere^3$ metric and the vielbein in coordinates $(\th,\phi_1,\phi_2)$
\begin{align}
g_{\mu\nu} &= \diag(1, \sin^2\!\th, \cos^2\!\th) \ ,
\\[1em]
E_\mu^A     &=  \diag(1, \sin\th\ , \cos\th\ ) \ .
\end{align}
From this we obtain the following non-zero Christoffel symbols
\begin{equation}
\G^{\phi_1}_{\phi_1\th} = \G^{\phi_1}_{\th \phi_1} = \cot\th \ , \qquad 
\G^{\phi_2}_{\phi_2\th} = \G^{\phi_2}_{\th \phi_2}  = - \tan\th \ , \qquad 
\G^\th_{\phi_1 \phi_1} = - \G^\th_{\phi_2 \phi_2}  = - \cos\th\ \sin\th \ ,
\end{equation}
and putting all this together, the only non-zero spin connection components are
\begin{align}
\omega_{\phi_1}^{1 2} &= - \omega_{\phi_1}^{2 1} = - \cos\th \ ,
\\[1em]
\omega_{\phi_2}^{3 1} &= - \omega_{\phi_2}^{1 3} =  - \sin\th \ .
\end{align}
Note that the tangent space indices corresponding to 
$(\th,\phi_1,\phi_2)$ run from $1$ to $3$.

\section{String solutions on \texorpdfstring{$\R \!\times\! \Sphere^3 \!\times\! \Sphere^3$}{R x S3 x S3}}
\label{FZM:AppRxS3xS3solutions}

In this appendix we describe a simple procedure that yields string 
solutions on  $\R \!\times\! \Sphere^3 \!\times\! \Sphere^3$ from two 
$\R \!\times\! \Sphere^3$ strings. Let us denote the maps from the
decompactified worldsheet to the two spheres $\Sphere^3_\pm$ by
\begin{equation} 
	\Z_\pm \ : \qquad \mathcal{M} \to \Sphere^3_\pm \ . 
\end{equation}
We can think of such maps more explicitly in terms of Hopf coordinates
\eqref{FZM:HopfS3S3} as 
\begin{equation}
\Z_\pm(\mathbf{x}) = \big( \th^\pm (t,x), \phi_1^\pm (t,x), \phi_2^\pm (t,x) \big) \ .
\end{equation}

The static conformal gauge string action on 
$\R\! \times\! \Sphere^3\! \times\! \Sphere^3$ is a sum of two single-sphere
actions \eqref{FZM:HopfActionS3}, weighted by the squared radii of the spheres
\begin{equation}
	S[\Z_+, \Z_-] = \fc2\, S_1[\Z_+] + \fs2\, S_1[\Z_-]\ .
\end{equation}
The equations of motion decouple for $\Z_+$ and $\Z_-$, taking the same form 
\eqref{FZM:HopfBosEom} on both. The Virasoro constraints \eqref{FZM:VirasoroS3},
however, now read
\begin{align}
\begin{split}
&\fc2\, V_1[\Z_+]  + \fs2\, V_1[\Z_-] = \k^2 \ ,
\\[0.5em]
&\fc2\, V_2[\Z_+]  + \fs2\, V_2[\Z_-] =0 \ ,
\end{split}
\end{align}
and connect the two spheres.

If $\Z_1(\mathbf{x})$, $\Z_2(\mathbf{x})$ are two string solutions on 
$\R \times\! \Sphere^3$, i.e. they satisfy \eqref{FZM:HopfBosEom} and
\eqref{FZM:VirasoroS3}, it is a simple exercise to see that 
\begin{equation}
	\Z_{+}(\mathbf{x}) = \Z_1(A \mathbf{x})\ ,
\quad
	\Z_{-}(\mathbf{x}) = \Z_2(B \mathbf{x})\ ,
\end{equation}
constitute a solution on $\R\! \times\! \Sphere^3\! \times\! \Sphere^3$,
as long as the constants $A, B$ satisfy
\begin{equation}\label{FZM:AppABVirasoro}
\frac{A^2}{\c2}  + \frac{B^2}{\s2} = 1 \ ,
\end{equation}
which is nothing but the combined Virasoro constraint.

\section{Pullback of the vielbein and spin connection to the worldsheet}
\label{FZM:AppPullBacks}

Putting the giant magnon \eqref{FZM:s3s3StationaryMagnon} as background, one 
finds the following components for the pulled-back vielbein 
$e_a^A = E_\mu ^A (X)\p_a X^\mu$
\begin{align}
		e_0^0 &= 1 \ , 
	&	e_1^0 &= 0 \ ,
	\\[0.5em]
		e_0^3 &= - \cos\vp\ \frac{u \g^2 \left( \qt^2 - u^2 \right) \tanh\cY }
											{\sqrt{\qt^2 \sinh^2\!\cY + u^2}} \ ,
	&	e_1^3 &=   \cos\vp\ \frac{\g^2 \left( \qt^2 - u^2 \right) \tanh\cY }
											{\sqrt{\qt^2 \sinh^2\!\cY + u^2}} \ ,
	\\[0.5em]
		e_0^4 &=   \cos\vp\ \frac{\left( q^2 u^2 \g^2 + \qt^2 \sinh^2\!\cY\right)\sech\cY }
											{\qt\sqrt{\qt^2 \sinh^2\!\cY + u^2}} \ ,
	&	e_1^4 &=   \cos\vp\ \frac{u \g^2 \left( \qt^2 - u^2 \right) \sech\cY }
											{\qt\sqrt{\qt^2 \sinh^2\!\cY + u^2}} \ ,
	\\[0.5em]	
		e_0^5 &=   \cos\vp\ \frac{q u \g^2 \sqrt{\qt^2 - u^2}\ \sech\cY }
											{\qt} \ ,
	&	e_1^5 &= - \cos\vp\ \frac{q  \g^2 \sqrt{\qt^2 - u^2}\ \sech \cY }
											{\qt} \ ,
	\\[0.5em]
		e_0^7 &= \sin\vp \ , 
	& 	e_1^7  &= 0 \ ,
\end{align}
while the only non-zero components of the spin connection (pulled back to 
the worldsheet) are
\begin{align}
	\omega_0^{34}    &=  - \omega_0^{43}  
			 =   -	\frac{\c2\, \sqrt{\qt^2 - u^2}}{\qt } 
			 		\frac{ \left( q^2 u^2 \g^2 
			 							+ \qt^2 \sinh^2\!\cY\right)\sech\cY }
							{\qt^2 \sinh^2\! \cY + u^2} ,
	\\[1em]
	\omega_1^{34}    &=  - \omega_1^{43}  
			 =  -	\frac{\c2\, u \g^2 \left(\qt^2 - u^2\right)^{3/2}}{\qt } 
			  		\frac{\sech \cY }
							{\qt^2 \sinh^2\! \cY + u^2}  ,
	\\[1em]
		\omega_0^{35}    &=  - \omega_0^{53}  
			 =\ \,   	\frac{\c2\, q u \g^2}{\qt } \
			 		\sqrt{\qt^2 \sinh^2\!\cY + u^2}\ \sech\cY  ,
	\\[1em]
		\omega_1^{35}    &=  - \omega_1^{53}  
			 =  -	\frac{\c2\, q  \g^2}{\qt } \
			 		\sqrt{\qt^2 \sinh^2\!\cY + u^2}\ \sech\cY .
\end{align}

\section{Gamma matrices}
\label{FZM:AppDirac}

In Section \ref{FZM:subsecKappa}--\ref{FZM:subsecDelta0} we demonstrate 
that it is beneficial to work with a set of boosted gamma matrices, related to 
our original 10d Dirac matrices $\G^A, A=0,1,...,9$, by
\begin{align}
\Gh^0 = \sec\vp \left(  \G^0 - \sin\vp\ \G^7   \right) \ ,
\quad
\Gh^7 = \sec\vp \left(  \G^7 - \sin\vp\ \G^0   \right) \ ,
\quad
\Gh^A = \G^A  \quad \text{for } A \neq 0,7  \ .
\end{align}
We pick the representation of $\G^A$ that yields the following forms for $\Gh^A$:
\begin{align}
\Gh^\mu         & = \sigma^1 \otimes \g ^\mu \otimes \Id \otimes \sigma ^2 \otimes \Id \,,\qquad \mu=0,1,2 
\\
\Gh^{n}          & = \sigma^1 \otimes  \Id \otimes \Id \otimes \sigma ^1 \otimes \g ^{n} \,,\qquad n=3,4,5 
\\
\Gh^{\dot{n}} & = \sigma^1 \otimes  \Id \otimes \g ^{\dot{n}} \otimes \sigma ^3 \otimes \Id \,,\qquad \dot{n}=6,7,8 
\\
\Gh^{9}          & = - \sigma^2 \otimes \Id \otimes \Id \otimes \Id \otimes \Id \ ,
\end{align}
where, in terms of the Pauli matrices $\sigma^i$, the 3d gamma matrices $\g ^i$ are given by
\begin{equation}\label{3ddirac}
\g^\mu         =  (-i\sigma^3,   \sigma^1,   \sigma^2) \ , \qquad 
\g^n             =  (  \sigma^1,   \sigma^2,   \sigma^3)  \ ,\qquad
\g^{\dot{n}} =  (  \sigma^2, - \sigma^3, - \sigma^1)  \ .
\end{equation}
In this basis,
\begin{align}
\Gh                 & =   \sigma^3 \otimes \Id \otimes \Id \otimes \Id \otimes \Id \ , \\
\Gh^{12}        & =  \Id \otimes (i \sigma^3) \otimes \Id \otimes \Id \otimes \Id \ , \\
\Gh^{68}        & =  \Id \otimes \Id \otimes (i\sigma^3) \otimes \Id \otimes \Id \ , \\
\Gh^{012345}& =  \Id \otimes \Id \otimes \Id  \otimes \sigma^3 \otimes \Id \ , \\[1em] 
\Gh^{34}        & =  \Id  \otimes \Id \otimes \Id \otimes \Id \otimes ( i \sigma^3) \ , \\
\Gh^{35}        & =  \Id  \otimes \Id \otimes \Id \otimes \Id \otimes (-i \sigma^2) \ , 
\end{align}
and in particular we see that $\Gh$ (Weyl matrix), $\Gh^{12}$, $\Gh^{68}$
 and the projectors $\hat{P}_{\pm} = \frac{1}{2} \left(\Id \pm \Gh^{012345}\right)$ 
 are simultaneously diagonalized. 

Note that in this representation, instead of $\G^A$, it is $\Gh^A$ that have definite 
hermiticity: $\Gh^0$ is anti-hermitian, while $\Gh^i$ is hermitian for $i=1,2,...,9$. 
Accordingly, for the intertwiners $B$, $T$ and $C$, defined by the relations\footnote{
These relations must hold for $\G^A$, not the boosted $\Gh^A$.}
\begin{align}
( \G^{A} )^*				&=  \ \ \, 	B\ \G^{A}\ B^{-1} \ ,
\\[0.5em]
( \G^{A} )^\dagger 	&= 			 - 	T\ \G^{A}\ T^{-1} \ ,
\\[0.5em]
( \G^{A} )^T 				&= 			 - 	C\ \G^{A}\ C^{-1} \ ,
\end{align}
we have
\begin{equation}
B = \G^{1469} \ ,
\quad
T = \Gh^{0} \ ,
\quad
C = T\ B \ .
\end{equation}

\section{Fermion derivatives}
\label{FZM:AppFermionDerivatives}

Looking at equation \eqref{FZM:EoMexpanded} we can define the following fermion derivatives
\begin{align}
	\begin{split}
		\DD   \ &= \ \frac{(1- u) \g}{\c2\, \zeta} 
							\left( \D_1 - \D_0  - \frac{1}{8}(\slashed{H}_0 - \slashed{H}_1) \right)
							_{\p_\cS \to 0}  \ ,
		\\[1em]
		\DDT  \ &= \ \frac{(1+ u) \g}{\c2\, \zeta}
							\left( \D_1 + \D_0  - \frac{1}{8}(\slashed{H}_0 + \slashed{H}_1) \right)
							_{\p_\cS \to 0}  \ ,
	\end{split}
\end{align}
where $\D_a =\p_a+\frac{1}{4}\omega_a^{AB}\G_{AB}$, and $\c2\, \zeta (1\pm u) \g$ 
were introduced to normalize the $\p_\cY$ term.
The NS-NS flux appears as $\slashed{H}_a \equiv\ e_a^A H_{ABC} \G^{BC}$, 
which we can rewrite
\begin{align}\label{FZM:Hslasha}
	\begin{split}
	\slashed{H}_a  &=\ 
			\frac{1}{3} H_{ABC} \left( e_a^A \G^{BC} + e_a^B \G^{CA} 
													 + e_a^C \G^{AB} \right)
	\\[1em]
						  	&=  \sum_{ABC} \frac{1}{6} H_{ABC} 
						   	\Big( e_a^A ( \G_A \G^{ABC}  + \G^{BCA} \G_A) 
						   				+ e_a^B ( \G_B \G^{BCA} +  \G^{CAB} \G_B )
		\\ & \qquad \qquad  \qquad  \qquad 
						 				+ e_a^C (\G_C \G^{CAB} + \G^{ABC} \G_C )\Big)
	\\[0.5em]
							&=  \sum_{ABC} \frac{1}{6} H_{ABC} \sum_{D \in {A,B,C}}  
										e_a^D ( \G_D \G^{ABC}  + \G^{ABC} \G_D)
	\\[1em]
							&=\ \frac{1}{6} H_{ABC} \sum_{D}  
										e_a^D ( \G_D \G^{ABC}  + \G^{ABC} \G_D) 	
	\\[0.5em] 
							&=\ \frac{1}{6} ( \rho_a \slashed{H}  + \slashed{H} \rho_a)\ .	
    \end{split}		
\end{align}
On the first line we used the antisymmetry of $H$, going to the second that 
$\G_A \G^A = \Id$ (no summation), on the third the antisymmetry of 
$\G^{ABC}$, and lastly on the fourth line the fact that for $D \notin \{A,B,C\}$
\begin{equation}
	\G_D \G^{ABC}  + \G^{ABC} \G_D = 0 \ .
\end{equation}
Hence we have
\begin{align}\label{FZM:DDapp1}
	\begin{split}
		\DD   &=  \p_\cY + \ha G\ \G_{34} + \ha Q\ \G_{35}  
						-  \frac{(1-u) \g}{48 \c2\, \zeta} 
							\left( \slashed{H} (\rho_0 - \rho_1) 
									+ (\rho_0 - \rho_1) \slashed{H} \right)\ ,
	\\[1em] 
		\DDT &=  \p_\cY + \ha \tilde{G}\ \G_{34}  + \ha Q\ \G_{35} 
						-  \frac{(1+u) \g}{48 \c2\, \zeta} 
							\left( \slashed{H} (\rho_0 + \rho_1) 
									+ (\rho_0 + \rho_1) \slashed{H} \right)\ .
	\end{split}					
\end{align} 
with
\begin{align}\label{FZM:ExplGQ}
	\begin{split}
		G	&= \frac{\omega_1^{34} - \omega_0^{34}}{\c2\, \zeta (1+ u) \g} 
			=	\ \ \frac{\qt^2(1-u) \cosh^2\!\cY - \qt^2 + u^2}
					{\qt \left(\qt^2 \sinh^2\!\cY + u^2\right)} \sech\cY \ ,
	\\[1em]
		\tilde{G} &= \frac{\omega_1^{34} + \omega_0^{34}}{\c2\, \zeta (1- u) \g} 
			=		- \frac{\qt^2(1+u) \cosh^2\!\cY - \qt^2 + u^2}
					{\qt \left(\qt^2 \sinh^2\!\cY + u^2\right)} \sech\cY \ ,
	\\[1em]                     
		Q  &=\frac{\omega_1^{35} \mp \omega_0^{35}}{\c2\, \zeta (1\pm u) \g}
				=  - \frac{q}{\qt \sqrt{\qt^2 - u^2}} \
			 			\sqrt{\qt^2 \sinh^2\!\cY + u^2}\ \sech\cY \ .
	\end{split}
\end{align}

The next step is to take \eqref{FZM:slashRewrite}
\begin{equation}
\slashed{H} = 24 q \cos\vp \left( \Gh_{*}\  P_{+}  -  \Delta\ \Gh^{12} \right)
\end{equation}
and substitute into \eqref{FZM:DDapp1}, with the further restriction that 
the derivatives act on kappa fixed spinors, as in \eqref{FZM:EomEquivFixed}. 
For $\DD K_1$ the relevant term is
\begin{align}
\begin{split}
			\frac{1}{48 \c2} \Big( \slashed{H} &(\rho_0 - \rho_1) 
												+ (\rho_0 - \rho_1) \slashed{H} \Big)\, K_1  
	\\[1em] &=\ 
			- \frac{1}{24 \cos\vp} \Big( \slashed{H} \Gh^0 K_2 + \Gh^0 K_2 \slashed{H} \Big) K_1
	\\[1em] &=\  
			- q\ \Big( \Gh_{*} P_{+} \Gh^0 K_2 K_1+ \Gh^0 K_2 \Gh_{*} P_{+} K_1 
	\\[0.5em] & \qquad  \qquad
							-\Delta \Gh^{12} \Gh^{0} K_2 K_1 - \Gh^{0} K_2 \Delta \Gh^{12} K_1	 \Big)		 
	\\[1em] &=\ 
			  q\ \Big( P_{-} \Gh^{12} K_2 K_1 +  \left( - K_2 + \Id \right) P_{+} \Gh^{12} K_1 
	\\[0.5em] & \qquad  \qquad
							-\left( \Delta_0 + \Delta_7\, \Gh^{07} \right) \Gh^{12}  K_2 K_1 
							+ \left( \Delta_0 + \Delta_7\, \Gh^{07} \right) \Gh^{12} K_2 K_1
							- \Delta_0 \,\Gh^{12} K_1  \Big)		  
	\\[1em] &=\ 
			- q \left( R P_{-} - \left( R + \Gh_{12}\right) P_{+} + \Delta_0\, \Gh_{12} \right) K_1,	
    \end{split}	
\end{align}
where we have also used the definition of the kappa projectors 
\eqref{FZM:KappaProj}, the form of $\Delta$ in \eqref{FZM:DeltaDef}, 
the relation $K_J \Gh^0 = - \Gh^0 K_J + \Gh^0$, and \eqref{FZM:KKR}.
Similarly, for $\DDT K_2$ we have
\begin{align}
	\frac{1}{48 \c2} \Big( \slashed{H} &(\rho_0 + \rho_1) 
												+ (\rho_0 + \rho_1) \slashed{H} \Big)\, K_2
	\\[1em] &=
	- q \left( R P_{-} - \left( R + \Gh_{12}\right) P_{+} + \Delta_0\, \Gh_{12} \right) K_2,	
\end{align}
and with this, the fermion derivatives take the final form
\begin{align}
\begin{split}
\DD   &=  \p_{\cY} + \ha G\ \Gh_{34} + \ha Q\ \Gh_{35}  
					+ \frac{q (1-u) \g}{\zeta}
						\left(  R P_{-} - (R + \Gh_{12})\,  P_{+}  + \Delta_0\, \Gh_{12} \right)\ ,
\\[1em] 
\DDT &=  \p_{\cY} + \ha \tilde{G}\ \Gh_{34}  + \ha Q\ \Gh_{35} 
					+ \frac{q (1+u) \g}{\zeta} 
						\left(  R P_{-} - (R + \Gh_{12})\,  P_{+}  + \Delta_0\, \Gh_{12}\right)\ .
\end{split}					
\end{align} 
Let us stress one last time, that these forms are only valid when acting on kappa-fixed spinors.

\section{No normalizable solutions for \texorpdfstring{$\Delta \neq 0$}{nonzero Delta}}
\label{FZM:AppDeltaNonZero}

In section \ref{FZM:SecMixedFlux} we found the expected number of 
normalizable solutions in an analytic form for $\Delta = 0$. 
However, to complete the counting argument for fermion zero modes, 
it is necessary to demonstrate that there are no normalizable solutions 
at all for $\Delta \neq 0$ . This happens for the maximally SUSY \AdsSSS 
giant magnon, on the $\G_{1268} = +1$ spinor subspace:
\begin{align}
\begin{split}
	&\Delta = - \tan^2\!\vp\ \Gh^0 - \tan\vp\sec\vp\ \Gh^7 
		= \left(\k^2 - \k\td{\k}\ \Gh_{07} \right) \Gh_0 \ ,
	\\[1em]
	& \qquad\qquad \k = \tan\vp \ , \quad \td{\k}= \sqrt{1+\k^2} = \sec\vp \ .
\end{split}
\end{align}
The equations of motion are
\begin{align}  \label{FZM:AppDeltaEom}
	\begin{split}
		\zeta (1+ u)\g \DD\ \Psi^1  
			+ \qt \left ( R\ P_{-}  -  K_1 \Delta\, \Gh_{*} \right) \Psi^2  &= 0 \ ,
	\\[1em] 
		\zeta (1- u)\g  \DDT\ \Psi^2
		 	-  \qt \left( R\ P_{-} -  K_2 \Delta\, \Gh_{*} \right) \Psi^1   &= 0 \ .
\end{split}
\end{align}
with fermion derivatives
\begin{align}
\begin{split}
\DD   &=  \p_{\cY} + \ha G\ \Gh_{34} + \ha Q\ \Gh_{35}  
					+ \frac{q (1-u) \g}{\zeta}
						\left(  R P_{-} - (R + \Gh_{12})\,  P_{+}  - \k^2\, \Gh_{12} \right)\ ,
\\[1em] 
\DDT &=  \p_{\cY} + \ha \tilde{G}\ \Gh_{34}  + \ha Q\ \Gh_{35} 
					+ \frac{q (1+u) \g}{\zeta} 
						\left(  R P_{-} - (R + \Gh_{12})\,  P_{+}  - \k^2\, \Gh_{12}\right)\ .
\end{split}					
\end{align} 
Our approach will be similar to section \ref{FZM:SecMixedFlux}. 
First we write  down general kappa-fixed spinors, which we then 
substitute into the equations of motion to get a system of simpler ODEs.

\paragraph*{Kappa fixing.}
The main difference from $\Delta = 0$ is that the solutions will not have definite 
$P_\pm$ chirality, since $\Delta$ mixes the $P_+$ and $P_-$ subspaces.
Accordingly, the kappa-fixed ansatz generalizing \eqref{FZM:KappaFixedAnsatz} 
will have to relate the two projections. This is achieved by
\begin{equation}\label{FZM:KappaFixedAnsatzDelta}
\Psi^{J} = \sum_{\lambda = \pm} \left[ 
				\left( \alpha^{J}_{+} + \alpha^{J}_{-}\, \Gh_{45} \right) f_{J}(\cY)
			+	\left( \bar{\alpha}^{J}_{+} + \bar{\alpha}^{J}_{-}\, \Gh_{45} \right) g_{J}(\cY)\, \Gh_{07}
			\right]\, U_{\lambda} \ ,
\end{equation}
where the constant spinor $U_\lambda$ is shared between $I=1$ and $2$, 
and has eigenvalues $\G_{34} U_\lambda = +i U_\lambda$, 
$P_{-} U_\lambda = U_\lambda$ and $\Gh_{12} U_\lambda = i \lambda U_\lambda$.
The functions $f_{J}$, $g_{J}$ represent the parts of the solution 
on the $P_-$ and $P_+$ subspaces respectively, and $\Gh_{07}$ transforms 
$U_\lambda$ between the two. We take $\alpha^{J}_{\pm}$ to be defined 
by \eqref{FZM:Alphas1}--\eqref{FZM:Alphas2}, and
\begin{equation}
	\bar{\alpha}^{J}_{\pm} \ \equiv\ \alpha^{J}_{\pm} \arrowvert_{\lambda \to -\lambda} \ .
\end{equation}
This is because the definition of $\lambda$ here 
differs from that in section \ref{FZM:SecMixedFlux}, the two agree 
on the $P_-$ subspace, while on $P_+$ they are related by a minus sign.

\paragraph*{The $\mathbf{K\Delta\Gh_{*}}$ terms.}
For the most part, substitution yields equations that are familiar from 
section \ref{FZM:SecMixedFlux}, the only new terms being 
$K_1 \Delta\, \Gh_{*} \Psi^2$ and  $K_2 \Delta\, \Gh_{*} \Psi^1$.
It is easy to see that
\begin{align}
	\Delta\, \Gh_{*} \Psi^J \ =\
	\sum_{\lambda = \pm} \bigg[ 
				&\left( \alpha^{J}_{+} + \alpha^{J}_{-}\, \Gh_{45} \right) 
					i\lambda \left( \k^2 - \k\td{\k}\ \Gh_{07} \right) f_{J}
	\\[1em]			
			+	&\left( \bar{\alpha}^{J}_{+} + \bar{\alpha}^{J}_{-}\, \Gh_{45} \right) 
					i\lambda \left( \k^2\ \Gh_{07} - \k\td{\k}\right) g_{J}
			\bigg]\, U_{\lambda} \ .
\end{align}
On the other hand, from \eqref{FZM:KappaRewrite} and the definitions 
\eqref{FZM:Alphas1}--\eqref{FZM:Alphas2} one can derive the action 
of the kappa-projectors on a general spinor $V = V_+ + V_-$ 
on the $P_-$ subspace, with components $\Gh_{34} V_\pm= \pm i V_\pm$
\begin{align}
\begin{split}
	K_1 V = \left( \alpha^{1}_{+} + \alpha^{1}_{-}\, \Gh_{45} \right) 
			&\bigg[ \ha e^{-i \chi} \sqrt{1 - \lambda Q_{+}\, \sech\cY}\ V_+
	\\ &\quad 
					- \ha \lambda e^{i \chi} \sqrt{1 + \lambda Q_{+}\, \sech\cY}\ 
					\Gh_{45} V_- \bigg]\ ,
	\\[1em]
	K_2 V = \left( \alpha^{2}_{+} + \alpha^{2}_{-}\, \Gh_{45} \right) 
			&\bigg[ \ha e^{-i \td\chi} \sqrt{1 + \lambda Q_{-}\, \sech\cY}\ V_+
	\\ &\quad 
					+ \ha \lambda e^{i \td\chi} \sqrt{1 - \lambda Q_{-}\, \sech\cY}\ 
					\Gh_{45} V_- \bigg]\ .
\end{split}
\end{align}
The corresponding expressions for the $P_+$ subspace are obtained by 
sending $\lambda \to -\lambda$. 
Putting these together we get
\begin{align}
\begin{split}
K_1 \Delta\, \Gh_{*} \Psi^2 \ =\
\sum_{\lambda = \pm} \bigg[ 
				&\left( \alpha^{1}_{+} + \alpha^{1}_{-}\, \Gh_{45} \right) 
					\left[i \lambda \qt \k^2 \delta_1\, f_2 - i \lambda \qt \k \td\k \delta_2\, g_2 \right]	
	\\
			+	&\left( \bar{\alpha}^{1}_{+} + \bar{\alpha}^{1}_{-}\, \Gh_{45} \right) 
					\left[i \lambda \qt \k^2 \bar{\delta}_1\, g_2 - i \lambda \qt \k \td\k \bar{\delta_2}\, f_2 \right]
					\Gh_{07} \bigg]\, U_{\lambda} \ ,
\\[1em]
K_2 \Delta\, \Gh_{*} \Psi^1 \ =\
\sum_{\lambda = \pm} \bigg[ 
				&\left( \alpha^{2}_{+} + \alpha^{2}_{-}\, \Gh_{45} \right) 
					\left[- i \lambda \qt \k^2 \bar{\delta}_1\, f_1 - i \lambda \qt \k \td\k \delta_2\, g_1 \right]	
	\\
			+	&\left( \bar{\alpha}^{2}_{+} + \bar{\alpha}^{2}_{-}\, \Gh_{45} \right) 
					\left[- i \lambda \qt \k^2 \delta_1\, g_1 - i \lambda \qt \k \td\k \bar{\delta_2}\, f_1 \right]
					\Gh_{07} \bigg]\, U_{\lambda} \ ,					
\end{split}
\end{align}
with
\begin{align}
\begin{split}
	\delta_1  = \ha \Bigg( 
		&e^{i (\td\chi - \chi)} \sqrt{\left(1 - \lambda Q_{+}\, \sech\cY\right)
														\left( 1 + \lambda Q_{-}\, \sech\cY \right)}
		\\
	& \quad - e^{-i (\td\chi - \chi)} \sqrt{\left(1 + \lambda Q_{+}\, \sech\cY\right)
														\left( 1 - \lambda Q_{-}\, \sech\cY \right)}  \Bigg) \ ,
\\[1em]
	\delta_2  = \ha \Bigg( 
		&e^{i (\td\chi - \chi)} \sqrt{\left(1 - \lambda Q_{+}\, \sech\cY\right)
														\left( 1 - \lambda Q_{-}\, \sech\cY \right)}
		\\
	& \quad + e^{-i (\td\chi - \chi)} \sqrt{\left(1 + \lambda Q_{+}\, \sech\cY\right)
																	\left( 1 + \lambda Q_{-}\, \sech\cY \right)}  \Bigg)\ ,
\end{split}
\end{align}
and $\bar\delta_J = \delta_J \arrowvert_{\lambda \to - \lambda}$.

\paragraph*{Reduced equations.}

Substituting \eqref{FZM:KappaFixedAnsatzDelta} into \eqref{FZM:AppDeltaEom} we get
\begin{align}
\begin{split}
\sum_{\lambda = \pm} \Bigg[ 
				\left( \alpha^{1}_{+} + \alpha^{1}_{-}\, \Gh_{45} \right) 
					&\bigg[ \p_\cY f_1 + \left( C_{11} - i\lambda q \k^2 \frac{(1-u)\g}{\zeta} \right) f_1
			\\[0.5em] & \
							+ \left( C_{12}  - i \lambda \qt \k^2 \delta_1 \frac{(1-u)\g}{\zeta} \right) f_2
							+ i \lambda \qt \k \td\k \delta_2 \frac{(1-u)\g}{\zeta} \, g_2 \bigg]
	\\[1em]
			+ \left( \bar{\alpha}^{1}_{+} + \bar{\alpha}^{1}_{-}\, \Gh_{45} \right) 
					&\bigg[ \p_\cY g_1 - \left( C_{+} + i\lambda q \k^2 \frac{(1-u)\g}{\zeta} \right) g_1
			\\[0.5em] & \
							+ i \lambda \qt \k \td\k \bar{\delta_2}\frac{(1-u)\g}{\zeta}\, f_2
							- i \lambda \qt \k^2 \bar{\delta}_1 \frac{(1-u)\g}{\zeta} \, g_2 \bigg]
					\Gh_{07} \Bigg]\, U_{\lambda}\ = \ 0 \ ,
\end{split}
\\[1em]
\begin{split}
\sum_{\lambda = \pm} \Bigg[ 
				\left( \alpha^{2}_{+} + \alpha^{2}_{-}\, \Gh_{45} \right) 
					&\bigg[ \p_\cY f_2 + \left( C_{21} - i\lambda q \k^2 \frac{(1+u)\g}{\zeta} \right) f_2
			\\[0.5em] & \
							+ \left( C_{22} - i \lambda \qt \k^2 \bar{\delta}_1 \frac{(1+u)\g}{\zeta} \right) f_1
							- i \lambda \qt \k \td\k \delta_2 \frac{(1+u)\g}{\zeta} \, g_1 \bigg]
	\\[1em]
			+ \left( \bar{\alpha}^{2}_{+} + \bar{\alpha}^{2}_{-}\, \Gh_{45} \right) 
					&\bigg[ \p_\cY g_2 - \left( C_{-} + i\lambda q \k^2 \frac{(1+u)\g}{\zeta} \right) g_2
			\\[0.5em] & \
							-i \lambda \qt \k \td\k \bar{\delta_2}\frac{(1+u)\g}{\zeta}\, f_1
							- i \lambda \qt \k^2 \delta_1 \frac{(1+u)\g}{\zeta} \, g_1 \bigg]
					\Gh_{07} \Bigg]\, U_{\lambda}\ = \ 0 \ ,			
\end{split}
\end{align}
where $C_\pm$ and $C_{ij}$ are as defined in \eqref{FZM:Cpm}, \eqref{FZM:Cij}. 
If we make the ansatz
\begin{align}
\begin{split}
&f_1 = \frac{e^{- \int C_{11}\d\cY}}{\sqrt{1+u}}\ \td f_1 \ ,
\qquad
g_1 = \frac{e^{ \int C_{+}\d\cY}}{\sqrt{1+u}}\ \td g_1 \ ,
\\[1em]
&f_2 = \frac{e^{- \int C_{21}\d\cY}}{\sqrt{1-u}}\ \td f_2 \ ,
\qquad
g_2 = \frac{e^{ \int C_{-}\d\cY}}{\sqrt{1-u}}\ \td g_2 \ ,
\end{split}
\end{align}
we get the following four equations
\begin{align}
\begin{split}
		& \left( \p_\cY  - i\lambda q \k^2 \frac{(1-u)\g}{\zeta} \right) \td f_1
							+ \left( e^{+i 2 \lambda \xi \cY} \left( \lambda \tanh\cY - i \xi \right)
								 - i \lambda \qt \k^2 \frac{\delta_1}{\zeta} e^{ \int \left(C_{11}-C_{21}\right) \d\cY} \right) \td f_2
		\\[0.5em] & \qquad \qquad
							+ i \lambda \qt \k \td\k \frac{\delta_2}{\zeta}e^{ \int \left(C_{11}+C_{-}\right) \d\cY} \, \td g_2 
							\ = \ 0 \ ,
\end{split}
\\[1em]
\begin{split}
		& \left( \p_\cY  - i\lambda q \k^2 \frac{(1+u)\g}{\zeta} \right) \td f_2
							+ \left( e^{-i 2 \lambda \xi \cY} \left( \lambda \tanh\cY + i \xi \right)
								 - i \lambda \qt \k^2 \frac{\bar\delta_1}{\zeta} e^{ \int \left(C_{21}-C_{11}\right) \d\cY} \right) \td f_1
		\\[0.5em] & \qquad \qquad
							- i \lambda \qt \k \td\k \frac{\delta_2}{\zeta} e^{ \int \left(C_{21}+C_{+}\right)\d\cY} \, \td g_1
							\ = \ 0 \ ,
\end{split}
\\[1em]
\begin{split}
		& \left( \p_\cY  -  i\lambda q \k^2 \frac{(1-u)\g}{\zeta}\right)  \td g_1
							+ i \lambda \qt \k \td\k \frac{\bar\delta_2}{\zeta} e^{-\int \left(C_{21}+C_{+}\right)\d\cY}\, \td f_2
				\\[0.5em] & \qquad \qquad
							- i \lambda \qt \k^2 \frac{\bar\delta_1}{\zeta}e^{\int \left(C_{-} - C_{+}\right)\d\cY} \, \td g_2 
							\ = \ 0\ ,
\end{split}
\\[1em]
\begin{split}
		& \left( \p_\cY  -  i\lambda q \k^2 \frac{(1+u)\g}{\zeta}\right)  \td g_2
							- i \lambda \qt \k \td\k \frac{\bar\delta_2}{\zeta} e^{-\int \left(C_{11}+C_{-}\right)\d\cY}\, \td f_1
				\\[0.5em] & \qquad \qquad
							- i \lambda \qt \k^2 \frac{\delta_1}{\zeta}e^{\int \left(C_{+} - C_{-}\right)\d\cY} \, \td g_1 
							\ = \ 0\ .
\end{split}
\end{align}
The first thing to observe is that setting $\k =0$ the functions $\td f_1, \td f_2$ 
decouple from $\td g_1, \td g_2$, and indeed we recover the $\Delta = 0$
solutions found in section \ref{FZM:SecMixedFlux}. 

\paragraph*{Pure R-R background.}
We have not been able to find exact solutions at general values of $q$ and $\k$, 
nonetheless, we can give an argument for their non-normalizability if we 
consider an expansion in powers of $q$ and $\k$.
It turns out we can already see non-normalizability at 
leading order in $q$, i.e. at $q=0$, with the equations simplifying to
\begin{align}\label{FZM:NonNormKEom}
\begin{split}
&\p_{\cY}\td f_1\ +\     \lambda \td\k^2 \tanh\cY\ \td f_2\ + \ 
		i\ \lambda \k \td\k\ \sech\cY\ \td g_2\   =\ 0 \ ,
\\[0.5em]
&\p_{\cY}\td f_2\ +\    \lambda \td\k^2 \tanh\cY\ \td f_1\ -\
		 i\ \lambda \k \td\k\ \sech\cY\ \td g_1\   =\ 0 \ ,
\\[0.5em]
&\p_{\cY}\td g_1\ +\  \lambda \k^2 \tanh\cY\ \td g_2\ +\ 
		i\ \lambda\k \td\k\ \sech\cY\ \td f_2\   =\ 0 \ ,
\\[0.5em]
&\p_{\cY} \td g_2\ +\  \lambda \k^2 \tanh\cY\ \td g_1\ -\ 
		i\ \lambda \k \td\k\ \sech\cY\ \td f_1\   =\ 0 \ .
\end{split}
\end{align}

\paragraph*{Zeroth order in $\k$.} 
The first thing to observe is that setting $\k =0$ leads to a significant simplification of the equations. 
$\td f_1, \td f_2$ decouple from $\td g_1, \td g_2$, and the solutions take the form
\begin{align}
\begin{split}
	&\td f_1 = \phantom{\lambda} c_1 \sech\cY  + \phantom{\lambda} c_2 \cosh\cY  \ ,
	\quad 
	\td g_1 = c_3 \ ,
	\\[0.5em]
	&\td f_2 = \lambda  c_1 \sech\cY  - \lambda c_2  \cosh\cY  \ ,
	\quad
	\td g_2 = c_4 \ .
\end{split}
\end{align}
This limit corresponds to the case of $\Delta = 0$, and the solutions match those 
found in section \ref{FZM:SecMixedFlux}, after we set $q=0$. 
Let us denote the only normalizable solution 
in the $\k \rightarrow 0$ limit by
\begin{equation}\label{FZM:kZeroNorm}
\begin{aligned}
&\td f^{(0)}_1 = \phantom{\lambda} C_0\, \sech\cY  \ , 
\qquad 
\td g^{(0)}_1= 0\ ,
\\
&\td f^{(0)}_2 = \lambda C_0\, \sech\cY \ ,
\qquad 
\td g^{(0)}_2 = 0 \ .
\end{aligned}
\end{equation}

\paragraph*{Expansion in $\k$.} 
Introducing the vector notation $\mathbf{f} = (\td f_1, \td f_2, \td g_1, \td g_2)^\top$,
the equations above, for general values of $\k$, can be written as
\begin{equation}
\p_\cY \mathbf{f} + M_\k(\cY) \mathbf{f} = \mathbf{0} \ .
\end{equation}
Since $M_\k(\cY)$ is regular at $\k=0$, we can make the ansatz
\begin{equation}
\mathbf{f} = \sum_{n=0}^\infty \k^n \mathbf{f}^{(n)} \ ,
\end{equation}
where $\mathbf{f}^{(n)} = (\td f^{(n)}_1,\td  f^{(n)}_2, \td g^{(n)}_1, \td g^{(n)}_2)^\top$
are independent of $\k$. 
Substituting this into the equations, then expanding in $\k$, we get a 
system of ODEs for each power of $\k$: for all $n$ the $\mathbf{f}^{(n)}$ 
equations will have the same homogeneous part as the $\k =0$ system, 
and the forcing terms will be given by some linear combination of lower order solutions 
\begin{equation}\label{FZM:kExpEom}
\p_\cY \mathbf{f}^{(n)} + M_0(\cY) \mathbf{f}^{(n)} 
			= \sum_{k=0}^{n-1} {F^n}_k\, \mathbf{f}^{(k)} \ .
\end{equation}
We need to solve these order-by-order, and for normalizability at 
generic values of $\k$, we would need all $\mathbf{f}^{(n)}$ to be normalizable.

\paragraph*{First order in $\k$.} 
At zeroth order we simply have the homogeneous $\k=0$ equations, and the normalizable $\mathbf{f}^{(0)}$ solution is \eqref{FZM:kZeroNorm}. The first subleading solution $\mathbf{f}^{(1)}$ is obtained from \eqref{FZM:kExpEom} with 
\begin{equation}
{F^1}_0 = i \lambda\, \sech\cY
	\begin{pmatrix}
		 0 &  0 &  0 & -1 \\
		 0 &  0 &  1 &  0 \\
		 0 & -1 &  0 &  0 \\
		 1 &  0 &  0 &  0 \\
	\end{pmatrix} \ ,
\end{equation}
and is given by
\begin{equation}\label{FZM:kFirstNorm}
\begin{aligned}
\td f^{(1)}_1\, =\, c_3 \sech\cY + c_4 \cosh\cY  \ , \qquad
\td g^{(1)}_1\, &= \, c_1 - \phantom{\lambda} i\ C_0 \tanh\cY\ ,
\\
\td f^{(1)}_2\, =\, c_3 \sech\cY - c_4 \cosh\cY\ , \qquad 
\td g^{(1)}_2\, &= \, c_2 - i\lambda \ C_0 \tanh\cY  \ .
\end{aligned}
\end{equation}
The terms with $C_0$ are fixed, they are the response to the zeroth order 
($\k=0$) solution \eqref{FZM:kZeroNorm}, while the integration constants
 $c_j$ for $j=1,...,4$ parametrize the homogeneous solution.
We see that there is no combination of $c_j$ that would make all components 
normalizable, in particular, $\td g^{(1)}_J$ can be chosen to decay at either 
$\cY \rightarrow \infty$ or $\cY \rightarrow -\infty$, but not both.

It is already impossible to find a decaying solution at first order in $\k$, 
and we conclude that there are no normalizable solutions for $\Delta \neq 0$.

\section{Phase identities}
\label{FZM:AppPhase}

The following formulae are useful when deriving the reduced equations of motion
\eqref{FZM:EomPplusU} and \eqref{FZM:EomPminU}. Using simple trigonometric and
hyperbolic identities and Euler's formula it is easy to see that \vspace{-0.5em}
\begin{align}
e^{i \arccot \left( \alpha \csch \cY \right)}\ &=\
		i \left( \frac{\sinh\cY - i \alpha}{\sinh\cY +  i \alpha}\right)^{1/2}\ ,
\\[1em]
e^{i \arcsin \left( \frac{\tanh\cY}{\sqrt{1-\alpha^2\, \sech^2\!\cY}}\right) }\ &=\
		i\, \frac{\tanh\cY - i \sqrt{1-\alpha^2} \sech\cY}{\sqrt{1-\alpha^2\, \sech^2\!\cY}}\ ,
\end{align}
and with these we have
\begin{align} 
\begin{split}
e^{i \chi}          \  & = \ \phantom{i} 
				\left( \frac{\qt \sinh \cY - i u}{\qt\sinh\cY + i u} \right)^{1/4} 
				\left( \frac{\tanh\cY + i \sqrt{1-Q_{+}^2} \sech\cY}
									{\sqrt{1-Q_{+}^2\, \sech^2\!\cY}}        \right)^{1/2} \ ,
\\[1em]
e^{i \td{\chi}}   \ & =  \ i 
				\left( \frac{\qt \sinh \cY - i u}{\qt\sinh\cY + i u} \right)^{1/4} 
				\left( \frac{\tanh\cY - i \sqrt{1-Q_{-}^2} \sech\cY}
									{\sqrt{1-Q_{-}^2\, \sech^2\!\cY}}        \right)^{1/2} \ ,
\end{split}
\end{align}
where, as defined in \eqref{FZM:logPhasesq},
\begin{align} 
	\begin{split}
		\chi	    & = \ha \left( 
								\arccot \left( \frac{u \csch \cY}{\qt} \right) 
							- 	\arcsin \left( \frac{\tanh\cY}
													{\sqrt{1-Q_+^2\, \sech^2\!\cY}}
							\right) \right) \ ,
	\\[1em] 
		\td{\chi} & = \ha \left( 
								\arccot \left( \frac{u \csch \cY}{\qt} \right) 
							+ 	\arcsin \left( \frac{\tanh\cY}
													{\sqrt{1-Q_-^2\, \sech^2\!\cY}}
							\right) \right) \ .
	\end{split}
\end{align}

\section{The \texorpdfstring{$\ce{\su(1|1)^2}$}{su(1|1)**2 c.e.} algebra and its representaions}
\label{App:S1Alg}

In this section we will build up the centrally extended $\su(1|1)^2$ algebra, 
and look at its short representations. This is the off-shell symmetry algebra 
of the light-cone gauge fixed superstring theory on \AdsSSS \cite{Borsato:2015mma}, 
and as such, is of great importance in understanding the $\AdS_3 / \CFT_2$ duality.

\subsection{The \texorpdfstring{$\su(1|1)$}{su(1|1)} algebra}

The algebra $\psu(1|1)$ consist of the anticommuting supercharges $\genQ$ 
and $\genS$ , and if we add the central charge $\genH$, i.e. introduce 
the non-trivial anticommutation relation
\begin{equation}
\acomm{\genQ}{\genS} = \genH \ ,
\end{equation}
we get the $\su(1|1)$ algebra.
In its simplest non-trivial representation a bosonic state $\ket{\phi}$ and 
a fermionic state $\ket{\psi}$ transform under the charges according to
\begin{equation}\label{alg:su(1|1)Rep}
  \begin{aligned}
   	\genQ \ket{\fixedspaceL{\psi}{\phi}} &= a \ket{\psi} \ , \quad &
   	\genS \ket{\fixedspaceL{\psi}{\phi}} &= 0 \ , \quad &
   	\genH \ket{\fixedspaceL{\psi}{\phi}} &= H \ket{\phi} \ ,
  \\
	\genQ \ket{\psi} &= 0 \ , \quad &
   	\genS \ket{\psi} &= b \ket{\phi} \ , \quad &
   	\genH \ket{\psi} &= H \ket{\psi} \ .
  \end{aligned}
\end{equation}
For closure of the algebra the eigenvalue of the central charge must be 
$H = ab$. In fact the representation is labelled by $H$ alone, the ratio of 
$a$ and $b$ is physically irrelevant, it only parametrizes the difference in 
normalization of the states $\ket{\phi}$ and $\ket{\psi}$. Let us denote 
this representation by $(\rep{1}|\rep{1})_H$.

\subsection{The \texorpdfstring{$\su(1|1)^2$}{su(1|1)**2} algebra}

In the direct product of two $\su(1|1)$ algebras we have two copies 
(left and right) of each charge, satisfying
\begin{equation}\label{alg:su(1|1)^2alg}
  \begin{aligned}
    \acomm{\genQ_{\sL}}{\genS_{\sL}} &= \genH_{\sL} , \qquad &
    \acomm{\genQ_{\sL}}{\genQ_{\sR}} &= 0 , \qquad &
    \acomm{\genQ_{\sL}}{\genS_{\sR}} &= 0 , \\
    \acomm{\genQ_{\sR}}{\genS_{\sR}} &= \genH_{\sR} , &
     \acomm{\fixedspaceL{\genQ_{\sL}}{\genS_{\sL}}}{\fixedspaceL{\genQ_{\sR}}{\genS_{\sR}}} &= 0 , &
    \acomm{\genQ_{\sR}}{\genS_{\sL}} &= 0 .
  \end{aligned}
\end{equation}
When coupling these two systems, we can introduce the total Hamiltonian 
$\genH$ and the angular momentum $\genM$
\begin{equation}
  \genH = \genH_{\sL} + \genH_{\sR} , \qquad
  \genM = \genH_{\sL} - \genH_{\sR} .
\end{equation}
In terms of these generators we have
\begin{equation}
  \acomm{\genQ_{\sL}}{\genS_{\sL}} = \tfrac{1}{2} \left(\genH + \genM\right) , \qquad
  \acomm{\genQ_{\sR}}{\genS_{\sR}} = \tfrac{1}{2} \left(\genH - \genM\right) .
\end{equation}

\paragraph*{Representations.}
Irreducible representations will be tensor products of a left-moving and a 
right-moving part, since the algebra is a direct product. For later convenience 
we take $\genS_{\sL}$ and $\genQ_{\sR}$ to be raising operators, while 
$\genQ_{\sL}$ and $\genS_{\sR}$ will be lowering operators. 
A highest weight state then satisfies
\begin{equation}
\genS_{\sL} \ket{\text{h.w.}} = 0 , \qquad
\genQ_{\sR} \ket{\text{h.w.}} = 0 .
\end{equation}

In a \emph{short} representation a highest weight state will be annihilated 
by additional supercharges. For the $\su(1|1)^2$ algebra the two shortening 
conditions are $H_{\sL} = 0$ and $H_{\sR} = 0$. A representation where 
the h.w.~state has vanishing $H_{\sR}$, and is therefore annihilated by 
$\genS_{\sR}$, is called a \emph{left-moving} representation. The simplest 
non-trivial example is given by $(\rep{1}|\rep{1})_H \otimes \rep{1}$, with 
a bosonic state $\ket{\phi}$ and a fermionic state $\ket{\psi}$ transforming as
\begin{equation}\label{alg:su(1|1)^2Rep}
\begin{aligned}
   	\genQ_{\sL} \ket{\fixedspaceL{\psi}{\phi}} &= a \ket{\psi} \ , \qquad &
    \genS_{\sL} \ket{\fixedspaceL{\psi}{\phi}} &= 0 \ , \qquad &
    \genH_{\sL} \ket{\fixedspaceL{\psi}{\phi}} &= H \ket{\phi} \ ,
\\
    \genQ_{\sL} \ket{\psi} &= 0 \ , \qquad &
    \genS_{\sL} \ket{\psi} &= b \ket{\phi} \ , \qquad &
    \genH_{\sL} \ket{\psi} &= H \ket{\psi} \ ,
\\
    \genQ_{\sR} \ket{\fixedspaceL{\psi}{\phi}} &= 0 \ , \qquad &
    \genS_{\sR} \ket{\fixedspaceL{\psi}{\phi}} &= 0 \ , \qquad &
    \genH_{\sR} \ket{\fixedspaceL{\psi}{\phi}} &= 0 \ , 
\\
    \genQ_{\sR} \ket{\psi} &= 0 \ , \qquad &
    \genS_{\sR} \ket{\psi} &= 0 \ , \qquad &
    \genH_{\sR} \ket{\psi} &= 0 \ .
\end{aligned}
\end{equation}
with $H = ab$. We also have \emph{right-moving} representations with 
$H_{\sL}=0$, whose highest weight states are annihilated by 
$\genQ_{\sL}$. An example is $\rep{1} \otimes (\rep{1}|\rep{1})_H$, 
in which the right generators act on the two states  $\ket{\bar{\phi}}$ 
and $\ket{\bar{\psi}}$ as in (\ref{alg:su(1|1)Rep}), and all the left 
generators annihilate them. 

\subsection{The centrally extended \texorpdfstring{$\su(1|1)^2$}{su(1|1)**2} algebra}

We can extend the $\su(1|1)^2$ algebra by introducing two additional 
central charges $\genC$ and $\genCbar$. These appear in anticommutators 
between the two sectors, and we take the choice\footnote{
	Alternatively we could have taken $\acomm{\genQ_{\sL}}{\genS_{\sR}} = \genC$, 
	but this deformation was ruled out for the case of $\AdS_3 / \CFT_2$, 
	by considering the length-changing effects on the spin-chain \cite{Borsato:2012ud}.
}
\begin{equation}\label{alg:su(1|1)^2ceAlg}
  \begin{aligned}
    \acomm{\genQ_{\sL}}{\genS_{\sL}} &= \genH_{\sL} \ , \qquad &
    \acomm{\genQ_{\sL}}{\genQ_{\sR}} &= \genC \ , \qquad &
    \acomm{\genQ_{\sL}}{\genS_{\sR}} &= 0 \ , \\
    \acomm{\genQ_{\sR}}{\genS_{\sR}} &= \genH_{\sR} \ , &
    \acomm{\fixedspaceL{\genQ_{\sL}}{\genS_{\sL}}}{\fixedspaceL{\genQ_{\sR}}{\genS_{\sR}}} &= \genCbar \ , &
    \acomm{\genQ_{\sR}}{\genS_{\sL}} &= 0 \ .
  \end{aligned}
\end{equation}

Note that  $\ce{\su(1|1)^2}$ is not of direct product form, i.e. we cannot 
construct its irreducible representations from irreps of the two sectors. 
To make connection to the physics, from now on we use the subscript $p$ 
on the states and representation parameters, indicating that these depend 
on the momentum of the excitation. Let us now consider the short 
representations of this algebra.

\subsubsection{The left-moving representation}
The generalization of (\ref{alg:su(1|1)^2Rep}) compatible with the above 
deformation is given by
\begin{equation}\label{alg:chiral-rep}
  \boxed{\varrho_{\sL}:} \qquad\quad
  \begin{aligned}
    \genQ_{\sL} \ket{\phi_p^{\sL}} &= a_p \ket{\psi_p^{\sL}} , \qquad &
    \genQ_{\sL} \ket{\psi_p^{\sL}} &= 0 , \\
    \fixedspaceL{\genQ_{\sL}}{\genS_{\sL}} \ket{\phi_p^{\sL}} &= 0 , \qquad &
    \fixedspaceL{\genQ_{\sL}}{\genS_{\sL}} \ket{\psi_p^{\sL}} &= b_p \ket{\phi_p^{\sL}} , \\
    \genQ_{\sR} \ket{\phi_p^{\sL}} &= 0 , \qquad &
    \genQ_{\sR} \ket{\psi_p^{\sL}} &= c_p \ket{\phi_p^{\sL}} , \\
    \fixedspaceL{\genQ_{\sR}}{\genS_{\sR}} \ket{\phi_p^{\sL}} &= d_p \ket{\psi_p^{\sL}} , \qquad &
    \fixedspaceL{\genQ_{\sR}}{\genS_{\sR}} \ket{\psi_p^{\sL}} &= 0 ,
  \end{aligned}
  \qquad\qquad\quad
\end{equation}
with central charges
\begin{equation}
  \begin{aligned}
    \genH_{\sL} \ket{\phi_p^{\sL}} &= a_p b_p \ket{\phi_p^{\sL}} , \quad &
    \genC \ket{\phi_p^{\sL}} &= a_p c_p \ket{\phi_p^{\sL}} , \\
    \genH_{\sR} \ket{\phi_p^{\sL}} &= c_p d_p \ket{\phi_p^{\sL}} , \quad &
    \genCbar \ket{\phi_p^{\sL}} &= b_p d_p \ket{\phi_p^{\sL} } . 
  \end{aligned}
\end{equation}

\paragraph*{Shortening condition.} The highest weight state 
$\ket{\phi_p^{\sL}}$ is annihilated by the raising operators 
$\genS_{\sL}$ and $\genQ_{\sR}$, but also satisfies the condition
\begin{equation}
  (\genH_{\sR} \genQ_{\sL} - \genC \genS_{\sR}) \ket{\phi_p^{\sL}} 
  = (a_p c_p d_p - a_p c_p d_p) \ket{\psi_p^{\sL}} = 0.
\end{equation}
Since this particular combination of the lowering operators $\genQ_{\sL}$ 
and $\genS_{\sR}$ annihilates the h.w.~state, the representation is short.
The state $\ket{\phi_p^{\sL}}$ must also be annihilated by the anticommutator 
$\acomm{\genS_{\sL}}{\genH_{\sR} \genQ_{\sL} - \genC \genS_{\sR}}  = \genH_{\sL} \genH_{\sR} - \genC\genCbar$, 
but this is a central charge, implying that
\begin{equation}\label{alg:shortening-left-su(1|1)^2}
  (\genH_{\sL} \genH_{\sR} - \genC\genCbar) \ket{\chi_p^{\sL}} = 0 
\end{equation}
for all states $\chi_p^{\sL} = \phi_p^{\sL}$, $\psi_p^{\sL}$ in the 
representation. This shortening condition, when applied to physical states, 
will play the role of the dispersion relation.

\subsubsection{The right-moving representation}
For this representation the role of $\genQ_{\sL}$, $\genQ_{\sR}$ 
and $\genS_{\sL}$, $\genS_{\sR}$ is exchanged, and the right-movers 
$\ket{{\phi}_p^{\sR}}$ and $\ket{{\psi}_p^{\sR}}$ transform according to
\begin{equation}\label{alg:antichiral-rep}
  \boxed{\varrho_{\sR}:} \qquad\quad
  \begin{aligned}
    \genQ_{\sR} \ket{{\phi}_p^{\sR}} &= a_p \ket{{\psi}_p^{\sR}}, \qquad &
    \genQ_{\sR} \ket{{\psi}_p^{\sR}} &= 0 , \\
    \fixedspaceL{\genQ_{\sR}}{\genS_{\sR}} \ket{{\phi}_p^{\sR}} &= 0 , \qquad &
    \fixedspaceL{\genQ_{\sR}}{\genS_{\sR}} \ket{{\psi}_p^{\sR}} &= b_p \ket{{\phi}_p^{\sR}} , \\
    \genQ_{\sL} \ket{{\phi}_p^{\sR}} &= 0 , \qquad &
    \genQ_{\sL} \ket{{\psi}_p^{\sR}} &= c_p \ket{{\phi}_p^{\sR}} , \\
    \fixedspaceL{\genQ_{\sL}}{\genS_{\sL}} \ket{{\phi}_p^{\sR}} &= d_p \ket{{\psi}_p^{\sR} } , \qquad &
    \fixedspaceL{\genQ_{\sL}}{\genS_{\sL}} \ket{{\phi}_p^{\sR}} &= 0 ,
  \end{aligned}
  \qquad \qquad \quad
\end{equation}
with the central charges acting as
\begin{equation}
  \begin{aligned}
    \genH_{\sL} \ket{{\phi}_p^{\sR}} &= c_p d_p \ket{{\phi}_p^{\sR}} , \quad &
    \genC \ket{{\phi}_p^{\sR}} &= a_p c_p \ket{{\phi}_p^{\sR}} , \\
    \genH_{\sR} \ket{{\phi}_p^{\sR}} &= a_p b_p \ket{{\phi}_p^{\sR}} , \quad &
    \genCbar \ket{{\phi}_p^{\sR}} &= b_p d_p \ket{{\phi}_p^{\sR}} .
  \end{aligned}
\end{equation}

\paragraph*{Shortening condition.}
The highest weight state, which is $\ket{{\psi}_p^{\sR}}$ in this case, 
again satisfies the condition
\begin{equation}
  (\genH_{\sR} \genQ_{\sL} - \genC \genS_{\sR}) \ket{{\psi}_p^{\sR}} = 0,
\end{equation}
and the representation is short. The state $\ket{{\psi}_p^{\sR}}$ must 
also be annihilated by the anticommutator $\acomm{\genS_{\sL}}
{\genH_{\sR} \genQ_{\sL} - \genC \genS_{\sR}}  
= \genH_{\sL} \genH_{\sR} - \genC\genCbar$, and we have the same 
shortening condition in terms of the central charges as for the left-movers
\begin{equation}\label{alg:shortening-right-su(1|1)^2}
  (\genH_{\sL} \genH_{\sR} - \genC\genCbar) \ket{{\chi}_p^{\sR}} = 0 
\end{equation}
for all states ${\chi}_p^{\sR} = {\phi}_p^{\sR} , {\psi}_p^{\sR}$.

\section{The \texorpdfstring{$\psu(1|1)^4_\text{c.e.}$}{psu(1|1)**4 c.e.} algebra and its representations}
\label{App:T4Alg}

The centrally extended $\psu(1|1)^4$ superalgebra is of particular interest 
to us, as it is the symmetry algebra of the ground state in the instance of 
$\AdS_3 / \CFT_2$ duality where the string background is \AdsST 
\cite{Borsato:2014hja}. As it already appeared in the study of the massive 
sector of the theory in \cite{Borsato:2013qpa}, the algebra can be obtained 
from two copies of the centrally extended $\su(1|1)^2$. In this section we 
briefly review this construction.

\subsection{From \texorpdfstring{$\ce{\su(1|1)^2}$}{su(1|1)**2 c.e.} to \texorpdfstring{$\ce{\psu(1|1)^4}$}{psu(1|1)**4 c.e.}}

If we take two copies of the $\ce{\su(1|1)^2}$ algebra 
\eqref{alg:su(1|1)^2ceAlg} that share the four central charges, we get the 
centrally extended $\psu(1|1)^4$ with generators
\begin{equation}
  \bigl\{ 	\genQ_{\sL}^{\ \dot{a}},\genS_{\sL \dot{a}},
  			\genQ_{\sR \dot{a}}, \genS_{\sR}^{\ \dot{a}},
  			\genH_{\sL},\genH_{\sR}, \genC, \genCbar \bigr\}_{\dot{a} = 1, 2} 
\end{equation}
satisfying the anticommutation relations
\begin{equation}\label{alg:psu(1|1)^4ceAlg}
  \begin{aligned}
    \acomm{\genQ_{\sL}^{\ \dot{a}}}{\genS_{\sL \dot{b}}} &= \delta^{\dot{a}}_{\ \dot{b}}\ \genH_{\sL} \ , \qquad &
    \acomm{\genQ_{\sL}^{\ \dot{a}}}{\genQ_{\sR \dot{b}}} &= \delta^{\dot{a}}_{\ \dot{b}}\ \genC \ ,  \\
    \acomm{\genQ_{\sR \dot{a}}}{\genS_{\sR}^{\ \dot{b}}} &= \delta_{\dot{a}}^{\ \dot{b}}\ \genH_{\sR} \ , &
    \acomm{\fixedspaceL{\genQ_{\sL}^{\ \dot{a}}}{\genS_{\sL \dot{a}}}}
    	  {\fixedspaceL{\genQ_{\sR \dot{a}}}{\genS_{\sR}^{\ \dot{b}}}} &= \delta_{\dot{a}}^{\ \dot{b}}\ \genCbar  \ .
  \end{aligned}
\end{equation}
In other words, we are dealing with
\begin{equation}
   \psu(1|1)^4 \ltimes \algU(1)^4 \ .
\end{equation}
Equivalently, we can consider a tensor product of two copies of \eqref{alg:su(1|1)^2ceAlg}
\begin{equation}\label{eq:supercharges-tensor-product}
\begin{aligned}
\genQ_{\sL}^{\ 1}	= \genQ_{\sL} \otimes \1 \ , \qquad &
\genS_{\sL 1}    	= \genS_{\sL} \otimes \1 \ , \qquad &
\genQ_{\sL}^{\ 2}	= \1 \otimes \genQ_{\sL} \ , \qquad &
\genS_{\sL 2}    	= \1 \otimes \genS_{\sL} \ ,
\\[0.5em]
\genQ_{\sR 1}			= \genQ_{\sR} \otimes \1 \ , \qquad &
\genS_{\sR}^{\ 1}	= \genS_{\sR} \otimes \1 \ , \qquad &
\genQ_{\sR 2}			= \1 \otimes \genQ_{\sR} \ , \qquad &
\genS_{\sR}^{\ 2}	= \1 \otimes \genS_{\sR} \ ,
\end{aligned}
\end{equation}
also for the central elements
\begin{equation}\label{eq:supercharges-tensor-product-C}
\begin{aligned}
\genH_{\sL}^{\ 1}	= \genH_{\sL} \otimes \1 \ , \qquad &
\genH_{\sL}^{\ 2}	= \1 \otimes \genH_{\sL} \ , \qquad &
\genC^{\ 1}	= \genC \otimes \1 \ , \qquad &
\genC^{\ 2}	= \1 \otimes \genC \ , 
\\[0.5em]
\genH_{\sR}^{\ 1}	= \genH_{\sR} \otimes \1 \ , \qquad &
\genH_{\sR}^{\ 2}	= \1 \otimes \genH_{\sR} \ , \qquad &
\genCbar^{\ 1}	= \genCbar \otimes \1 \ , \qquad &
\genCbar^{\ 2}	= \1 \otimes \genCbar \ . 
\end{aligned}
\end{equation}
After identifying the central charges as
\begin{equation}
\gen{H}_{\sL}^{\ 1}=\gen{H}_{\sL}^{\ 2},
\quad
\gen{H}_{\sR}^{\ 1}=\gen{H}_{\sR}^{\ 2},
\quad
\gen{C}^1=\gen{C}^2,
\quad
\overline{\gen{C}}{}^1=\overline{\gen{C}}{}^2,
\end{equation}
and consequently dropping the indices $1, 2$, we are left with $\ce{\psu(1|1)^4}$.
Looking at the algebra this way will be helpful in constructing its 
short representations.

\subsection{Bi-fundamental representations}

It was shown, first for the spin-chain and later for the \AdsSSS superstring, 
that the massive off-shell excitations in both of the left- and right-moving
sectors transform in \textit{short} (four-dimensional) bi-fundamental 
representations of the centrally extended $\psu(1|1)^4$. That is, we can 
obtain the relevant representations by tensoring the fundamental 
representations $\varrho_{\sL}$ \eqref{alg:chiral-rep} and $\varrho_{\sR}$ 
\eqref{alg:antichiral-rep} of $\ce{\su(1|1)^2}$.

\paragraph*{Left module.} Borrowing notation from \cite{Borsato:2014hja}, 
the four left-movers can be written as
\begin{equation}
	Y^{\sL} 		= \phi^{\sL} \otimes \phi^{\sL} \ , \quad
    \eta^{\sL 1} 	= \psi^{\sL} \otimes \phi^{\sL} \ , \quad
    \eta^{\sL 2} 	= \phi^{\sL} \otimes \psi^{\sL} \ , \quad
    Z^{\sL} 		= \psi^{\sL} \otimes \psi^{\sL} \ ,
\end{equation}
and they transform under the tensor product of two left-moving 
representations $\varrho_{\sL}$:
\begin{equation}\label{alg:repr-double-left}
  \boxed{\varrho_{\sL}\otimes\varrho_{\sL}:} \qquad \quad
  \begin{aligned}
	\gen{Q}_{\sL}^{\ \dot{a}} \ket{Y_p^{\sL}} 
    	&= a_p \ket{\eta^{\sL \dot{a}}_p} \ , \qquad
    &\gen{Q}_{\sL}^{\ \dot{a}} \ket{\eta^{\sL \dot{b}}_p} 
    	&= \epsilon^{\dot{a}\dot{b}} \, a_p \ket{Z_p^{\sL}} \ , 
    \\
    \genS_{\sL \dot{a}} \ket{Z_p^{\sL}} 
    	&=  - \epsilon_{\dot{a}\dot{b}}  \, b_p \ket{\eta^{\sL \dot{b}}_p} \ , \qquad
    &\genS_{\sL \dot{a}} \ket{\eta^{\sL \dot{b}}_p}
    	& =  \delta_{\dot{a}}^{\ \dot{b}}  \, b_p \ket{Y_p^{\sL}} \ , 
    \\
    \gen{Q}_{\sR \dot{a}} \ket{Z^{\sL}_p} 
    	&= - \epsilon_{\dot{a}\dot{b}} \,  c_p \ket{\eta^{\sL \dot{b}}_p} \ , \qquad
    &\gen{Q}_{\sR \dot{a}} \ket{\eta^{\sL \dot{b}}_p} 
    	&= \delta_{\dot{a}}^{\ \dot{b}} \, c_p \ket{Y^{\sL}_p} \ ,
    \\
    \genS_{\sR}^{\ \dot{a}} \ket{Y^{\sL}_p} 
    	&= d_p \ket{\eta^{\sL \dot{a}}_p} \ , \qquad
    &\genS_{\sR}^{\ \dot{a}} \ket{\eta^{\sL \dot{b}}_p} 
    	&= \epsilon^{\dot{a}\dot{b}} \,  d_p \ket{Z^{\sL}_p} \ .
  \end{aligned}
  \qquad \qquad 
\end{equation}
The representation coefficients of the two $\varrho_{\sL}$ must match, 
since the central charges are shared, and we get a minus sign when charges 
of the second type act on a state with a fermion in the first part of the tensor 
product. Each central charge acts uniformly across all states
\begin{equation}
  \begin{aligned}
    \genH_{\sL} \ket{{\chi }^{\sL}} &= a_p b_p \ket{{\chi }^{\sL}} , \qquad\qquad &
    \genC \ket{{\chi }^{\sL}} &= a_p c_p \ket{{\chi }^{\sL}} , \\
    \genH_{\sR} \ket{{\chi }^{\sL}} &= c_p d_p \ket{{\chi }^{\sL}} , \qquad\qquad &
    \genCbar \ket{{\chi }^{\sL}} &= b_p d_p \ket{{\chi }^{\sL} } . 
  \end{aligned}
\end{equation}

\paragraph*{Right module.} Similarly we can introduce the right-moving excitations
\begin{equation}
    Y^{\sR} 			= {\phi}^{\sR} \otimes {\phi}^{\sR} \ , \quad
    \eta^{\sR}_{\ 1} 	= {\psi}^{\sR} \otimes {\phi}^{\sR} \ , \quad
    \eta^{\sR}_{\ 2} 	= {\phi}^{\sR} \otimes {\psi}^{\sR} \ , \quad
    Z^{\sR} 			= {\psi}^{\sR} \otimes {\psi}^{\sR} \ ,
\end{equation}
and these will transform in the representation
\begin{equation}\label{alg:repr-double-right}
  \boxed{\varrho_{\sR}\otimes\varrho_{\sR}:} \qquad \quad
  \begin{aligned}
	\gen{Q}_{\sR \dot{a}} \ket{Y_p^{\sR}} 
		&=  \epsilon_{\dot{a}\dot{b}} \,  a_p \ket{\eta^{\sR \dot{b}}_p} \ ,  \qquad
    &\gen{Q}_{\sR \dot{a}} \ket{\eta^{\sR \dot{b}}_p} 
    	&= \delta_{\dot{a}}^{\ \dot{b}} \,  a_p \ket{Z_p^{\sR}} \ , 
    \\
    \genS_{\sR}^{\ \dot{a}} \ket{Z_p^{\sR}} 
    	&= b_p \ket{\eta^{\sR \dot{a}}_p} \ ,  \qquad
    &\genS_{\sR}^{\ \dot{a}} \ket{\eta^{\sR \dot{b}}_p} 
    	&= - \epsilon^{\dot{a}\dot{b}}  \, b_p \ket{Y_p^{\sR}} \ ,
    \\
    \gen{Q}_{\sL}^{\ \dot{a}} \ket{Z_p^{\sR}} 
    	&=  c_p \ket{\eta^{\sR \dot{a}}_p} \ , \qquad
    &\gen{Q}_{\sL}^{\ \dot{a}} \ket{\eta^{\sR \dot{b}}_p} 
    	&=- \epsilon^{\dot{a}\dot{b}} \,  c_p \ket{Y_p^{\sR}} \ ,
    \\
    \genS_{\sL \dot{a}} \ket{Y_p^{\sR}} 
    	&= \epsilon_{\dot{a}\dot{b}} \, d_p \ket{\eta^{\sR \dot{b}}_p} \ , \qquad
    &\genS_{\sL \dot{a}} \ket{\eta^{\sR \dot{b}}_p} 
    	&= \delta_{\dot{a}}^{\ \dot{b}} \,  d_p \ket{Z_p^{\sR}} \ ,
  \end{aligned}
  \qquad \qquad 
\end{equation}
and for all right-movers
\begin{equation}
  \begin{aligned}
    \genH_{\sL} \ket{{\chi }^{\sR}} &= c_p d_p \ket{{\chi }^{\sR}} , \qquad\qquad &
    \genC \ket{{\chi }^{\sR}} &= a_p c_p \ket{{\chi }^{\sR}} , \\
    \genH_{\sR} \ket{{\chi }^{\sR}} &= a_p b_p \ket{{\chi }^{\sR}} , \qquad\qquad &
    \genCbar \ket{{\chi }^{\sR}} &= b_p d_p \ket{{\chi }^{\sR}} .
  \end{aligned}
\end{equation}

\paragraph*{Shortening condition.} Naturally extending the choice made for 
$\su(1|1)^2$, we take $\genS_{\sL \dot{a}}$ and $\genQ_{\sR \dot{a}}$ 
as our raising operators, while $\gen{Q}_{\sL}^{\ \dot{a}}$ and 
$\genS_{\sR}^{\ \dot{a}}$ will be lowering operators. The highest weight 
states for $\varrho_{\sL}\otimes\varrho_{\sL}$ and $\varrho_{\sR}\otimes\varrho_{\sR}$ 
are $\ket{Y_p^{\sL}}$ and $\ket{Z_p^{\sR}}$ respectively, but they are 
also annihilated by two combinations of lowering operators, as should be 
the case for short representations
\begin{equation}
\begin{aligned}
  (\genH_{\sR} \gen{Q}_{\sL}^{\ \dot{a}} - \genC \genS_{\sR}^{\ \dot{a}}) 
  	\ket{Y_p^{\sL}} &= (a_p c_p d_p - a_p c_p d_p) \ket{\eta^{\sR \dot{a}}_p}= 0 \ ,
  \\[1em]
  (\genH_{\sR} \gen{Q}_{\sL}^{\ \dot{a}} - \genC \genS_{\sR}^{\ \dot{a}}) 
  	\ket{Z_p^{\sR}} &= (a_p b_p c_p - a_p b_p c_p)\, \ket{\eta^{\sR \dot{a}}_p}=0 \ .
\end{aligned}
\end{equation}
Similarly to the case of $\ce{\su(1|1)^2}$, the anticommutator of this 
with $\genS_{\sL \dot{b}}$ still annihilates the highest weight states, 
and in fact any state across both sectors, since it is a central element of 
the algebra
\begin{equation}\label{alg:shortening_psu(1|1)^4}
  (\genH_{\sL} \genH_{\sR} - \genC\genCbar) \ket{\chi_p^{\sL,\sR}} = 0.
\end{equation}
Note that this is the same as \eqref{alg:shortening-left-su(1|1)^2} 
and \eqref{alg:shortening-right-su(1|1)^2}.

\section{\texorpdfstring{$\SU (2)$}{SU(2)} currents for the \texorpdfstring{$q=1$}{q=1} giant magnon}
\label{FZM:AppQ1Currents}

Using the usual $\SU (2)$ embedding \eqref{FZM:SU2explicit}, it is a relatively 
simple exercise to derive the left- and right-currents for the $q=1$ giant magnon
\eqref{FZM:Q1magnon}:
\begin{align}
\begin{split}
	& \Jfrak_+ = 
	\begin{pmatrix}
	    i a     &  b  \\
	    -b^*  & -i a 
	\end{pmatrix}\ ,
\quad
	\Jfrak_- = 
	\begin{pmatrix}
	    i c    &  d  \\
	    -d^*  & -i c 
	\end{pmatrix}\ ,
\\[1em]
	& \kfrak_+ = 
	\begin{pmatrix}
	    i e     &  f  \\
	    -f^*  & -i e 
	\end{pmatrix}\ ,
\quad
	\kfrak_- = 
	\begin{pmatrix}
	    i   &  0  \\
	    0  &  -i 
	\end{pmatrix}\ ,
\end{split}
\end{align}
where
\begin{align}
\begin{split}
&a = 1- 2 \beta^2  \sin^2\!\tfrac{p}{2}\, \sech^2\!\cY \ ,
\\[0.5em]
&b = 2 i  \beta  \sin^2\!\tfrac{p}{2}\, \sech\cY 
				(\sec\rho -\beta\tan \rho - i \beta \tanh\cY)
				e^{-2 i (1-\beta  \sin\rho) x^+} \ ,
\\[0.5em]
&c = 1- 2  \sin^2\!\tfrac{p}{2}\, \sech^2\!\cY\ ,
\\[0.5em]
&d = 2 i \sin\tfrac{p}{2}\, \sech\cY 
		\sqrt{1 -  \sin^2\!\tfrac{p}{2}\, \sech^2\!\cY}
		e^{-2 i (1-\beta  \sin\rho) x^+ - i \arctan\left(\tan\tfrac{p}{2} \tanh\cY \right)}\ ,
\\[0.5em]
&e = 1- 2  \cos^2\!\rho\, \sech^2\!\cY\ ,
\\[0.5em]
&f = 2 \cos^2\!\rho\, \sech\cY (\tanh\cY - i \tan\rho)
		 e^{2 i \left( \beta \sin\rho x^+  +  x^- \right)}\ .
\end{split}
\end{align}

\section{Terms appearing in the \texorpdfstring{$q=1$}{q=1} fermion equations}
\label{FZM:AppQ1}

With the bosonic solution from section \ref{FZM:SecQ1} as background, the 
following are the components of the pulled-back vielbein $e_a^A = E_\mu ^A (X)\p_a X^\mu$
\begin{align}
	e_0^0 &= 1 \ , 
	&  
	e_1^0 &= 0 \ ,
\\[0.5em]
	e_0^3 &= \frac{\beta \cos\rho\, \tanh\cY}{\sqrt{b^2 + (1+b^2)\sinh^2\!\cY}} \ ,
	&  
	e_1^3 &=  \frac{\beta \cos\rho\, \tanh\cY}{\sqrt{b^2 + (1+b^2)\sinh^2\!\cY}} \ ,
\\[0.5em]
	e_0^4 &=   \frac{b \beta \cos\rho + b^2 + (1+b^2)\sinh^2\!\cY}{\sqrt{b^2 + (1+b^2)\sinh^2\!\cY}}\,
						\frac{\sech\cY}{\sqrt{1+b^2}} \ ,
	&
	e_1^4 &=   \frac{b \beta \cos\rho}{\sqrt{b^2 + (1+b^2)\sinh^2\!\cY}}\,
						\frac{\sech\cY}{\sqrt{1+b^2}} \ ,
\\[0.5em]
	e_0^5 &=     \beta\sin\rho       \frac{ \sech\cY}{\sqrt{1+b^2}} \ ,
	& 
	e_1^5  &=   ( \beta\sin\rho - 1)\frac{ \sech\cY}{\sqrt{1+b^2}} \ ,
\end{align}
while the only non-zero components of the spin connection (pulled back to the worldsheet) are
\begin{align}
\omega_0^{34}    &=  - \omega_0^{43}   
								=   - \frac{b \beta \cos\rho + b^2 + (1+b^2)\sinh^2\!\cY}{b^2 + (1+b^2)\sinh^2\!\cY}\,
										\frac{\sech\cY}{\sqrt{1+b^2}} \ , 
\\[1em]
\omega_1^{34}    &=  - \omega_1^{43}   
								=  - \frac{b \beta \cos\rho}{b^2 + (1+b^2)\sinh^2\!\cY}\,
									 	\frac{\sech\cY}{\sqrt{1+b^2}} \ ,
\\[1em]
\omega_0^{35}    &=  - \omega_0^{53}   
								=   \beta\sin\rho       \sqrt{b^2 + (1+b^2)\sinh^2\!\cY}\, \frac{ \sech\cY}{\sqrt{1+b^2}}  \  ,
\\[1em]
\omega_1^{35}    &=  - \omega_1^{53}   
								=   (\beta\sin\rho - 1)\sqrt{b^2 + (1+b^2)\sinh^2\!\cY}\, \frac{ \sech\cY}{\sqrt{1+b^2}}  \  .
\end{align} 
Note that $\cY = 2 \beta \cos\rho \, x^{+}$ and the three parameters are related by $\beta = -(b \cos\rho - \sin\rho)$.
The combinations appearing in the fermion derivatives are
\begin{align}
G            &= \omega_1^{34} - \omega_0^{34} 
				 = \frac{ \sech\cY}{\sqrt{1+b^2}}  \ ,   
\\[1em]     
Q           &= \omega_1^{35} - \omega_0^{35} 
				= - \sqrt{b^2 + (1+b^2)\sinh^2\!\cY} \frac{ \sech\cY}{\sqrt{1+b^2}} \ ,
\\[1em]
\tilde{G} &= \omega_1^{34} + \omega_0^{34} 
				= - \frac{2 b \beta \cos\rho + b^2 + (1+b^2)\sinh^2\!\cY}{b^2 + (1+b^2)\sinh^2\!\cY}\,
										\frac{\sech\cY}{\sqrt{1+b^2}} \ , 
\\[1em]										
\tilde{Q} &= \omega_1^{35} + \omega_0^{35} 
				=  (2\beta\sin\rho - 1)\sqrt{b^2 + (1+b^2)\sinh^2\!\cY}\, \frac{ \sech\cY}{\sqrt{1+b^2}}  \  .
\end{align}


\bibliographystyle{nb}
\bibliography{./bibliography} 

\end{document}